\documentclass{aa} 
\usepackage{graphicx}

\usepackage{savesym}
\usepackage{txfonts}
\savesymbol{iint}
\savesymbol{iiint}
\savesymbol{iiiint}
\savesymbol{idotsint}
\usepackage{amsmath}
\restoresymbol{AMS}{iint}
\restoresymbol{AMS}{iiint}
\restoresymbol{AMS}{iiiint}
\restoresymbol{AMS}{idotsint}

\newcommand*\diff{\mathop{}\!\mathrm{d}}

\usepackage{bbm}
\usepackage{aas_macros}
\usepackage{dsfont}
\usepackage{fixltx2e}

\usepackage{natbib,twoopt}
\usepackage[breaklinks=true]{hyperref} 
\bibpunct{(}{)}{;}{a}{}{,} 
\makeatletter
\newcommandtwoopt{\citeads}[3][][]{\href{http://adsabs.harvard.edu/abs/#3}%
{\def\hyper@linkstart##1##2{}%
\let\hyper@linkend\@empty\citealp[#1][#2]{#3}}}
\newcommandtwoopt{\citepads}[3][][]{\href{http://adsabs.harvard.edu/abs/#3}%
{\def\hyper@linkstart##1##2{}%
\let\hyper@linkend\@empty\citep[#1][#2]{#3}}}
\newcommandtwoopt{\citetads}[3][][]{\href{http://adsabs.harvard.edu/abs/#3}%
{\def\hyper@linkstart##1##2{}%
\let\hyper@linkend\@empty\citet[#1][#2]{#3}}}
\newcommandtwoopt{\citeyearads}[3][][]%
{\href{http://adsabs.harvard.edu/abs/#3}
{\def\hyper@linkstart##1##2{}%
\let\hyper@linkend\@empty\citeyear[#1][#2]{#3}}}
\makeatother
\usepackage{color}

\begin{document}

\title{The Effelsberg--Bonn \ion{H}{i} Survey: Milky Way gas.}
\subtitle{First data release}

\author{B. Winkel\inst{1,2}
        \and
        J. Kerp\inst{1}
        \and
        L. Fl\"{o}er\inst{1}
        \and
        P.~M.~W. Kalberla\inst{1}
        \and
        N. Ben Bekhti\inst{1}
        \and
        R. Keller\inst{2}
        \and
        D. Lenz\inst{1}
       }

\institute{Argelander-Institut f\"{u}r Astronomie (AIfA),
              Auf dem H\"{u}gel\,71, 53121 Bonn, Germany
            \and
            Max-Planck-Institut f\"{u}r Radioastronomie (MPIfR),
              Auf dem H\"{u}gel\,69, 53121 Bonn, Germany\\
              \email{bwinkel@mpifr.de}
}

\date{Received July 20, 2015; accepted October 26, 2015}
\abstract {The Effelsberg--Bonn \ion{H}{i} Survey (EBHIS) is a new 21-cm survey performed with the 100-m telescope at Effelsberg. It covers the whole northern sky out to a redshift of $z\sim0.07$ and comprises \ion{H}{i} line emission from the Milky Way and the Local Volume.} {We aim to substitute the northern-hemisphere part of the Leiden/Argentine/Bonn Milky Way \ion{H}{i} survey (LAB) with this first EBHIS data release\thanks{EBHIS data sets are available in electronic form (FITS) at the CDS via anonymous ftp to cdsarc.u-strasbg.fr (130.79.128.5) or via\newline\url{http://cdsweb.u-strasbg.fr/cgi-bin/qcat?J/A+A/585/A41}}, which presents the \ion{H}{i} gas in the Milky Way regime.} {The use of a seven-beam L-band array made it feasible to perform this all-sky survey with a 100-m class telescope in a reasonable amount of observing time. State-of-the-art fast-Fourier-transform spectrometers provide the necessary data read-out speed, dynamic range, and spectral resolution to apply software radio-frequency interference mitigation. EBHIS is corrected for stray radiation and employs frequency-dependent flux-density calibration and sophisticated baseline-removal techniques to ensure the highest possible data quality.} {Detailed analyses of the resulting data products show that EBHIS is not only outperforming LAB in terms of sensitivity and angular resolution, but also matches the intensity-scale of LAB extremely well, allowing EBHIS to be used as a drop-in replacement for LAB. Data products are made available to the public in a variety of forms. Most important, we provide a properly gridded Milky Way \ion{H}{i} column density map in HEALPix representation. To maximize the usefulness of EBHIS data, we estimate uncertainties in the \ion{H}{i} column density and brightness temperature distributions, accounting for systematic effects.}{}


\maketitle

\section{Introduction}

In 2009 we initiated a new northern-hemisphere \ion{H}{i} survey with the 100-m telescope at Effelsberg, Germany, to succeed the Leiden/Dwingeloo Survey \citep[LDS;][]{hartmann97} done with the 25-m Dwingeloo telescope. Outstanding in terms of sensitivity and sky coverage compared to any prior endeavor, the LDS was later merged with the Instituto Argentino de Radioastronom\'{ı}a Survey \citep[IAR;][for $\delta<-27\fdg5$]{arnal00,bajaja05} to form the Leiden/Argentine/Bonn Survey \citep[LAB;][]{bajaja85,kalberla05} -- the first full-sky Milky Way \ion{H}{i} survey that was corrected for stray radiation (SR). The high-quality SR correction made the LAB survey one of the most important \ion{H}{i} data bases to date. The references to the seminal article by \citet{kalberla05} reveal the tremendous legacy value of the LAB: one of its most frequent uses is to correct high-energy observations for galactic foreground extinction. However, the coarse angular resolution of LAB may cause significant uncertainties when it is applied to evaluating the intensity attenuation of unresolved sources.

With the development of 21-cm multibeam receivers in the late 1990s, it finally became feasible for the 100-m class single-dish observatories to survey significant portions of the sky with the resulting better angular resolution in reasonable amounts of observing time. The first such project was the very successful \ion{H}{i} Parkes All-Sky Survey \citep[HIPASS,][]{barnes01}, which mapped the southern sky out to a radial velocity of $\sim$$13\,000~\mathrm{km\,s}^{-1}$ with the 64-m Parkes telescope. Limitations in the spectrometer meant that the velocity resolution is only $18~\mathrm{km\,s}^{-1}$, so too coarse to be useful for many studies of the Milky Way (MW) and its halo. Therefore, a second large-area Parkes \ion{H}{i} survey was initiated in 2005: the Galactic All-Sky Survey \citep[GASS;][]{mcclure09,kalberla10,kalberla15}. It recorded data with narrow bandwidth, hence higher spectral resolution.

Another recent project is the GALFA-HI survey \citep{peek11} currently going on with the Arecibo 300-m dish. The sheer size of the telescope makes GALFA-HI unique in terms of sensitivity. For the Galactic velocity regime, a dedicated backend is delivering spectral resolution up to $0.2~\mathrm{km\,s}^{-1}$. The downside is the limited area on the sky that is accessible to the Arecibo telescope.

In late 2008 a seven-beam 1.4-GHz (L-band) receiver was installed at the 100-m telescope at Effelsberg.
Attached to the seven-feed array are state-of-the-art digital FFT-type spectrometers \citep[FFTS,][]{stanko05,klein12}, allowing to observe Galactic and extra-galactic \ion{H}{i} simultaneously for the first time with sufficient spectral resolution. The FFTS not only offers great spectral resolution ($\sim$$1~\mathrm{km\,s}^{-1}$) but also allows high-speed ($\sim$1~s) storage of spectra. Both properties are beneficial for removing time-dependent radio frequency interference (RFI) from the data during post-processing.

\begin{figure*}[!t]
\centering%
\includegraphics[width=0.95\textwidth,viewport=40 50 1040 1050,clip=]{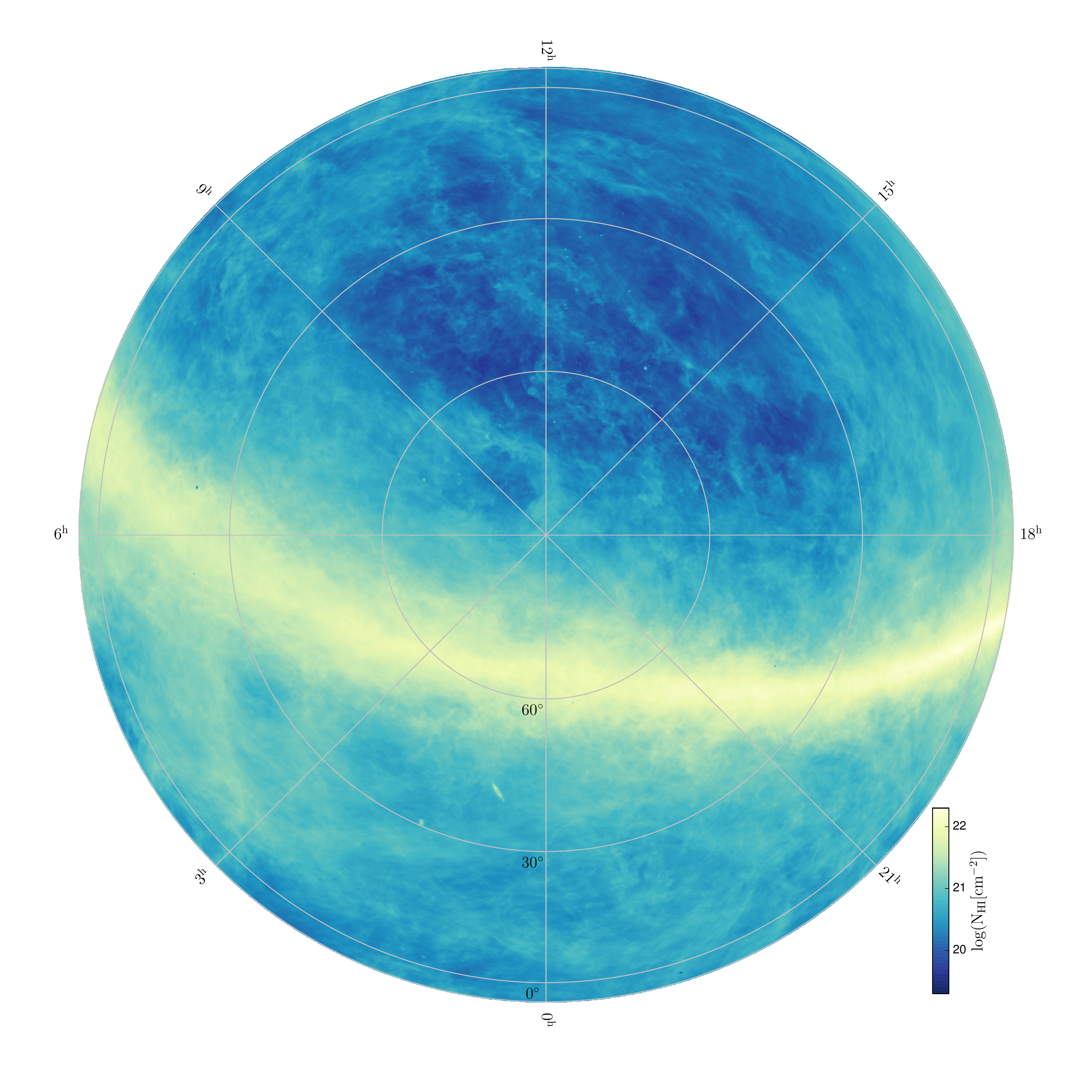}\\[0ex]
\caption{EBHIS \ion{H}{i} column density map, as integrated over the velocity interval $-600\leq v_\mathrm{lsr}\leq600~\mathrm{km\,s}^{-1}$.}%
\label{fig:allsky_nhi}%
\end{figure*}

With this new receiving system, we conducted a major \ion{H}{i} survey of the sky north of declination $\delta\gtrsim-5\degr$, called Effelsberg--Bonn \ion{H}{i} Survey (EBHIS). To allow for an early first data release, survey observations were divided into two runs, each mapping the northern sky completely with a net integration time of about 35~s per beam. In April 2013 we finished first-coverage observations. This data has now been processed and is made available to the scientific community. Meanwhile, measurements for the second coverage are ongoing. Once finished, EBHIS sensitivity will improve by about $30\%$.

The main science goals of EBHIS have already been discussed in detail in \citet{kerp11}. Here, we concentrate on assessing the quality of the final data products and on the way these will be made available to the astronomical community.

In Section~\ref{sec:surveydescription} we briefly recapitulate the properties of EBHIS. The data processing pipeline was explained in detail in \citet{winkel10}; however, several steps were further improved, such as RFI flagging, baseline-fitting, and correction for stray radiation, which we present in Section~\ref{sec:datareduction}. Section~\ref{sec:dataquality} comprises the data quality evaluation of EBHIS. We compare EBHIS with the two recent Galactic \ion{H}{i} \textit{all-sky} surveys, LAB and GASS, and discuss current limitations of the data. In Section~\ref{sec:uncertainties} we study ensemble uncertainties of the observed column density and brightness temperature distributions. Section~\ref{sec:dataproducts} explains how EBHIS data are made accessible to the astronomical community, most importantly the all-sky \ion{H}{i} column density map (Fig.~\ref{fig:allsky_nhi}). We conclude with a summary and outlook.

\section{Survey description}\label{sec:surveydescription}

Observations of the first coverage were performed in on-the-fly mapping mode, scanning in right ascension, $\alpha$, along lines of constant declination, $\delta$.
Since the parallactic angle changes during each scan line, the seven-feed array\footnote{One dual-circular central feed and six dual-linear off-axis feeds in a hexagonal layout. The separation of the off-axis feeds from the optical axis is $15'$.} needs to be rotated accordingly to ensure a regular scan pattern of the offset feeds. Furthermore, the feed array is rotated by an additional 19\fdg1 with respect to the scanning direction such that the scan tracks of the individual feeds have equal separations.

The sky north of $\delta>-5\degr$ was divided into $25~\deg^2$ sections for a total of 915 individual maps. The polar cap $\delta\ga85\degr$ was observed in the Galactic coordinate frame, because near the equatorial pole, the Az--El mount of the 100-m telescope would significantly affect the possible mapping speed in the equatorial system. Hereafter we follow the Effelsberg observatory parlance, where one observation/map is called `scan', consisting of several `subscans' (the individual scan lines).

Strong observing time constraints were applied for each field:

\begin{enumerate}
\item The southeastern quadrant is preferred in azimuth to maximize the angular distance to a mobile-communication broadcast tower in the vicinity of the telescope (at $\mathrm{Az}\approx290\degr$).
\item We need to avoid large differential changes of parallactic angle during a scan line because the rotation speed of the receiver box at the 100-m dish is limited.
\item The angular distance to the Sun has to be kept as large as possible during daytime observations to avoid ``solar ripples'' in the spectral baselines \citep[see also][and Section~\ref{subsubsec:solar_ripples}]{barnes05}. These are most likely caused by reflections of solar emission off the telescope structure. The exact paths of reflection are difficult to predict, and it is therefore nearly impossible to avoid solar interference by means of observation scheduling. Because of that, several fields had to be observed again for our survey.
\end{enumerate}

The FFT spectrometers allow storing a full spectrum every 500~ms, which is an advantage to identifying time-variable RFI. Fast read-out is also beneficial, because the dump time restricts the maximum mapping speed because each independent beam area on the sky needs to be sampled with at least two data points along the scanning direction \citep{shannon49}. To allow for a reasonably fast completion of the first sky coverage, we chose a relatively high scan speed of $240\arcsec$ per second, pushing the azimuthal engines of the 100-m almost to their limits for sources above an elevation of $60\degr$. We spread the observed bandwidth of 100 MHz over 16384 channels, yielding a spectral resolution of 6.1 kHz. For the Milky Way data release, only the velocity range of $\vert v_\mathrm{lsr}\vert\leq600~\mathrm{km\,s}^{-1}$ is considered, which is about the frequency interval from 1417.5~MHz to 1423.5~MHz.

After gridding, the median noise level in the data cubes is approximately $90~\mathrm{mK}$ but differs significantly from field to field (see Section~\ref{subsec:noisemap}). This is mainly a consequence of different elevation angles during the observations causing different levels of stray radiation from the ground and atmosphere. A second important effect is the seasonal changes in ambient temperature \citep[compare][]{winkel10}.

\begin{table}
\caption{Comparison of basic parameters of recent \ion{H}{i} surveys of the Milky Way. The table quotes the declination range,  $\delta$, angular resolution $\vartheta_\mathrm{fwhm}$, velocity interval, $v_\mathrm{lsr}$, channel separation, $\Delta v$, spectral resolution, $\delta v$, brightness temperature noise level, $\sigma_\mathrm{rms}$, as well as the normalized noise level, $\sigma_\mathrm{rms}^\mathrm{norm}$, the data would have at a common spectral resolution of $1~\mathrm{km\,s}^{-1}$. The main beam sensitivity, $\Gamma_\mathrm{mb}$, is the conversion factor between brightness temperature and flux density, i.e., $T_\mathrm{B}=\Gamma_\mathrm{mb}S$. The two bottom rows quote theoretical $5\sigma$ detection limits (velocity-integrated intensity) of an object with a Gaussian profile of $20~\mathrm{km\,s}^{-1}$ line width (FWHM).}
\label{tab:survey_comparison}
\centering
\begin{tabular}{l c c c c l}
\hline\hline
\rule{0ex}{3ex} & LAB & GASS & GALFA$^\ast$ & EBHIS \\
\hline
\rule{0ex}{3ex}$\delta$ & Full & $\leq1\degr$ & $-1\degr\ldots38\degr$ & $\geq-5\degr$ \\
$\vartheta_\mathrm{fwhm}$ & $36\arcmin$ & $16\farcm1$ & $4\farcm0$ & $10\farcm8$ \\
$\vert v_\mathrm{lsr}\vert$ & $\leq460^\mathrm{\dagger}$ & $\leq470$     & $\leq750$ & $\leq600$ & $\mathrm{km\,s}^{-1}$\\
$\Delta v$ & $1.03$ & $0.82$    & $0.18$ & $1.29$ & $\mathrm{km\,s}^{-1}$\\
$\delta v$ & $1.25$ & $1.00$    & $0.18$ & $1.44$& $\mathrm{km\,s}^{-1}$\\
$\sigma_\mathrm{rms}$ & $80$ & $57$    & $325$ & $90$ & $\mathrm{mK}$\\
$\sigma_\mathrm{rms}^\mathrm{norm}$ & $89$ & $57$    & $140$ & $108$& $\mathrm{mK}$\\
$\Gamma_\mathrm{mb}$ & $0.132$ & $0.649$    & $10.52$ & $1.434$ & $\mathrm{K\,Jy}^{-1}$\\
\hline
\rule{0ex}{3ex}$N_\ion{H}{i}^\mathrm{lim}$ & $3.9$ & $2.5$    & $6.1$ & $4.7$ & $10^{18}~\mathrm{cm}^{-2}$\\
\rule{0ex}{3ex}$S_\ion{H}{i}^\mathrm{lim}$ & $16.1$ & $2.1$    & $0.3$ & $1.8$ & $\mathrm{Jy~km\,s}^{-1}$\\
\hline
\end{tabular}
\begin{list}{}{}
\item[$^\ast$] Numbers given are for the shallowest mode. Effective integration time varies across the survey area. Fields observed in commensal mode with AGES \citep[][$\sigma_\mathrm{rms}^\mathrm{norm}=33~\mathrm{mK}$]{minchin07} and ALFALFA \citep[][$\sigma_\mathrm{rms}^\mathrm{norm}=60~\mathrm{mK}$]{giovanelli05,haynes11} are much deeper (J. Peek priv. comm.).
\item[$^\dagger$] Northern (LDS) part only goes out to $v_\mathrm{lsr}\leq 400~\mathrm{km\,s}^{-1}$.
\end{list}
\end{table}

In Table~\ref{tab:survey_comparison} we compile important survey parameters of EBHIS in comparison with the most relevant other recent single-dish Milky Way \ion{H}{i} surveys LAB, GASS, and GALFA-HI. Apart from the $T_\mathrm{B}$ noise level, which is usually quoted for full spectral resolution data, we also provide a normalized noise value, $\sigma_\mathrm{rms}^\mathrm{norm}$, scaled to a velocity resolution of $1~\mathrm{km\,s}^{-1}$. The pure noise level, however, only quantifies the instrument's ability to detect gas, which entirely fills the observing beam. For the many unresolved or point-like objects in the surveys, a better sensitivity proxy is the velocity-integrated flux density, $S_\ion{H}{i}^\mathrm{int}$. In the table we quote numbers for the $5\sigma$ flux-density detection limit, $S_\ion{H}{i}^\mathrm{lim}$, of an unresolved object with a Gaussian profile having a line width (FWHM) of $w_{50}=20~\mathrm{km\,s}^{-1}$. Likewise, a limiting column density, $N_\ion{H}{i}^\mathrm{lim}$, was calculated without assuming point likeness. Both quantities, $N_\ion{H}{i}^\mathrm{lim}$ and $S_\ion{H}{i}^\mathrm{lim}$, scale with $\sqrt{w_{50}}$ and $\sigma_\mathrm{rms}^\mathrm{norm}$.

\section{Data reduction revisited}\label{sec:datareduction}

In the following we report on major modifications of the EBHIS data reduction software over the procedures reported in \citet{winkel10}.

\subsection{Improved RFI flagging}\label{subsec:rfiflagger}

\subsubsection{Automated flagging algorithm}\label{subsubsec:autorfiflagger}
\citet{floeer10} described how to make the best use of the fact that all seven feeds of the receiver are exposed to the same RFI environment. Following their approach improves RFI rejection efficiency significantly. We briefly review the basic RFI flagging workflow. Our automated flagging pipeline is optimized for three distinct types of RFI encountered at the Effelsberg 100-m telescope:

\begin{enumerate}
    \item Near-constant narrow-band spikes, typically affecting one or two spectral channels;
    \item Intermittent broad-band events affecting a hundred up to thousands of spectral channels;
    \item Extremely strong RFI caused by the L3 mode of the GPS satellite system.
\end{enumerate}

To detect the first two types of RFI, we use very similar detection strategies. Because we have 14 independent measurements (7 feeds with 2 polarization channels each) of the RFI environment at any given time, we use coincidence flagging to distinguish man-made interference from astronomical signals. Here we assume that a signal being present in more than one feed at the same time does not have an astronomical origin, because all seven feeds are pointed at different locations on the sky. This assumption is violated if an astronomical source is very extended, as for diffuse \ion{H}{i} emission from the Milky Way disk. We therefore also use matched filtering adapted to the typical appearance of RFI in a time-frequency plot. This enables a reliable RFI detection with the exception of the innermost part of the bright Galactic emission, where contributions at all spatial frequencies are present that cannot be reliably suppressed by matched filtering.

To process each individual observation, we first average the two orthogonal polarizations for each feed. Since RFI is often strongly polarized, either by reflection or intrinsically, there are cases where an RFI event is only present in a single polarization channel. Keeping the polarizations separate would then weaken our assumption that all feeds are exposed to the same RFI environment. While reducing the sensitivity gained from requiring signal coincidence, we still gain a factor of $\sqrt{2}$ in sensitivity from the averaging. After averaging, the subscans of each observation are processed independently, allowing multicored CPUs to be exploited for processing multiple subscans in parallel.

\begin{figure}[!t]
\centering
\includegraphics[width=0.49\textwidth,clip=]{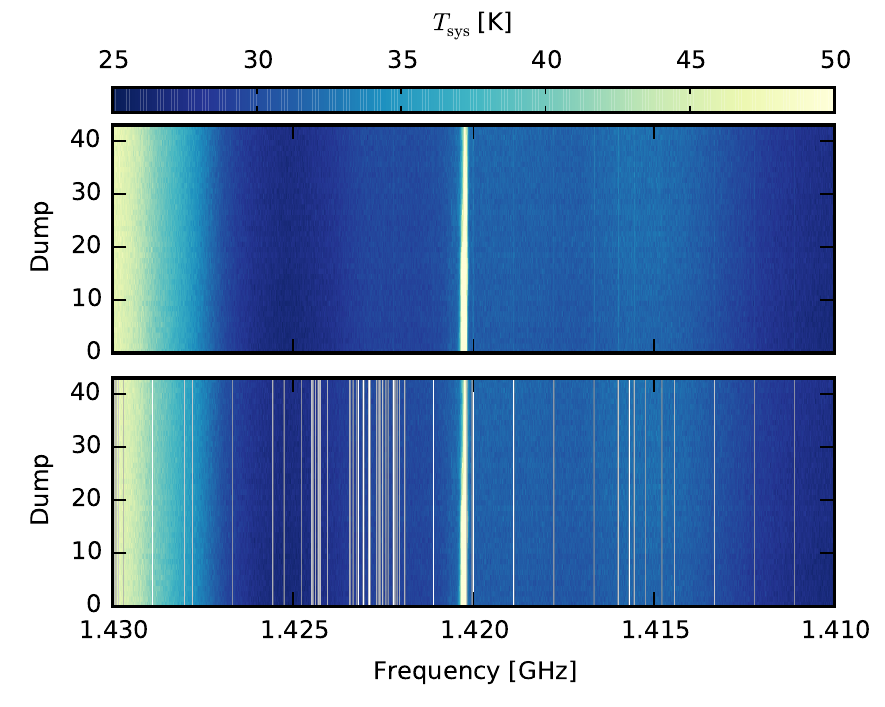}\\[0ex]
\caption{Narrow-band flagging result. \textit{Top panel:} spectrogram of raw data, \textit{bottom panel:} same, but with flags visualized as vertical lines.}
\label{fig:narrowband_flagger}
\end{figure}

To detect narrow-band RFI, we average each subscan in time, yielding seven spectra per subscan. Because narrow-band RFI events are typically confined to fewer than three channels at full spectral resolution, we remove any large scale spectral component by subtracting a median-filtered version of each spectrum. To flag a certain channel as contaminated, we use combinatorial thresholding. Instead of marking a channel as containing RFI once its level exceeds a certain threshold $t_1$, we require that a channel exceeds a lower threshold $t_N$ in $N$ feeds simultaneously. Assuming the noise in the data is Gaussian with standard deviation $\sigma$, the thresholds for $t_N$ that have the same statistical significance as an individual threshold $t_1$ are calculated by solving the equation
\begin{equation}
    1-\Phi(t_1) = \left[1-\Phi(t_N)\right]^N
\end{equation}
where $\Phi(x)$ is the cumulative distribution function of the standard normal distribution. Here we also make use of the fact that all RFI events have positive intensities.

Using this type of combinatorial thresholding, we can impose the coincidence requirement and with that make better use of the available data to optimize our detection efficiency. We choose a lowest required coincidence level of $N=4$. To flag a certain channel, we compare the data from all feeds to all successive thresholds for $N=4...7$ and flag the channel if the criterion is met for any $N$. An example for the RFI detection quality is provided in Fig.~\ref{fig:narrowband_flagger}. It shows a time--frequency plot (spectrogram) of one subscan of raw data (one feed) in the top panel. The bottom panel shows the same data but with all narrow-band events, detected by the RFI flagger, marked.

To detect broad-band interference, we apply a three-point median filter in the time domain of the data to suppress all persistent signals. We furthermore smooth the data in the spectral domain with a Gaussian filter adapted to the typical extent of the broad-band events. With the prepared data we again perform combinatorial thresholding across the seven feeds.

Since the GPS L3 mode is extremely bright and radiates always at a center frequency of $1381.05$\,MHz, we simply compare the RMS in a $1$\,MHz window around this frequency to the RMS in neighboring frequencies. If the RMS differs by a factor of two, we flag a 10-MHz-wide window around this frequency. This removes the brightest part of the interference, but GPS L3 features extended spectral side lobes, which can reduce the fidelity of the data over a large percentage of the 100-MHz band (about 50\%).

The data affected by broad-band and GPS interference is excluded from further processing, i.e., flagged. Typically, this does not limit the sensitivity of the \textit{Galactic} survey, because broadband interference mostly occurs at the low-frequency end of the 100-MHz-wide observing band. Also, GPS interference usually does not affect the radial velocities $-600 < v_\mathrm{LSR} < +600\,\mathrm{km\,s^{-1}}$. Because narrow-band interference in EBHIS data is numerous and often has near constant intensities, we attempt to subtract the RFI contribution instead of completely flagging an affected channel. This removes most of the narrow-band interference in the Galactic survey with the exception of a few narrow-band signals that mix with the bright \ion{H}{i} emission from the Milky Way disk. Here, a reliable subtraction is not possible and the interference is left untouched. From our perspective, this is better than flagging the spectral channels in question completely because leftover narrow-band RFI signals have amplitudes (typically $\lesssim0.5~\mathrm{K}$) that are small compared to the bright MW disk emission.

In the final data cubes, the Doppler correction causes frequency shifts of narrow-band RFI signals. In the topocentric frame they are mostly fixed in frequency (a few kHz change during hours), whereas after applying the LSR correction, shifts of up to two spectral channels occur. Therefore, residual narrow-band RFI can appear as ``wave-like'' patterns when visually inspecting consecutive channel maps in the data cube (see Section~\ref{subsec:quality_rfi}).

\begin{figure*}[!t]
\centering
\includegraphics[width=0.9\textwidth,clip=]{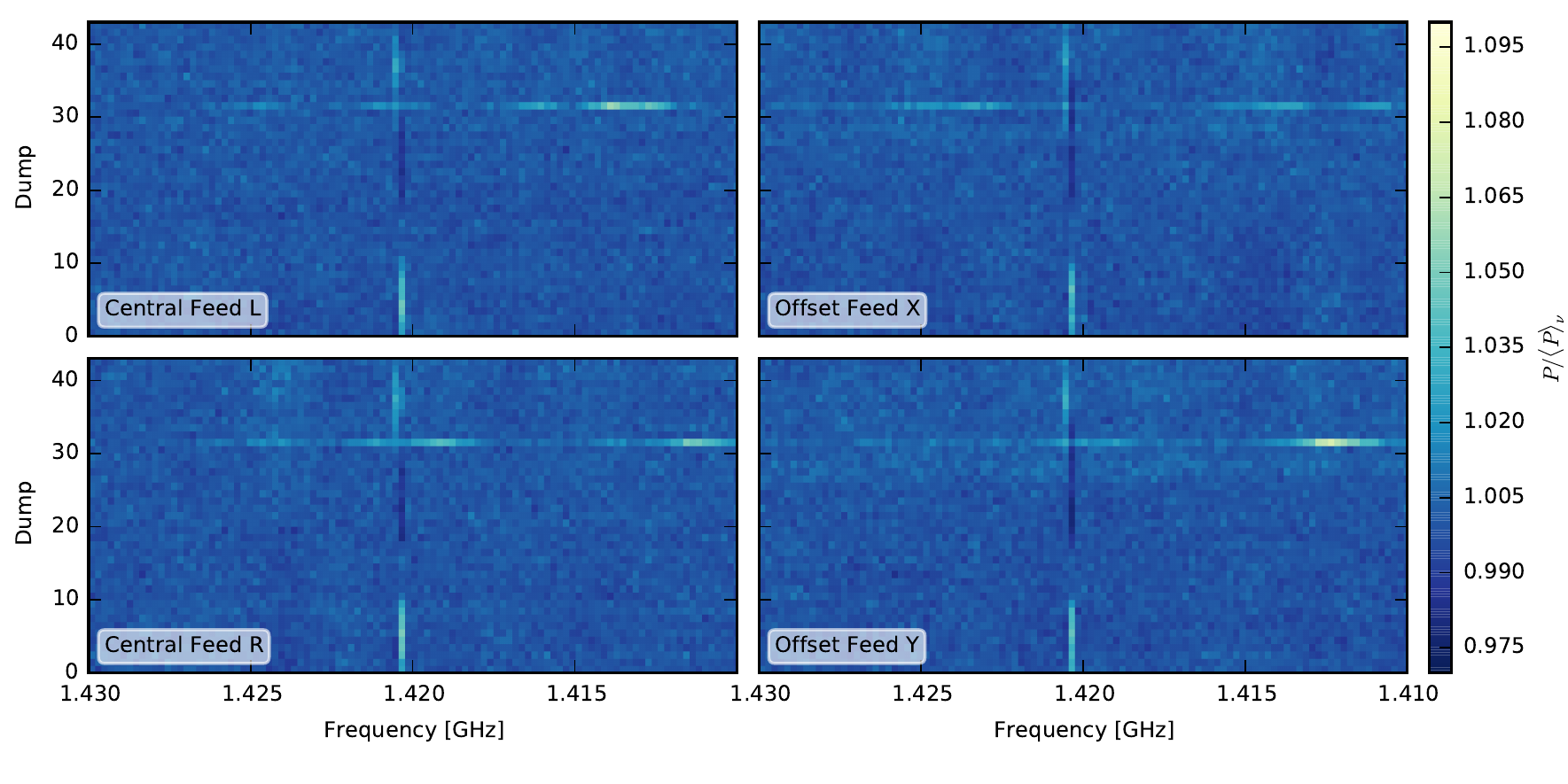}
\caption{Broadband RFI event, which is only visible in a single spectral dump (31). The left panels show the two polarizations (L and R) of the central feed, and right panels contain offset feed data (X and Y). For improved visualization, data is binned in frequency (32-fold, $\Delta\nu=195~\mathrm{kHz}$), and each spectral dump is divided by the median spectrum. The Milky Way \ion{H}{i} emission line appears slightly displaced from its rest frequency owing to the LSR Doppler correction, which was not applied here.}
\label{fig:rfi_manual_broadband2}
\end{figure*}

\begin{figure}[!t]
\centering
\includegraphics[width=0.49\textwidth,clip=]{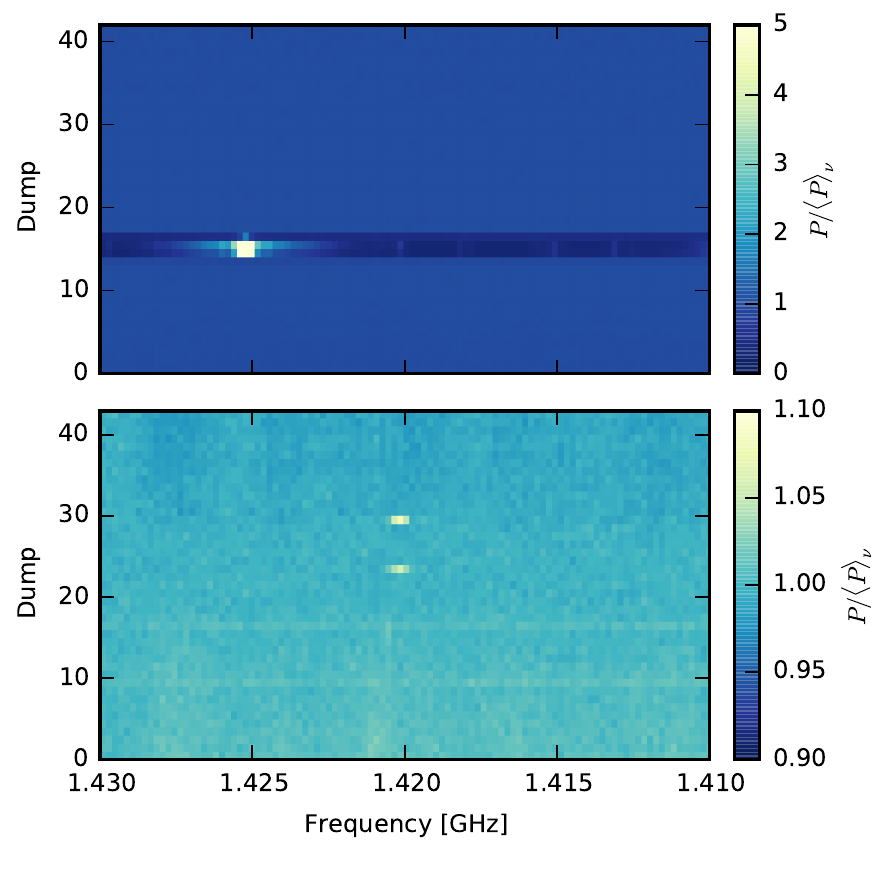}
\caption{Examples of RFI. Top panel: a very intense out-of-band event temporarily saturates the receiver and causes intermodulation products (e.g., at 1425~MHz). Bottom panel: two short bursts of RFI, likely radar pulses, with a width of a few 100 kHz.}
\label{fig:rfi_manual_broadband}
\end{figure}

\subsubsection{Manual RFI flagging}
The RFI described is present in all observations. Apart from these regular signals, there are additional glitches that are not easily recognized by automated algorithms. To ensure consistently high data quality, each observation was also inspected manually. A simple graphical user interface was developed to compute spectrograms (images of the time-frequency plane). The user can quickly iterate over the different subscans, feeds, and polarization channels to search for RFI and other defects in the data. By mouse-clicking on a spectral dump, an RFI flag (bad spectrum) is generated in a data base. To improve the RFI-to-noise ratio, data is binned in frequency (usually 32-fold, giving a channel width of 195~kHz). To remove the bandpass shape, which would otherwise dominate the spectrogram, each spectral dump is divided by the median spectrum of the current subscan. A side effect is that the Milky Way emission line is only shown as a residual signal. For fields closer to the Galactic disk, where the Milky Way is highly structured, this can make it difficult to distinguish faint broad-band RFI from astronomical features.

In the following we briefly discuss the three most frequent types of broad-band interference that were encountered in EBHIS Milky Way data. Figure~\ref{fig:rfi_manual_broadband2} shows an example of a wide-band burst that is polarized, varies with frequency, and couples differently into the individual feeds. It lasted less than 500~ms and affected the full 100~MHz band. Another glitch that is extending over the full 100~MHz frequency range is plotted in Fig.~\ref{fig:rfi_manual_broadband} (top panel). Here, most of the affected spectrum was more negative than the median spectrum, which is most likely caused by a high-intensity out-of-band RFI event, saturating the front end. Intermodulation products are detected all over the spectrum, for instance, at 1425~MHz. The bottom panel of Fig.~\ref{fig:rfi_manual_broadband} depicts two short bursts of events, which are a few 100-kHz wide, superposed on Milky Way emission. In this example, RFI and MW can be distinguished well, but closer to the Galactic plane this becomes more difficult. The time between both events is 12~s -- which is a typical delay for Radar pulses. Sometimes up to three consecutive bursts were observed, but never more. We speculate that the 100-m dish may have received reflected radar pulses, such as from airplanes. However, the original emitter was never identified.

\subsection{New 2-D baseline fitting}\label{subsec:baselines}
\citet{winkel10} proposed adaptive Gaussian smoothing to calculate baseline solutions for each individual calibrated spectrum. Even though this approach performed well in most cases, we improved the robustness of the baseline solution by extending the fitting procedure to the time-frequency domain. The revised algorithm made use of the fact that the system temperature in the baseline changes only slowly with time.

Further tests revealed, however, that faint sources were sometimes surrounded by negative baseline artifacts. The Gaussian-smoothing baseline algorithm relies on proper flagging or masking of sources. Otherwise, the flux contribution of sources will be smeared into the baseline solution itself. In this context, a source is thought to be anything that superposes the baseline, such as \ion{H}{i} clouds in the Milky Way and its halo or other galaxies (both in continuum and spectral line emission). The baseline algorithm itself tries to mask sources, for example, by iterative flagging of $3\sigma$ outliers. However, since baseline-fitting is done on the raw spectral data that have about 500~mK RMS, faint sources are not always reliably detected.

It turned out that in these cases, polynomials are better suited for baseline estimation. They can describe underlying fluctuations well, but produce shallower troughs when no mask is provided. Because the Gaussian-smoothing baseline method benefited from extension to the time--frequency plane ($t$, $f$), we implemented 2-D polynomial fitting, where the baseline, $y_\mathrm{b}$, is described by

\begin{equation}
y_\mathrm{b} = \sum_{i,j\geq0} \alpha_{i,j} f^i t^j\,.
\end{equation}

Baselines in EBHIS have a complicated structure, especially along the frequency axis. To avoid unreasonably high polynomial orders, we fit the data on tiles of 1024 spectral channels times the number of dumps per subscan (about 40). To suppress sharp gaps in the baseline solution at the tile edges, we interleave tiles (overlap: 512 channels) and interpolate between each of two adjacent solutions using sigmoid thresholds. Good results are achieved with polynomial orders of ten in spectral and two in temporal direction, while allowing only one cross term, $\alpha_{1,1}$, to describe mild tilts in the 2-D baseline.

To improve the baseline solution even further, we initially used the LAB survey to provide a spectral mask around MW emission. This is supplemented with information from the HyperLEDA \citep{makarov14} database\footnote{http://leda.univ-lyon1.fr/} containing extragalactic \ion{H}{i} objects and NVSS \citep{condon98} to flag strong continuum sources ($\geq$1.5~Jy). Weaker continuum sources are handled
well by subtracting the average $T_\mathrm{sys}$ level from each spectral dump prior to computing the baseline. Finally, a full-resolution datacube of EBHIS itself is used to mask \ion{H}{i} emission from the MW and its halo in an iterative scheme.

\subsection{Frequency dependency of $T_\mathrm{cal}$}
\citet{winkel10} explain the intensity calibration of EBHIS data incorporating the frequency-dependence of the system temperature, $T_\mathrm{sys}$. There, the temperature, $T_\mathrm{cal}$, of the noise diode, which is fed into the receiver, was assumed to be constant over frequency. In the meantime, based on the methods proposed in \citet{winkel12}, we could derive the frequency dependence of $T_\mathrm{cal}$ using continuum calibration sources (e.g., 3C48, NGC7027). Unfortunately, such a measurement is time-consuming, since each beam must be positioned individually onto the continuum source.

As a faster alternative, we used an absorber -- placed around the seven feeds -- to obtain a simultaneous and almost RFI-free measurement of all noise diodes. Using the hot absorber alone cannot provide absolute calibration of the $T_\mathrm{cal}$ spectra. However, it is well suited to assessing the frequency-dependent behavior. The absolute value of $T_\mathrm{cal}$ at 1.42~GHz can then be measured using observations of the IAU standard position/source \object{S\,7} at $v_\mathrm{lsr}\approx0~\mathrm{km\,s}^{-1}$, as explained in \citet{winkel10}. This procedure was repeated several times during the whole observing campaign and yielded consistent results for the  $T_\mathrm{cal}$ spectra of all 14 channels (7 feeds with 2 polarizations each).

\subsection{$T_\mathrm{sys}$-based weighting scheme}
For EBHIS data, the system temperatures vary between the different feeds and naturally also between the phases where the noise diode is switched on and off. It is possible to achieve a slightly lower final noise level of the data by weighting each spectrum with its associated system temperature \citep[see][]{winkel12}. Since we do not use frequency switching to remove the bandpass from our data, we can reconstruct the full frequency-dependent system temperature during calibration and baseline calculations.

One difficulty remains: the \ion{H}{i} emission line itself contributes to the system temperature at the appropriate frequencies. In theory, one had to add each individual \ion{H}{i} profile to the underlying continuum $T_\mathrm{sys}$ spectrum to obtain the proper weights. This would be disadvantageous in practice because the individual 500~ms-line profiles are very noisy; $\sigma_\mathrm{rms}\sim0.5~\mathrm{K}$. Such weighting would introduce an additional noise component. Therefore, we decided to compute the average \ion{H}{i} line profile per subscan and per beam and use it as the system temperature input for the weighting. Strictly speaking, this attempt is an oversimplification for maps where the $T_\mathrm{sys}$ level varies significantly during the subscan, for instance, owing to large changes in elevation. However, since all feeds experience the same gradient in system temperature, only a higher order effect is introduced into the weighting, and that can be neglected for our purpose. Introducing the $T_\mathrm{sys}$-based weighting scheme improves the final RMS level by about 1 to 2\%.

\subsection{Stray-radiation correction}\label{subsec:SR}

Galactic emission at 21-cm is seen in all directions, but predominantly from the Galactic plane extending across the whole sky. Antenna side lobes, pointing to such regions of high flux density, will eventually produce artificial signals unrelated to the intensity received by the primary beam. This so-called stray radiation varies with time and season \citep{vanwoerden62} and is most critical for observations of high-galactic latitude objects with faint \ion{H}{i} emission. In extreme cases, stray radiation may provide a larger contribution to a measured spectrum than the true brightness temperature of the observed sky position.

The side-lobe structure of a telescope depends on antenna type and design. For a paraboloidal reflector, the extended side lobes, called stray cones, are mainly caused by the support legs that carry the prime focus cabin and/or secondary mirror \citep[][their Fig.~6]{kalberla80a}. Radiation is also received from regions outside the rim of the reflector, the spillover region. Reflecting surfaces within the aperture can also cause additional side lobes (see Sect. \ref{subsubsec:farSL}).

The impact of stray radiation can be minimized by reducing the number of scattering surfaces within the telescope aperture. Prime examples are the Bell Labs horn reflector antenna \citep{stark92} and the Green Bank Telescope \citep[GBT;][]{prestage09}. For these telescopes the stray radiation is reduced significantly in comparison to a standard paraboloid; i.e., the main beam efficiency is increased from about 70\% to 90\%. However, the remaining 10\% of side-lobe efficiency can still be serious enough that a numerical correction of the observations is unavoidable \citep{boothroyd11}.

First attempts by \citet{vanwoerden62} to correct observations of the Dwingeloo telescope for stray radiation were limited by a lack of sufficient computing power. \citet{kalberla80a,kalberla80b} provided the first corrections for the Effelsberg 100-m telescope. \citet{lockman86} obtained similar results by bootstrapping from the Bell Labs Survey \citep{stark92}, which is affected only
a little by stray radiation. Subsequently, \citet{hartmann96} and \citet{kalberla05} developed a correction for the resurfaced Dwingeloo dish, \citet{higgs00} for the 26-m Telescope at the Dominion Radio Astrophysical Observatory, \citet{bajaja05} for the 30-m telescope at Villa Elisa, \citet{kalberla10} for the Parkes 64-m dish, and \citet{boothroyd11} for the 100-m GBT.

\subsubsection{Basics}

The antenna temperature $T_\mathrm{A}$ observed by a radio telescope is given for each of the individual receivers by a convolution of the true brightness temperature distribution $T$ on the sky with the beam pattern $P$ of the antenna:

\begin{equation}
T_\mathrm{A}(x,y)  =  \int P(x-x',y-y') T(x',y') \diff x' \diff y'\,.
\end{equation}
Here we use an approximation in Cartesian coordinates to simplify the expression of the convolution integrals. In general, $T_\mathrm{A}$ is time- and frequency-dependent, spherical coordinates have to be used, and the integration needs to take the observable part of the sky (i.e., the horizon) into account, as well as ground reflectivity. For the main beam and all sidelobes, atmospheric attenuation and refraction need to be considered, too.

For the pattern $P$ of the antenna we use the normalization
\begin{equation}
\int P(x,y) \diff x \diff y  = 1
\end{equation}
and split the antenna diagram into the main beam area (MB) and the stray
pattern (SP):

\begin{eqnarray}
\label{eq:TAsplit}
  T_\mathrm{A}(x,y) = \int_\mathrm{MB} P(x-x',y-y') T(x',y') \diff x' \diff y' + \nonumber \\
  \int_\mathrm{SP}P(x-x',y-y') T(x',y') \diff x' \diff y'\,.
\end{eqnarray}
By defining the main beam efficiency,
\begin{equation}
\eta_\mathrm{MB} \equiv \int_\mathrm{MB} P(x,y) \diff x \diff y\,,
\end{equation}
of the telescope, Eq.~(\ref{eq:TAsplit}) can be rewritten as
\begin{equation}
T_\mathrm{B}(x,y) =  \frac{T_\mathrm{A}(x,y)}{\eta_\mathrm{MB}}
- \frac{1}{\eta_\mathrm{MB}}\int_\mathrm{SP} P(x-x',y-y') T(x',y') \diff x' \diff y'\,.
\label{eq:deconvolve}
\end{equation}

To determine the main beam-averaged brightness temperature, $T_\mathrm{B}$, we need to know the antenna pattern, $P$, with sufficient accuracy, but also the true brightness temperature $T$ on the sky, which a priori is unknown. Apparently we need an approximation for $T$.

\subsubsection{How to solve Equation (\ref{eq:deconvolve})}\label{subsubsec:srhowto}

There are two unknowns in Eq.~(\ref{eq:deconvolve}), $T_\mathrm{B}$ and $T$. Furthermore, the
solution depends on the choice of the main beam area MB.
We first consider conditions for a solution of this equation. Replacing $T_\mathrm{B}$ on the lefthand side with $T$ leads to a Fredholm equation of the second kind, which can be solved unambiguously for $\eta_\mathrm{MB} > 0.5$ \citep{kalberla80a}. State-of-the-art telescopes have main beam efficiencies $\eta_\mathrm{MB} \gtrsim 0.7$, defined for the region within the first minimum of the antenna diagram. The existence of a solution of Eq.~(\ref{eq:deconvolve}) is therefore warranted. A standard procedure for finding the solution is to use successive approximations, as proposed by \citet{bracewell54}.

Knowing these basic limitations, we can develop a strategy for deriving correct brightness temperatures. Most important is to find ways to detect stray radiation effects in the observations and, likewise, to separate regions in the antenna diagram that cause contributions.  As far as possible, we measure parameters of the antenna diagram $P$, next we model near-side-lobe structures (Section~\ref{subsubsec:nearSL}). For side lobes far from the main beam, we study details of the telescope construction and employ ray-tracing to find the critical regions in the pattern (Section~\ref{subsubsec:farSL}).

In most cases it is impossible to determine accurate absolute side-lobe levels from ray tracing alone, but by varying individual free parameters (intensities for 69 side-lobe structures in the final version of the far side-lobe model), we get satisfactory solutions. A simple but powerful test for identifying a critical region in the antenna diagram is to correct the observations with an unreasonably high side-lobe level, which then causes overcorrections. The resulting spurious negative features in corrected profiles and data cubes can be easily spotted by eye.

Characteristic of stray radiation errors is the variability of spurious line components that also led to the detection of stray radiation in the first place \citep{vanwoerden62,kalberla80b}. We make use of this effect. An iterative determination of side-lobe levels requires a sufficiently large sample. We use EBHIS for declinations $0\degr < \delta < 60\degr$ and velocities $-100 < v_\mathrm{lsr} < +100\,\mathrm{km\,s^{-1}}$.

Coherent regions in $(l,\,b,\,v_\mathrm{lsr})$ are selected that contain little or no \ion{H}{i} emission but are sensitive to stray radiation effects. First we use LAB data to verify the absence of significant emission features. Next we determine peak deviations and independently the RMS scatter over larger emission-free $(l,\,b,\,v_\mathrm{lsr})$ regions with the aim of minimizing these errors. Our analysis is hampered by the fact that we only have a single sky coverage, so it is not possible to compare different EBHIS observations at the same position. Stray radiation problems are, however, recognizable as discontinuities between individual fields that cause a patchy or blocky structure (compare Fig.~\ref{fig:srcorrection_example}, top panel). Our strategy is to improve the SR corrections by successive approximations, modifying antenna parameters but also the correction algorithm itself (Sect. \ref{subsubsec:SRcorralgo}), as well as the estimate for the brightness temperature distribution $T$ in Eq.~(\ref{eq:deconvolve}).

\subsubsection{The multibeam antenna pattern: near side lobes}\label{subsubsec:nearSL}

To solve Eq.~(\ref{eq:deconvolve}), it is necessary to work out the complete antenna response, $P$, for each of the seven beams of the Effelsberg multifeed system. For practical reasons we distinguish between near side lobes around the main beam within a distance of $4\degr$ and far side lobes farther out. Both regions cover approximately half of the antenna's solid angle outside the main beam.

Using strong radio continuum sources such as \object{Cas\,A}, it is hardly possible to measure accurate side-lobe structures below $-40~\mathrm{dB}$. Such observations are typically restricted to distances $\lesssim$1~deg from the main beam but cover only $\lesssim$50\% of the near side-lobe beam's solid angle required in Eq.~(\ref{eq:deconvolve}). We therefore model the side lobes out to radial distance of $4~\mathrm{deg}$ from the main beam in a similar way to what was previously done for GASS.

\begin{figure}[!t]
\centering
\includegraphics[width=0.49\textwidth,viewport=18 37 413 827,clip=]{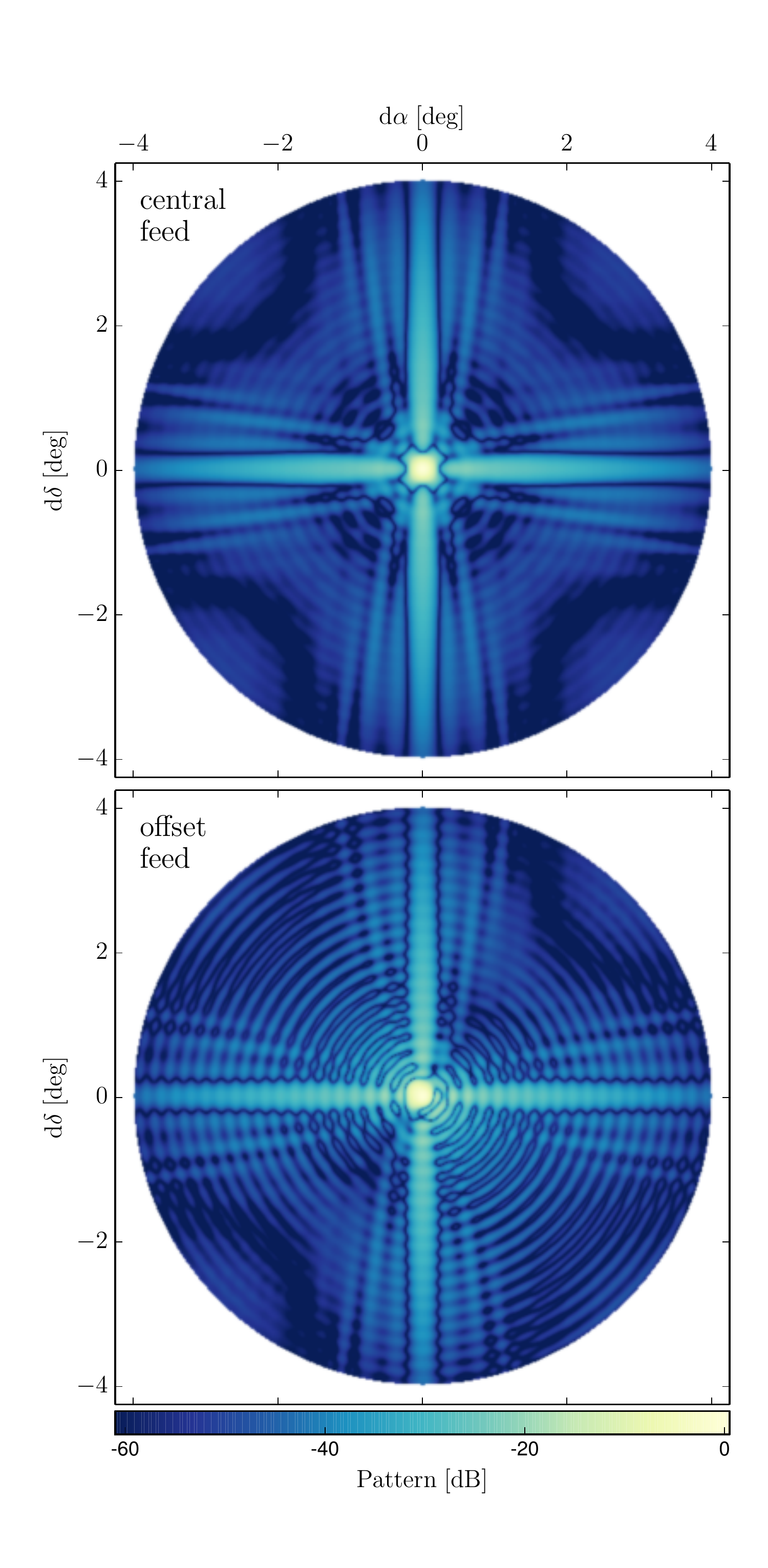}\\[0ex]
\caption{Synthetic antenna patterns within a radius of $4\degr$ for the central beam (top) and an offset beam (bottom) with a coma lobe at a position angle of $60\degr$.}
\label{fig:Pattern}
\end{figure}

The far-field pattern of an antenna is approximated by the auto-correlation of the aperture plane distribution. We used the measured feed-horn response pattern \citep{keller06} and the telescope geometry, including shadowing caused by the focus cabin and the feed support legs to derive the complex aperture distribution function for each feed. The pattern was then calculated by Fourier transformation in a similar way to \citet[][their Sect.~2.2]{baars07}.

For the seven beams, we distinguish between the central feed and the six surrounding feeds  in a hexagonal layout. The offset beams have a radial offset of 17.3~cm from the optical axis. As mentioned in Section~\ref{sec:surveydescription}, the seven-beam receiver is rotated during observations to account for the changing parallactic angle that would otherwise distort the resulting scan pattern in the equatorial coordinate frame. Unfortunately, this means that the feed horns rotate relative to the focus cabin's support legs. This causes the already complex aperture distribution function to change with time. For the sake of processing speed, we accounted for this rotation only in steps of $5\degr$. We also tested a less accurate stepping of $10\degr$, which was previously used to correct data from GASS \citep{kalberla10}, but differences are barely noticeable. We conclude that the approximation of the beam rotation within $\pm2\fdg5$ (EBHIS) or $\pm5\degr$ (GASS) is accurate enough to solve Eq.~(\ref{eq:deconvolve}) for the near side lobes without noticeable uncertainties. Utilizing the four-fold symmetry of the antenna aperture, in total 18 antenna diagrams for the offset feeds need to be provided during the solution of Eq.~(\ref{eq:deconvolve}). As an example, we plot in Fig.~\ref{fig:Pattern} the diagram for the central feed and an offset feed showing the coma lobe at a position angle of $60\degr$.

For the correction algorithm, we averaged the previously modeled side-lobe intensities on a fixed grid containing 2160 cells in cylindrical coordinates for the inner $4\degr$ of the antenna
diagrams. The cells have an azimuthal extent of $5\degr$ and a radial extent of $0\fdg125$, covering the radial range of $0\fdg25$ (first minimum) up to $4\degr$.

\subsubsection{The far side lobes}\label{subsubsec:farSL}

The far side lobes are determined by details of the telescope structure. It is difficult to measure these side lobes \citep[e.g.,][]{higgs67,hartsuijker72}, so we used ray-tracing to model them. While in the near side-lobe range, details of each of the individual antenna diagrams need to be considered, and it is sufficient to use a single common far side-lobe diagram for all of the receivers. For the subsequent discussion of far-side-lobe effects, \textit{we consider the telescope as a transmitting system.}

The most important side lobes that are far away from the main beam of the 100-m telescope, were determined by \citet[][see their Fig. 3]{kalberla80a}. These are the stray cones, caused by reflection of plane waves from the primary mirror at the feed support legs. The second most important structure is the spill-over at the edge of the main reflector. Its irradiation level depends on the edge taper of the primary feeds.

\begin{figure}[!t]
\centering
\includegraphics[width=0.49\textwidth,viewport=13 -2 415 428,clip=]{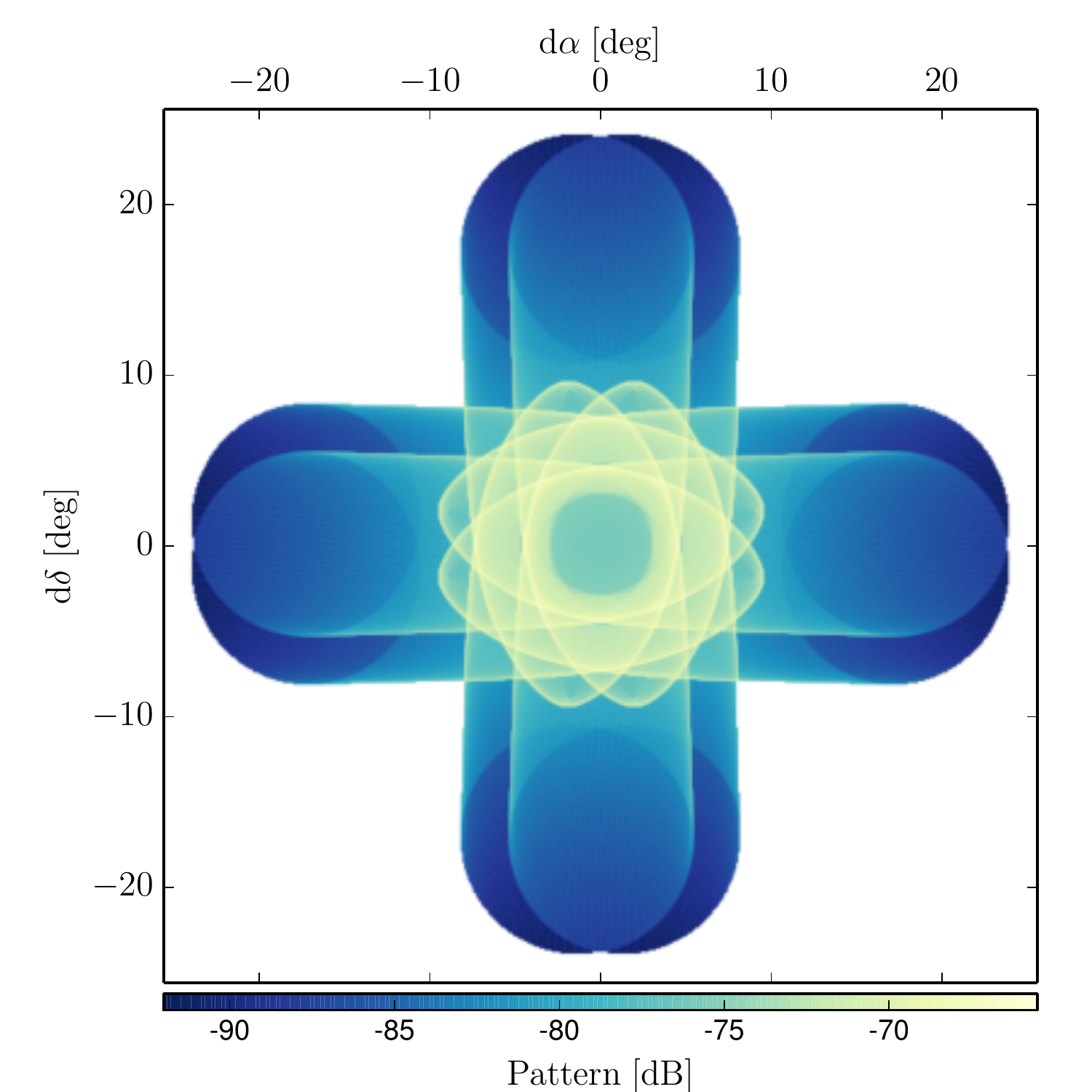}\\[0ex]
\caption{Side lobes caused by reflections of spherical waves from the feed at the support legs of the 100-m (intensities relative to main beam).}
\label{fig:spericalcones}
\end{figure}

In addition, we considered spherical waves originating in the feeds and determined their reflections at the support legs. Figure~\ref{fig:spericalcones} shows these structures. The eightfold symmetry is caused by the four legs. Each leg consists of four main tubes; two of them, separated by 1.98 m, are visible from the feed. These side lobes have an interesting structure, but it turned out that they are rather unimportant.

\begin{figure}[!t]
\centering
\includegraphics[width=0.49\textwidth,viewport=13 -2 415 428,clip=]{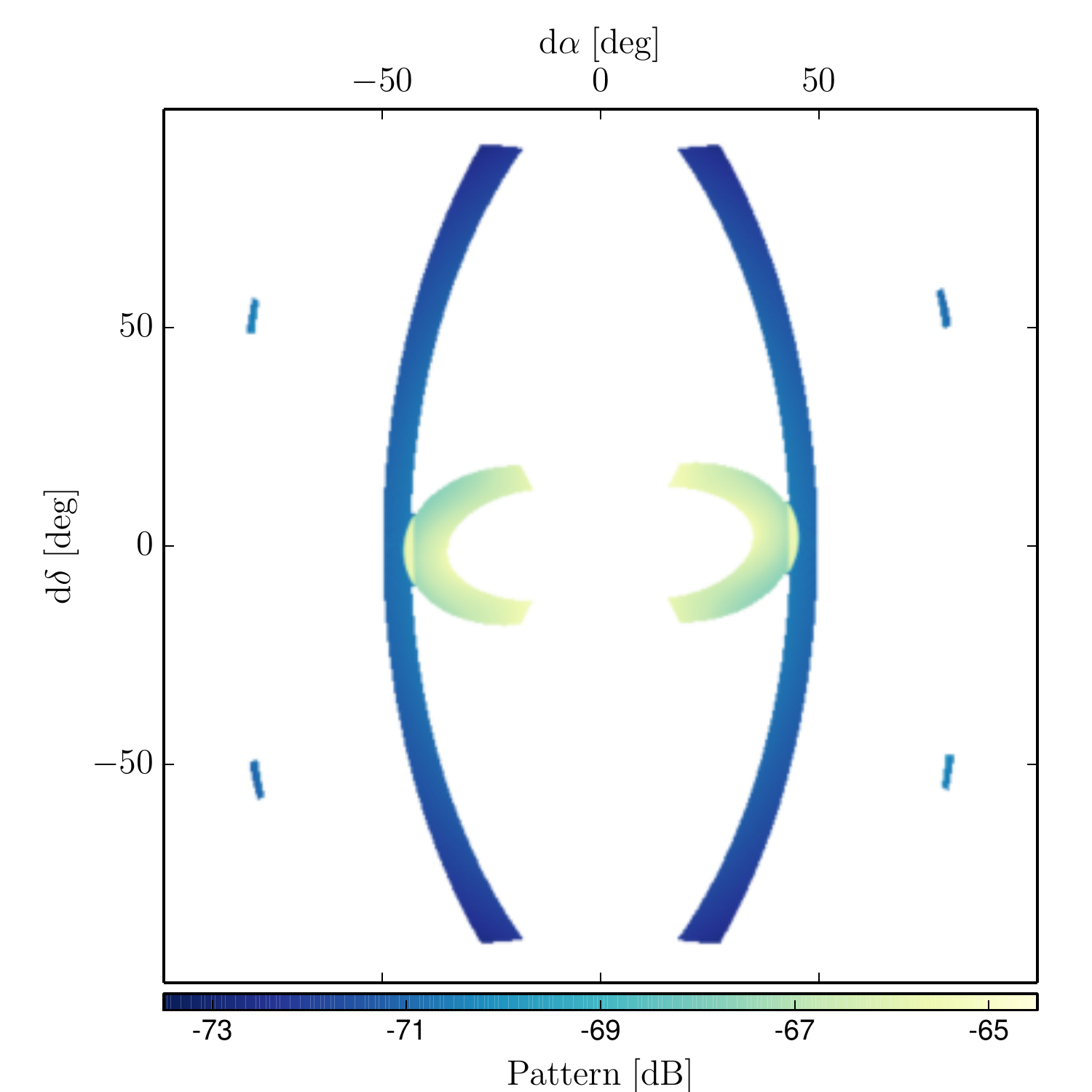}\\[0ex]
\caption{Side lobes caused by reflections from the roof of the secondary focus cabin onto the feed support legs (intensities relative to main beam). The large vertical feature is a secondary stray cone. The inner lobes are caustics caused after reflections from two tubes of the feed support legs back to the primary mirror. The four spot-like side lobes are caused by the fence around the rim of the paraboloid. }
\label{fig:apex}
\end{figure}

The Effelsberg 100-m telescope has a Gregorian secondary focus with several feed systems in an apex cabin. The seven-feed system is situated in primary focus, and the roof of the apex cabin is closed during EBHIS observations. This roof causes reflections that are offset from the main axis by $40\degr$. The side lobes are easily determined from the geometry of the apex roof. However, that these reflections are oriented in the east-west direction makes the situation very uncomfortable since it can produce multiple reflections. A part of the rays being reflected off the apex roof can hit feed support legs located east or west of the roof. This causes secondary stray cones. Only a fraction of these secondary stray cones are reflected onto the sky. The other part is reflected to the main mirror, causing caustics. A minor fraction of the secondary stray cones can hit the fence at the rim of the reflector before escaping. These features were determined by ray-tracing. Figure~\ref{fig:apex} shows the resulting pattern. We disregarded further reflections caused by support legs to the north and south of the focus cabin.

The apex lobes discussed here differ significantly from side-lobe properties of the original telescope design. Initially the roof of the apex cabin was designed so that reflections occurred in four triangular lobes in between the support legs \citep[see Fig. 3 of][]{kalberla80a}. The new construction allows fast switching between primary and secondary focus but causes extremely complicated side-lobe structures far off the main beam that can only partly be modeled. We also do not account for minor construction details like stairs, cable ducts, and cross ties of the prime-focus support legs.

The side-lobe levels from the spillover lobes were estimated from the edge taper of the receiver feed. For the stray cones we used previous results from \citet{kalberla80a}, but in the new software the radial side-lobe levels of the cones are modeled by a Gaussian ($2\degr$~FWHM). All far side-lobe components were adjusted individually in a way similar to what is described by \citet{kalberla05}, when we were searching for a consistent solution to Eq.~(\ref{eq:deconvolve}).

\subsection{The correction algorithm}\label{subsubsec:SRcorralgo}

Our correction algorithm is based on \citet{kalberla80a} and was extended later for multibeam systems \citep{kalberla10}. Some further improvements were made, the most important ones concerning the all-sky brightness temperature distribution, $T_\mathrm{B}$, that is needed for the deconvolution according to Eq.~(\ref{eq:deconvolve}). As previously mentioned, the basic strategy was to utilize the LAB data set to calculate an initial guess of the SR correction, i.e. to solve the side-lobe-pattern integral in Eq.~(\ref{eq:deconvolve}). After optimizing free parameters in the antenna pattern for the LAB-based SR correction, we switched over to use EBHIS itself and GASS data for the input sky. Again, several iterations were processed to find the optimal solution.

For a deconvolution on the sphere, an equal pixel-area representation of $T_\mathrm{B}$ is desirable. Here we used the Hierarchical Equal Area isoLatitude Pixelisation \citep[HEALPix][]{gorski05} scheme. HEALPix is a versatile structure for the pixelization of data on the sphere. Most important for our needs is that this scheme allows the spatial resolution of the $T_\mathrm{B}$ spectra to  be easily matched to the needs in accuracy during the correction. A single parameter, $N_\mathrm{side}$, is sufficient to define the resolution of $T_\mathrm{B}$ \citep[][Table 1]{gorski05}. Here $N_\mathrm{side} = 256$ corresponds to a size of $13\farcm7$ per pixel, which in most cases is adequate for near side-lobe correction. For $N_\mathrm{side} = 64,$ a pixel size of $55\farcm0$ is given, which would be more suitable for far side-lobe corrections.

\begin{figure}[!t]
\centering
\includegraphics[width=0.49\textwidth,viewport=80 58 1003 528,clip=]{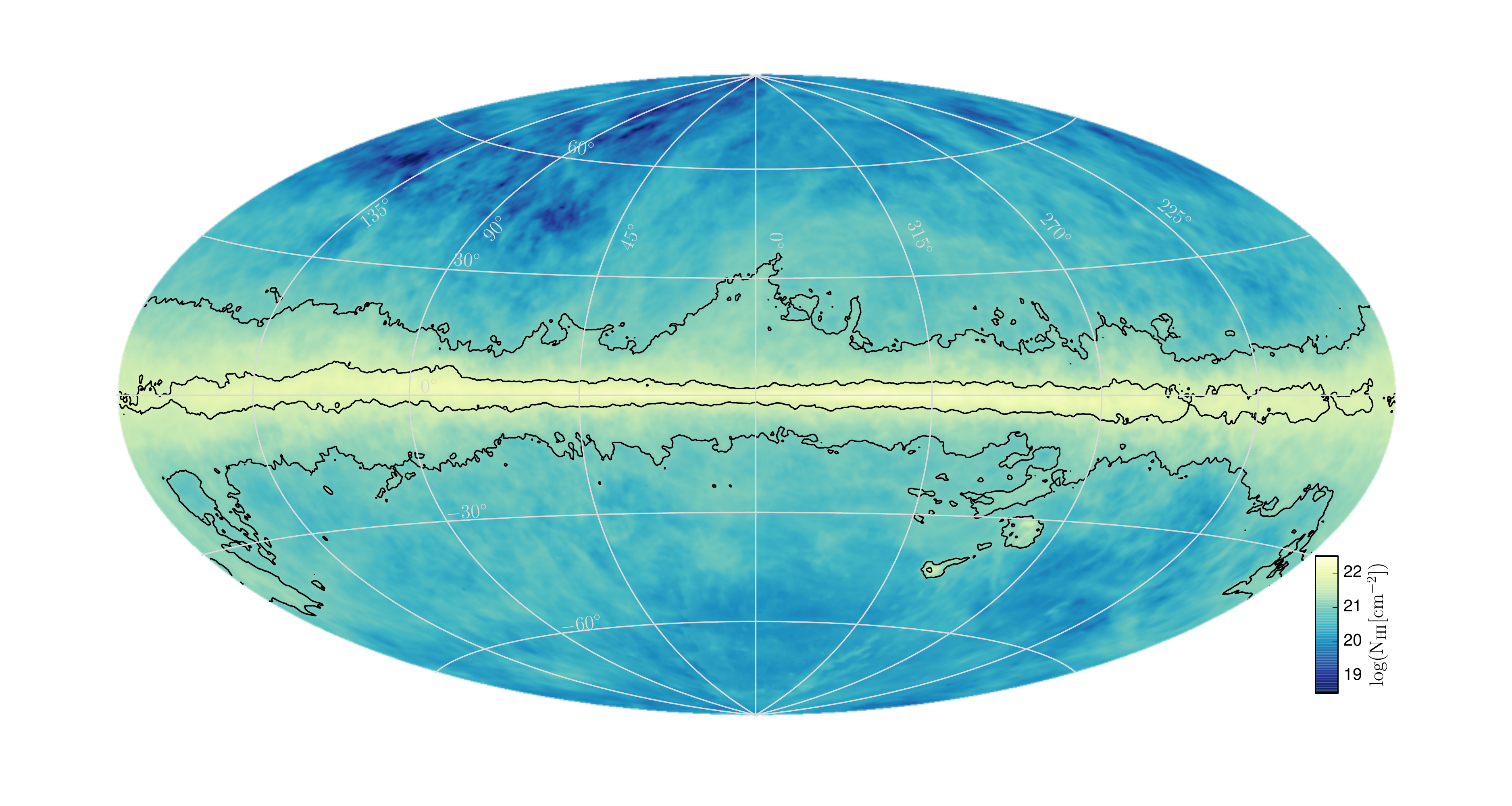}\\[0ex]
\caption{Column density distribution from the LAB survey. The isophotes indicate $N_\ion{H}{i} = 10^{21}~\mathrm{cm}^{-2}$ and $N_\ion{H}{i} = 5\cdot10^{21}~\mathrm{cm}^{-2}$, limiting ranges with different HEALPix resolutions that were used for far-side-lobe stray radiation correction.}
\label{fig:labnh}
\end{figure}

Side-lobe correction computing speed depends critically on the number of the processed $T_\mathrm{B}$ spectra. High-resolution single-dish data for the correction of the \textit{far side lobes} are only needed in regions with significant fluctuation in brightness temperature. The HEALPix scheme allows a flexible scaling, and we have chosen $N_\mathrm{side} = 128$ for regions with column densities $N_\ion{H}{i} > 5\cdot10^{21}~\mathrm{cm}^{-2}$, $N_\mathrm{side} = 64$ for $10^{21} < N_\ion{H}{i} < 5\cdot10^{21}~\mathrm{cm}^{-2}$, and $ N_\mathrm{side} = 32 $ for $ N_\ion{H}{i} < 10^{21}~\mathrm{cm}^{-2}$ (see Fig.~\ref{fig:labnh}).

A near side-lobe correction demands a high-resolution approximation of the true-sky brightness temperature $T_\mathrm{B}$ according to Eq.~(\ref{eq:deconvolve}). We initially used the LAB survey, resampled to $N_\mathrm{side} = 256$ and later to $N_\mathrm{side} = 512$. After this preliminary SR correction based on low-resolution data, we employed the SR-corrected EBHIS dataset itself ($N_\mathrm{side} = 1024$, i.e., full-angular resolution, $\theta_\mathrm{pix}=3\farcm4$). GASS data were supplemented on the southern sky on the same HEALPix grid. The last step was repeated several times in conjunction with a recomputation of the 2-D baseline models (Section~\ref{subsec:baselines}). Thus we increased the accuracy step-by-step for the near side-lobe correction. In addition we improved the accuracy of the deconvolving kernel function: for each beam from initially 396 to finally 2160 samples of the side-lobe structure. The beam rotation was traced initially within $\pm 5$ degrees, finally accurate to $\pm 2.5$ degrees.

All these improvements in the correction procedure led to a more precise and efficient algorithm but eventually did not improve the quality of the solution, as desired. Because we cannot model all parameters of the antenna diagram with sufficient detail, residual SR features appear in the data. Unaccounted for are substructures of the feed support legs such as stairways and cable channels. In calculating caustics (Fig.~\ref{fig:apex}), we only took three scattering surfaces into account and disregarded further scattering at the feed support legs. Also some substructures of the roof of the apex cabin were ignored. We did not attempt to model reflections from the subreflector mount.

In Appendix\,\ref{appsubsec:fullskymaps} we show maps of the near-side-lobe, far-side-lobe, and total  SR correction (in \ion{H}{i} column densities, Fig.\,\ref{fig:allsky_SRtot}), which was applied to the EBHIS data. To allow comparison, the three figures share the same colorbar scale as Fig.\,\ref{fig:allsky_nhi}. It is remarkable that the far-side-lobe correction in several low-column density regions at higher Galactic latitudes is greater than the reconstructed column densities in these fields, which underlines the importance of SR correction.

\section{Data quality}\label{sec:dataquality}

\begin{figure*}[!t]
\centering%
\includegraphics[width=0.32\textwidth,viewport=35 50 555 605,clip=]{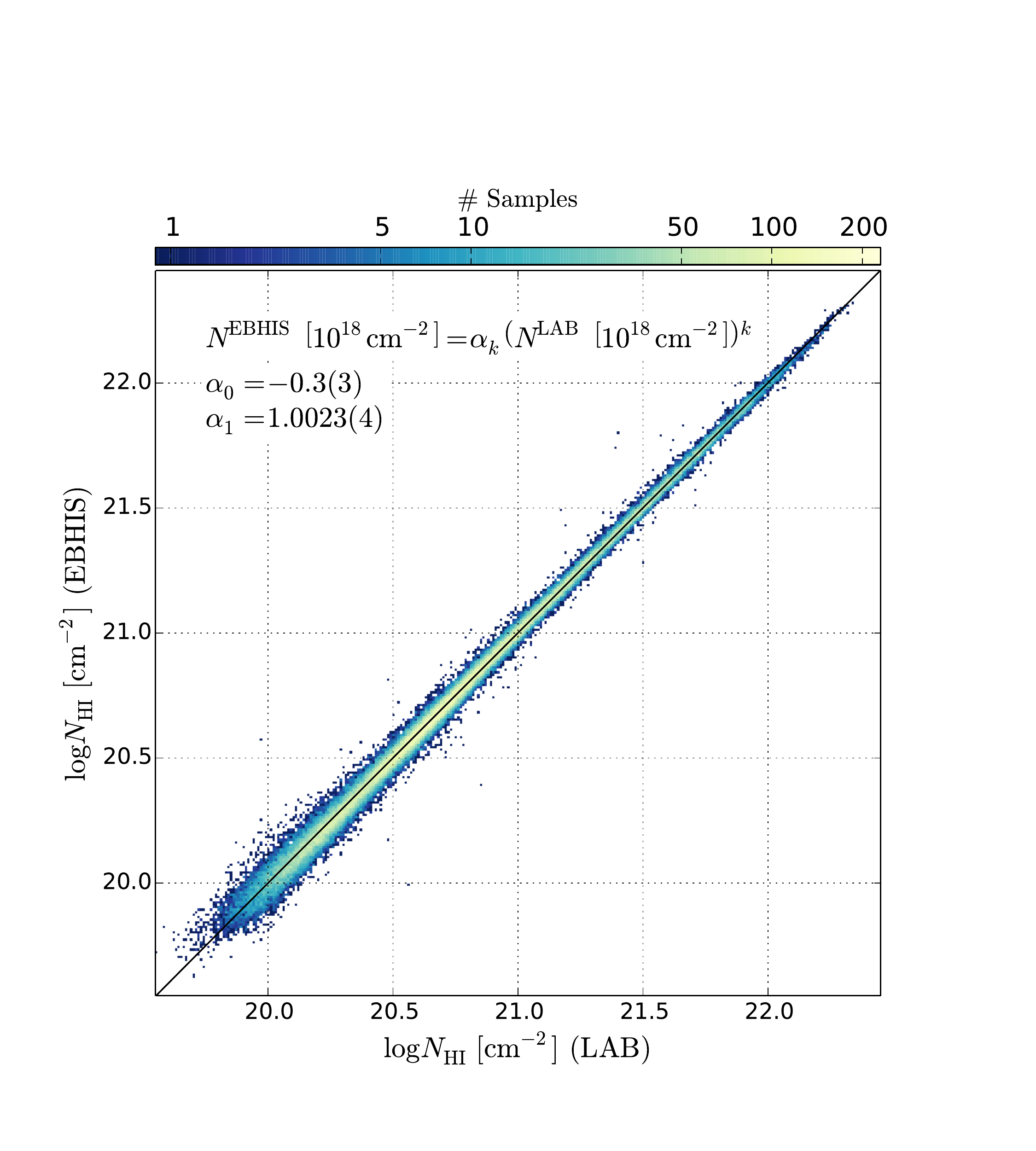}~
\includegraphics[width=0.32\textwidth,viewport=35 50 555 605,clip=]{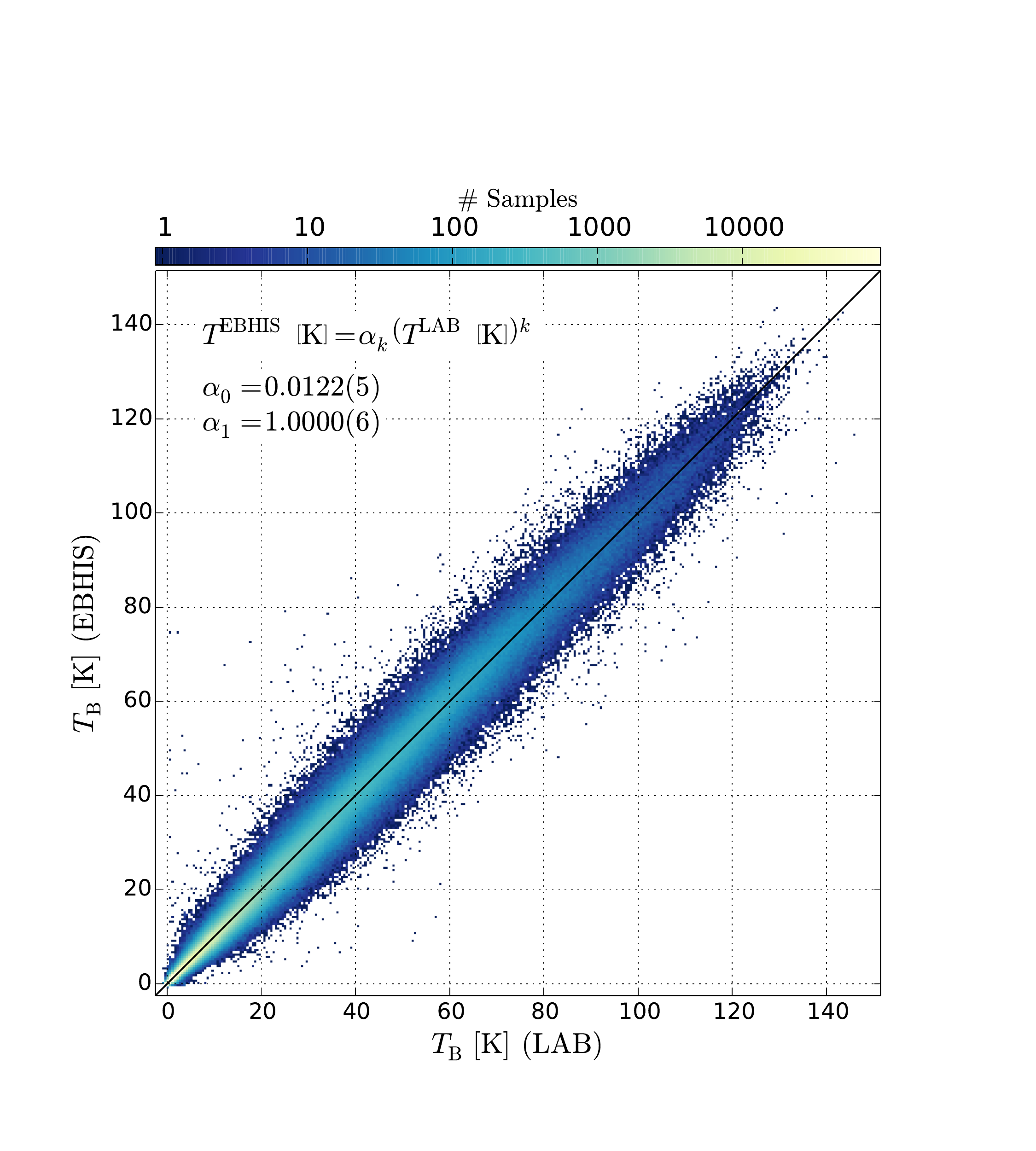}~
\includegraphics[width=0.32\textwidth,viewport=35 50 555 605,clip=]{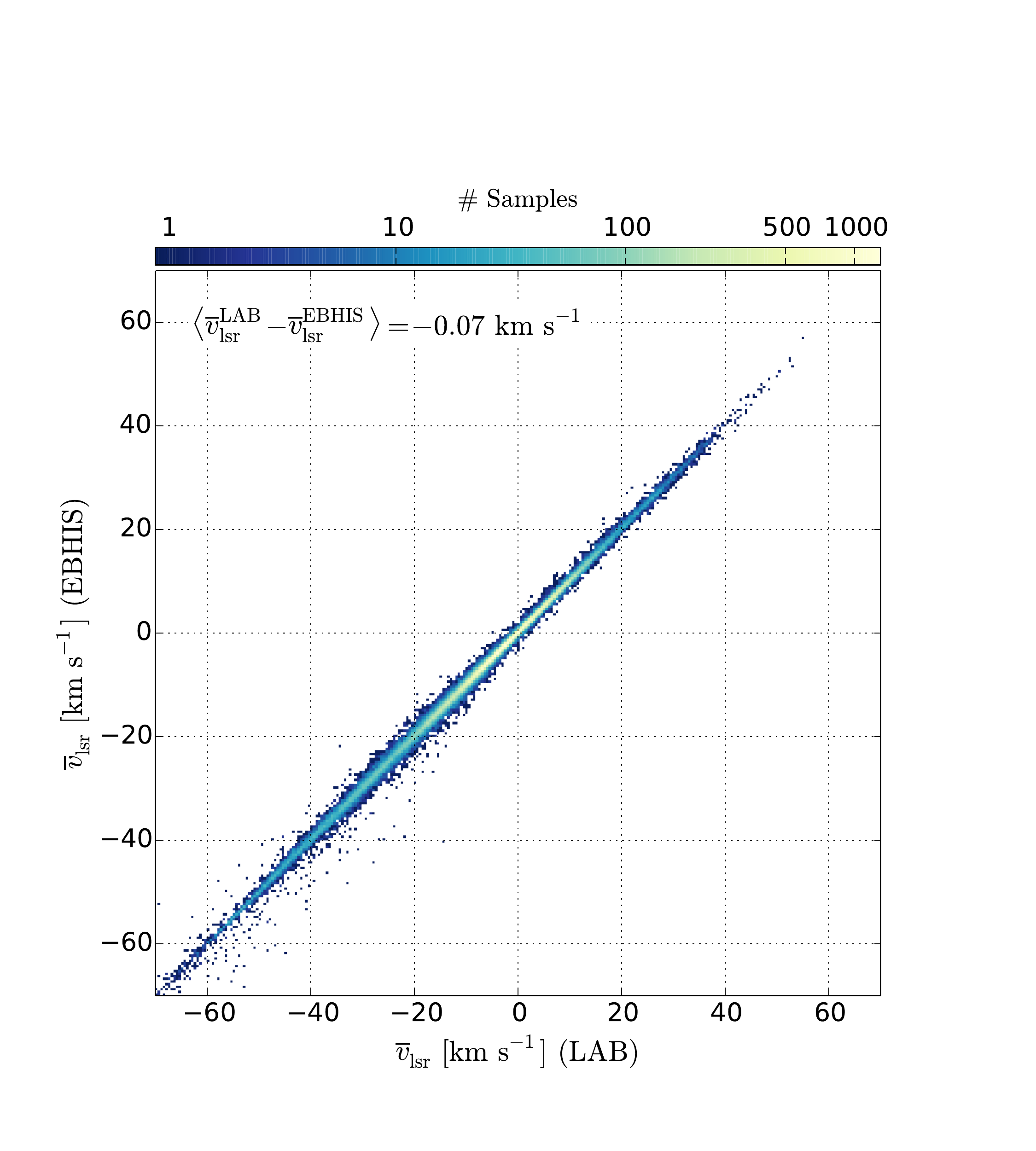}\\[0ex]
\caption{Comparison of the LAB survey and EBHIS. \textit{Left panel:} column densities; \textit{center panel:} brightness temperatures; and \textit{right panel:} intensity-weighted velocities (Moment-1). Black solid lines represent the best-fit relation (see text).}%
\label{fig:ebhis_vs_lab}%
\end{figure*}

In this section we assess the data quality of EBHIS in detail. As with many other large-scale single dish surveys, there are still residual artifacts visible in the data. One important reason is the sheer amount of data that makes automatic processing necessary. In \citet{winkel10} and Section~\ref{sec:datareduction}, we presented how many problems with the raw data had to be dealt with. In general the developed data reduction software proved to be very robust. In the following we comprehensively discuss the cases where our data processing pipeline reaches its limits. This is not because we consider the final data products lacking in quality compared to, for instance, GASS and LAB, but instead we want the potential users of EBHIS to be able to work efficiently and be well-informed with the data. Furthermore, we thoroughly cross-checked the EBHIS intensity calibration against LAB and GASS.

\subsection{Calibration: consistency check vs. LAB and GASS}\label{subsec:calibrationaccuracy}
To have an independent consistency check of the overall brightness temperature calibration, baseline solution, and residual SR, we compared the EBHIS data with the LAB survey. To avoid distortion of the results by gridding effects and angular undersampling of the sky in LAB, we took the original pointing positions of LAB and computed a weighted average of EBHIS dumps surrounding the associated positions.

In Fig.~\ref{fig:ebhis_vs_lab} (left panel) we show the comparison of total column densities between both surveys. The relation between the two is almost perfectly a one-to-one correlation, based on a linear model fit to the data:

\begin{equation}
N_\ion{H}{i}^\mathrm{EBHIS} = 1.0023(4) \cdot N_\ion{H}{i}^\mathrm{LAB} - 0.3(3) \cdot10^{18}~\mathrm{cm}^{-2}\,.
\end{equation}

To account for both errors in x and y, orthogonal distance regression \citep{boggs90} is used \citep[provided by the SciPy module odr;][]{SciPy}. Proper statistical column density errors, used as weights for the regression, are found using error propagation of the frequency-dependent RMS accumulated in the column density integral.

Likewise, the center panel of Fig.~\ref{fig:ebhis_vs_lab} displays the relationship of brightness temperatures between EBHIS and LAB. Again, a one-to-one correlation is found:
\begin{equation}
T_\mathrm{B}^\mathrm{EBHIS} = 1.0000(6) \cdot T_\mathrm{B}^\mathrm{LAB} + 12.2(5)~\mathrm{mK}\,.
\end{equation}

By computing the intensity-weighted velocity (Moment-1) for each spectrum, it is possible to test the correctness of the frequency/velocity scale (see. Fig.~\ref{fig:ebhis_vs_lab}, right panel). The median velocity mismatch between EBHIS and LAB is $0.07~\mathrm{km\,s}^{-1}$. This indicates a well-behaved relationship. For an insignificant number of positions, we observe outliers. We attribute this to residual RFI and other artifacts in either of the data sets, which can have a strong impact on the Moment-1 calculation. We minimize the potential impact of artifacts in the data on the Moment-1 calculation by applying a mask with a threshold of 2~K.

While the two surveys agree extremely well on average, it has to be noted that the brightness-temperature scatter about the linear relation is relatively high. We attribute this to the LAB survey because a comparison of EBHIS with GASS reveals much lower scatter; see Appendix~\ref{appsubsec:flux_comparison}, where we provide comparison plots for EBHIS, GASS, and LAB computed for the common data slice ($-4.5\degr\leq\delta\leq-0.2\degr$). Unfortunately, we could not find the cause of this enhanced scatter. One well-known problem of the LAB survey is that it is not fully sampled in the spatial domain. But since we compare spectra on the original pointing positions, the angular sampling is irrelevant here. For completeness, we also show similar comparison plots for GASS vs. LAB on the southern hemisphere in Appendix~\ref{appsubsec:flux_comparison}.

Another important finding, which we made during our tests, is that EBHIS and GASS (second data release), as well as LDS (the northern part of LAB) and GASS, showed a discrepancy in calibration by about 4\%. On the other hand, the IAR contribution to LAB ($\delta\leq-30\degr$) matches the GASS flux-density scale, which means that there is an inherent inconsistency in the LAB data between the northern and southern parts. This finding led to a revised intensity calibration scale for the third GASS data release \citep{kalberla15}. The rescaled GASS is very consistent with EBHIS and LDS in terms of column densities and brightness temperatures (see Figs.~\ref{fig:ebhis_vs_gass_vs_lab_16.2} and~\ref{fig:gass_vs_lds_16.2}). As a consequence, however, the IAR part of LAB became inconsistent with GASS (Fig.~\ref{fig:gass_vs_iar_16.2}).

\subsection{RFI}\label{subsec:quality_rfi}

\begin{figure*}[!t]
\centering%
\includegraphics[width=0.9\textwidth,viewport=50 25 890 590,clip=]{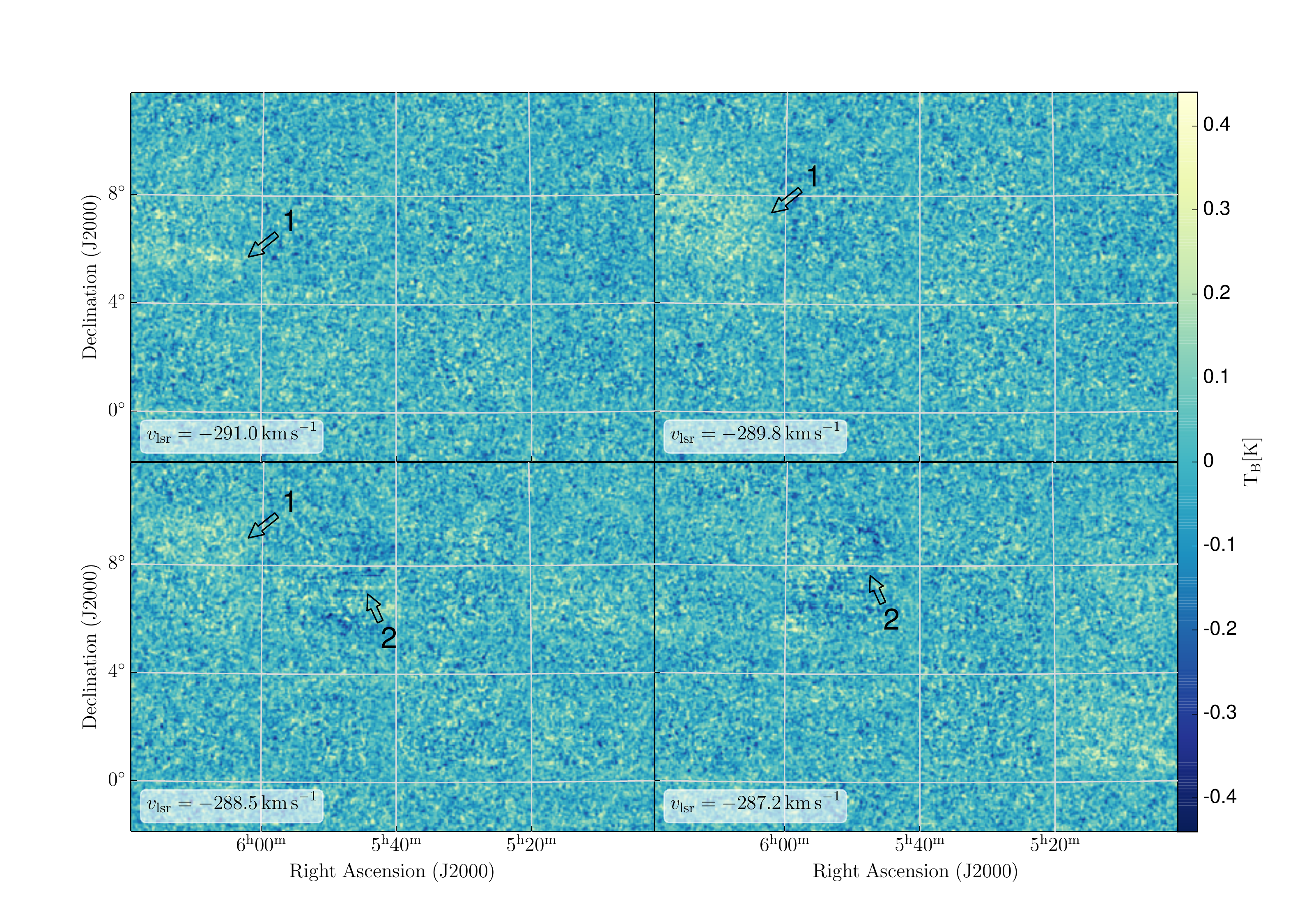}\\[0ex]
\caption{Example of residual artifacts in the data caused by imperfect subtraction of near-constant-amplitude narrowband RFI. Arrows (1) and (2) denote two different features that appear as ``wave-like'' patterns. The average residual intensities of both events are about $75~\mathrm{mK}$ and $-45~\mathrm{mK}$, respectively, which is below the $1\sigma$ noise level of EBHIS.}
\label{fig:quality_rfi_waves}%
\end{figure*}

According to Section~\ref{subsubsec:autorfiflagger}, the automatic RFI flagger is able to find even faint narrow-band interference. As described earlier, we attempt to subtract near-constant-amplitude narrow-band events from the data to minimize loss of samples caused by the flagging. This subtraction is done subscan-wise per receiver feed and polarization channel. For each flagged channel in the average spectrum, we calculated the difference to the mean value of adjacent channels and corrected for the offset. To distinguish whether an RFI event has constant intensity, we measured the RMS of the time series in the spectral channel relative to interference-free regions. Both procedures are, of course, affected by noise. In case of inaccurate RFI intensity estimates or wrong classification, residual artifacts will be visible in the data cubes. The LSR Doppler shift correction, which is subsequently applied, can shift these artifacts slightly along the spectral axis, with the size of the shift being a function of spatial position and observing time.

Figure~\ref{fig:quality_rfi_waves} shows the visual signature of the residual RFI that appears as ``wave-like'' structure along the frequency axis. We marked two distinct features in four consecutive spectral planes with arrows to emphasize the change in position. For visualization purposes we show two strong artifacts. For most of the residual RFI, an experienced astronomer might not even recognize the features in a single channel map, but when sliding through the data cube, the trained human brain can identify such faint changes.

\subsection{Stray radiation}

\begin{figure*}[!t]
\centering
\includegraphics[width=0.9\textwidth,viewport=66 10 970 458,clip=]{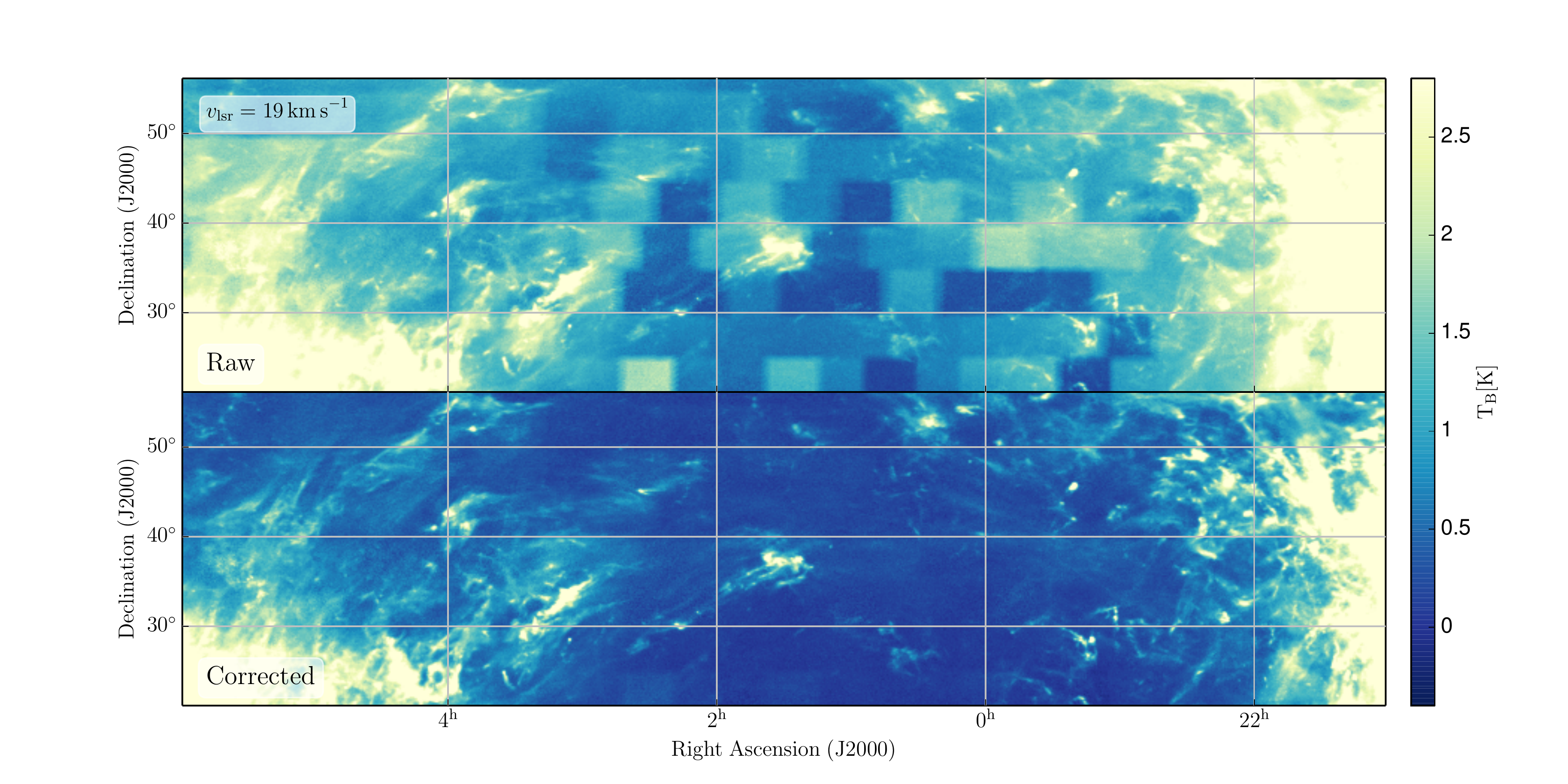}\\[0ex]
\caption{Example for a region that is significantly affected by stray radiation. Top panel: brightness temperature distribution without stray-radiation correction applied. Bottom panel: same as top panel after correction for stray radiation. The blocky appearance in the top panel is caused by the $5\degr\times5\degr$ observing grid.}
\label{fig:srcorrection_example}
\end{figure*}

The quality of the SR correction can in principle be assessed by repeated observations of each field and calculation of the scatter between the measurements \citep{kalberla80a}. To apply this method, we simply lack data, because each field was observed only once. It is possible, however, to use the small overlaps between the individual fields, which were usually observed on different dates or even seasons. Still, only two realizations of the SR are recorded per overlap, meaning that this method can only be used to measure the average scatter of the ensemble of all fields. Likewise, comparison to other datasets, such as the LAB survey or GASS, can be used to estimate deviations. However, the difference between SR correction and baseline fitting errors cannot be worked out sufficiently with such an all-sky statistical attempt, making this approach not well-suited to our SR-correction optimization. For an analysis of the ensemble distributions, we refer to Section\,\ref{sec:uncertainties} where the overall uncertainties of EBHIS data are studied in more detail.

Nevertheless, for the SR correction, a quality measure was needed during the iterative manual optimization. A more local approach was chosen, with visual inspection of larger regions with relatively weak line emission, as in the field shown in Fig.\,\ref{fig:srcorrection_example}. Here, residual far side-lobe SR errors manifest as patches associated with the individual observed fields (see Fig.\,\ref{fig:srcorrection_example} top panel). The amplitudes of these patches is then minimized during the optimization. It is helpful to look at the largest remaining amplitudes of the SR patches. Here, we find errors of up to 200~mK, which is only twice as large as the noise level, $\sigma_\mathrm{rms}$, of EBHIS data at full resolution. Furthermore, the nominal noise level of large emission-line free regions is used as a measure of correction quality. As SR artifacts are predominant close to MW disk velocities, the number of such regions is limited, though.

By looking at spectral-channel maps, it is easier to differentiate between SR errors and baseline effects, because the typical velocity width of SR errors is smaller than that of baseline errors, although there is a regime of a few $\mathrm{km\,s}^{-1}$ that could be explained by both. For the visual inspection, we mostly used data that is averaged over four velocity channels, reducing the system noise by a factor of about two.
\begin{figure*}[!t]
\centering%
\includegraphics[width=0.95\textwidth,viewport=40 2 730 220,clip=]{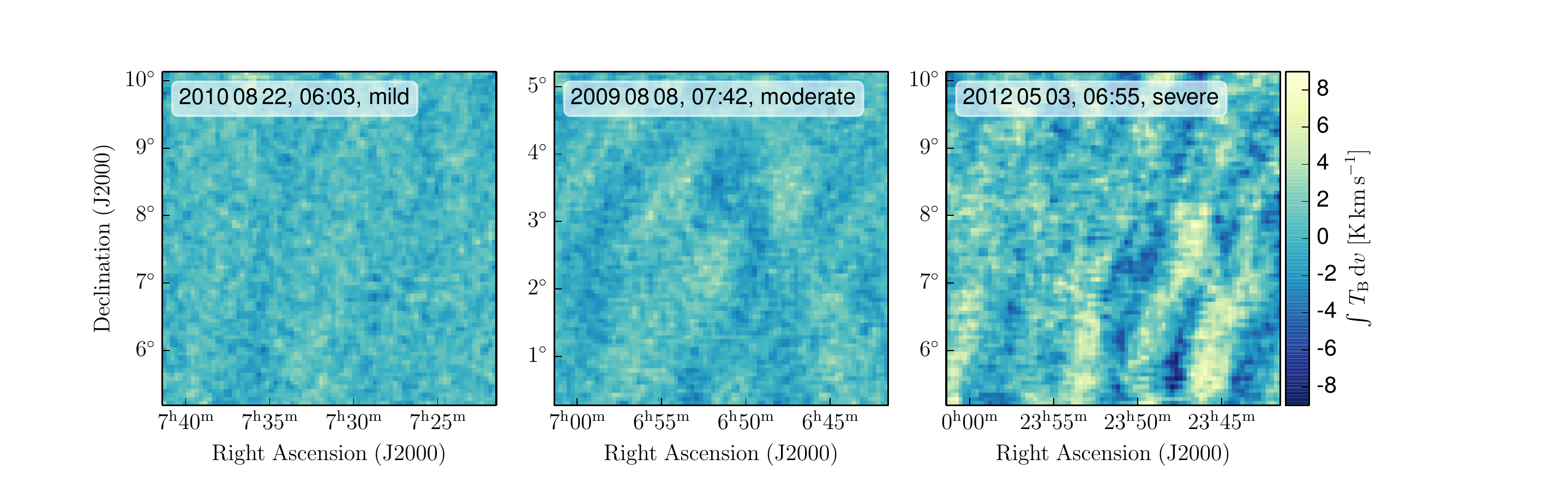}\\[0ex]
\caption{Examples for the classification of solar interference. Fields with moderate or severe ripples were observed again. To emphasize the effect, each panel shows the velocity-integrated $T_\mathrm{B}$ over 100 spectral channels.}%
\label{fig:quality_solar_interference}%
\end{figure*}

A thorough analysis of EBHIS SR correction quality is also done in \citet{martin15} in the context of their GHIGLS project (\ion{H}{i} mapping at intermediate Galactic latitude using the Green Bank Telescope). In total 15 selected relatively low-column-density fields were observed. For three of the targets \citep[NEP, NCPLEB, and DRACO; see][for definition]{martin15}, we provided EBHIS prerelease data cubes for comparison. The best-fit relation between EBHIS and GHIGLS flux-density scale is reported as
\begin{equation}
T_\mathrm{B}^\mathrm{EBHIS} = 1.009(11) \cdot T_\mathrm{B}^\mathrm{GHIGLS}\,.
\end{equation}

Using this scale factor, individual and averaged spectra (per field) are compared. In contrast to the GBT data, EBHIS can have substantial SR contribution even at high-velocity-cloud velocities owing to the different antenna type. For the SR-corrected spectra, however, both data sets agree very well. We refer the interested reader to \citet{martin15} for more details.

\subsection{Baselines}

As discussed in Section~\ref{subsec:baselines}, we initially had problems with baseline depression in regions around faint emission that was not correctly masked during the fitting process. Such artifacts could be largely suppressed using 2-D polynomial fits in combination with improved masking procedures. However, the nature of any baseline fitting process makes the baseline estimate within a masked region uncertain, because one can only interpolate between the surrounding data points. This becomes a serious problem if the mask covers several tens to hundreds of spectral channels and high polynomial orders are used. Owing to the complexity of the baseline, we have to apply relatively high polynomial orders (see Section~\ref{subsec:baselines}). From our experience, at low Galactic latitudes where the mask can be up to a few hundred spectral channels wide, uncertainties become significant. Not knowing the true baseline (and there is no way to measure it accurately\footnote{Using a hot or cold load would in principle allow removing some of the complexity in the underlying system temperature curve, allowing smaller polynomial orders. However, for the 21-cm seven-feed receiver, the load would need to be placed in front of the feeds -- because the first stages of the front end are thought to produce most of the variations -- which is simply not possible for practical reasons given the physical size of the feed assembly.}), we can only try to assess resulting baseline errors in a statistical sense (see Sect.~\ref{sec:uncertainties}).

There are two additional effects that can cause baseline problems: solar interference and strong continuum sources. We describe them in the following.

\subsubsection{Solar interference}\label{subsubsec:solar_ripples}
In the L band, the Effelsberg 100-m is vulnerable to solar radio emission, which can produce strong time-variable sinusoidal intensity variations, ``ripples'', in the spectra. To avoid this so-called solar interference, EBHIS observing sessions are mainly scheduled during night-time. Still, about a tenth of the maps have been recorded during the daytime (mainly sunsets/-rises at beginning/end of a session). For these, we maximized the angular distance to the Sun; however, in some cases the data is still affected to some extent. This is caused by scattering of the solar radiation at the telescope structure, i.e., entering the receiving system via far-side lobes.

In Fig.~\ref{fig:quality_solar_interference} we give an example of how the resulting ripples appear in the final data cubes. To make them visible, 100 spectral channel maps without line emission were summed up. Three different fields, all observed during sunrise, show various severity levels from mild (RMS increased by less than 50\%), moderate (50\% to 100\%), to severe distortion ($\geq$100\%). It is interesting to note, that -- by chance -- the distance of the three field centers to the Sun was about $40\degr$, and also the bearing was comparable ($\sim$55$\degr$ for left and right panels, $\sim$70$\degr$ for center panel). Despite the similar conditions, the ripple differs significantly in strength, demonstrating that the degradation depends sensitively on the exact path of incident of solar rays.

In the majority of the fields affected by solar interference, the ripples are difficult to spot in the full-resolution data cube. Despite that, we decided to re-observe all maps that were classified as `moderate' or `severe' cases.

\subsubsection{Standing-wave pattern}

\begin{figure}[!t]
\centering%
\includegraphics[width=0.49\textwidth,viewport=10 10 525 350,clip=]{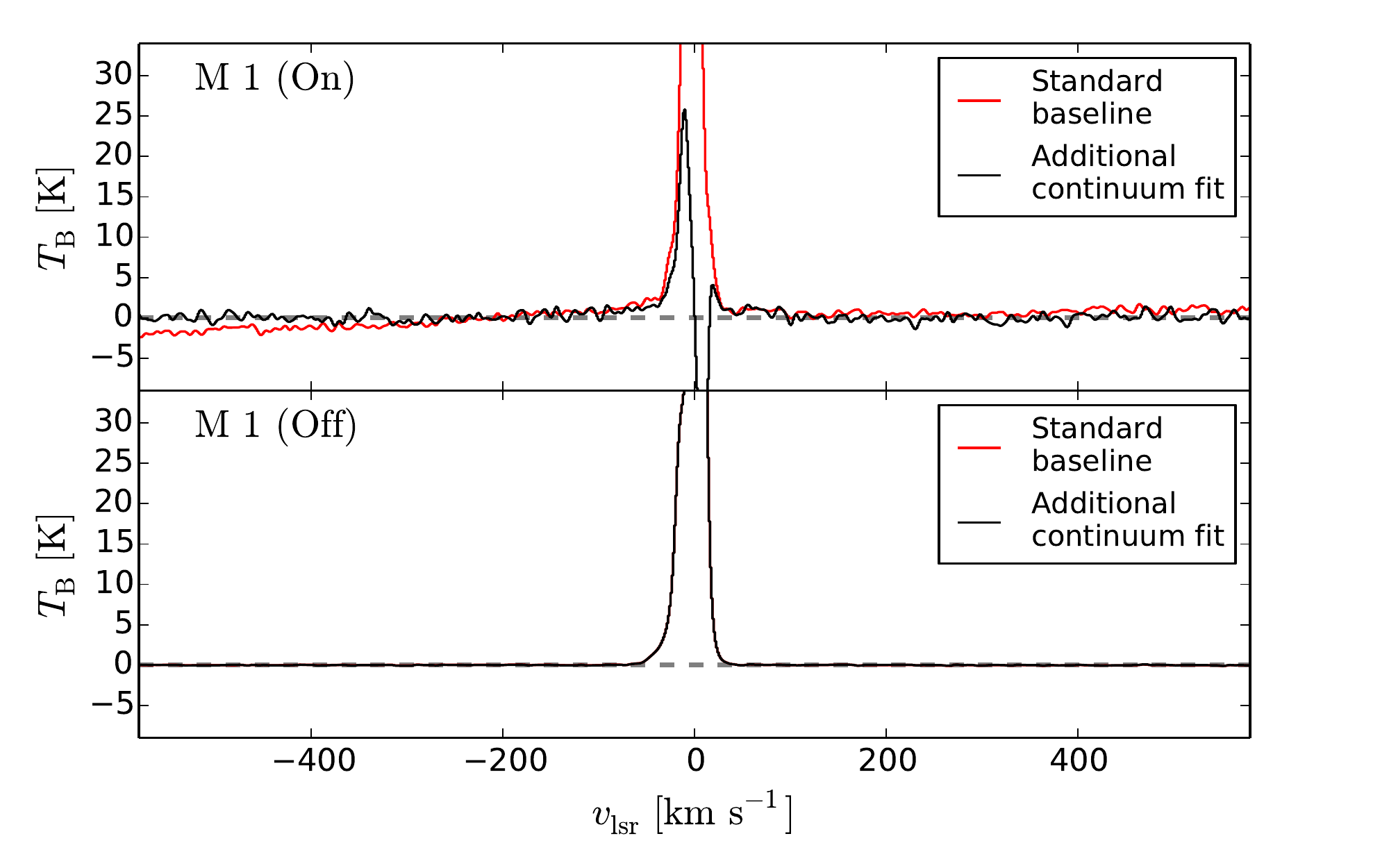}\\[0ex]
\includegraphics[width=0.49\textwidth,viewport=10 10 525 350,clip=]{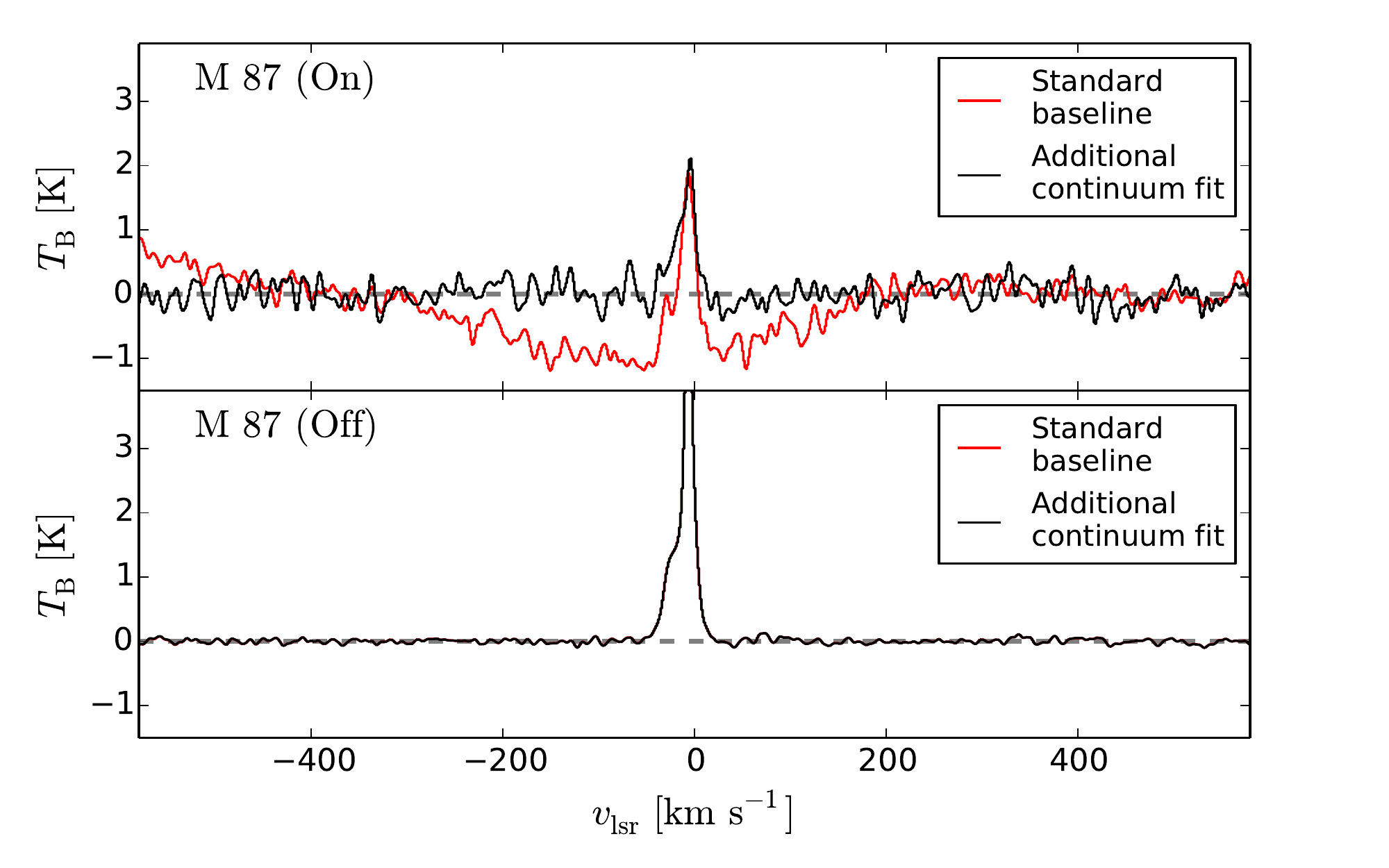}\\[0ex]
\caption{Two of the brightest continuum sources, \object{M\,1} (\object{Tau\,A}, top panel) and \object{M\,87} (\object{Vir\,A}, bottom panel). For comparison we also show \textsc{Off} positions below each \textsc{On} source spectrum. Such strong continuum sources can cause residual ripples in the final data (red curves). Therefore, we employ an additional adaptive low-order polynomial baseline fitting, producing sufficiently flat baselines (black curves).}%
\label{fig:quality_adaptive_contfit}%
\end{figure}

Like most modern receiving systems, the L-band seven-beam receiver at Effelsberg utilizes low-noise amplifiers (LNAs) that have a much higher bandwidth than what is processed by the intermediate-frequency system and the back-ends. Such broad-band LNAs can make a system vulnerable to ``out-of-band'' RFI at usually higher\footnote{This is because a waveguide acts as high-pass filter, i.e., all radiation above a certain cutoff frequency can enter the system.} frequencies than recorded.

Furthermore, the L-band receiver is also used for ongoing pulsar surveys. For these projects, a larger bandwidth of 250~MHz is desired \citep[e.g.,][]{kramer13,barr13}. Therefore, the seven-feed system was designed to process frequencies above 1260~MHz, and the 100~MHz bandpass filter required for EBHIS is only inserted at the very end of the IF chain.

As a result, the receiver feeds and front-end LNAs are sensitive to radiation in the frequency range of about 1260~MHz to $2+$~GHz. For the receiver designers, GSM at 1.8~GHz was a major concern, because it could easily drive the receiver into saturation. As a countermeasure, the seven-beam system makes use of waveguide resonance cavities, which provide a bandpass-like behavior around the frequency range of interest \citep[for details see][]{keller13}.

A side effect of this is a multimodal wave-like pattern across the spectrum. It can be well-fitted using our baseline algorithms, though the necessary polynomial order in the spectral domain needs to be relatively high (compare Section~\ref{subsec:baselines}). A more serious side effect, however, is that strong continuum sources cause a change in the pattern. The baseline solution, which is calculated for each subscan as a whole (2-D fit, see Section~\ref{subsec:baselines}), is then not working well. As a consequence, bright continuum sources (above a few Jy) produce a residual standing-wave-like pattern in the data cubes, since the average continuum level itself is subtracted from the spectral line cubes.

The residual ripples can fortunately be described by low-order polynomials. As a post-processing step, we usually apply a third-order polynomial fit across the full velocity range, i.e., not on the subtiles used for 2-D baselining. For NVSS sources in excess of 1.5~Jy, we use an adaptive scheme, where the polynomial order is increased (maximum: 8) until the residual is sufficiently flat; i.e., the RMS in the residual has converged. This approach produces very good baselines even for $\sim$100-Jy-class sources (see Fig.~\ref{fig:quality_adaptive_contfit}). For \object{M\,1} (top panel) and \object{M\,87} (bottom panel) and respective \textsc{Off} positions in the direct vicinity, the spectra with and without continuum-polynomial-fit post-processing are shown. In all cases, the additional treatment produces very flat baselines.

In Fig.~\ref{fig:quality_adaptive_contfit} two additional effects, which are associated with continuum radiation, can be studied. First, the brightness temperature noise is significantly higher toward bright
continuum sources, caused by the much higher system temperature. Second, for \object{M\,1} (\textsc{On}), the emission line itself appears completely different in the post-processed version. In fact, the EBHIS pipeline measures the (line-free) RMS in each spectral dump and flags outliers as bad. Therefore, in the final data product, some original spectral dumps were not accounted for in the gridding process -- those being located closest to the bright continuum source. As these contain the strongest absorption imprint, the associated red curve in Fig.~\ref{fig:quality_adaptive_contfit} is missing parts of the \ion{H}{i} absorption feature. The effect is also visible for the \object{M\,87} case but to a lesser degree.

\section{Uncertainties of EBHIS data}\label{sec:uncertainties}

Uncertainties of the EBHIS data can be subdivided into three main categories: pure thermal noise, calibration uncertainties, and errors in the baseline (which includes SR). Brightness temperature RMS can be determined easily from emission line-free regions in the \ion{H}{i} data under the assumption that the noise is normal-distributed. In Section~\ref{subsec:calibrationaccuracy} we have already shown that intensity calibration is unbiased with respect to the established LAB survey data. Here, we have a closer look at the typical scatter in the calibration and on systematic effects caused by baseline subtraction and SR correction, which is a more complicated task.

\subsection{Noise map}\label{subsec:noisemap}
\begin{figure}[!t]
\centering%
\includegraphics[width=0.48\textwidth,viewport=35 40 545 550,clip=]{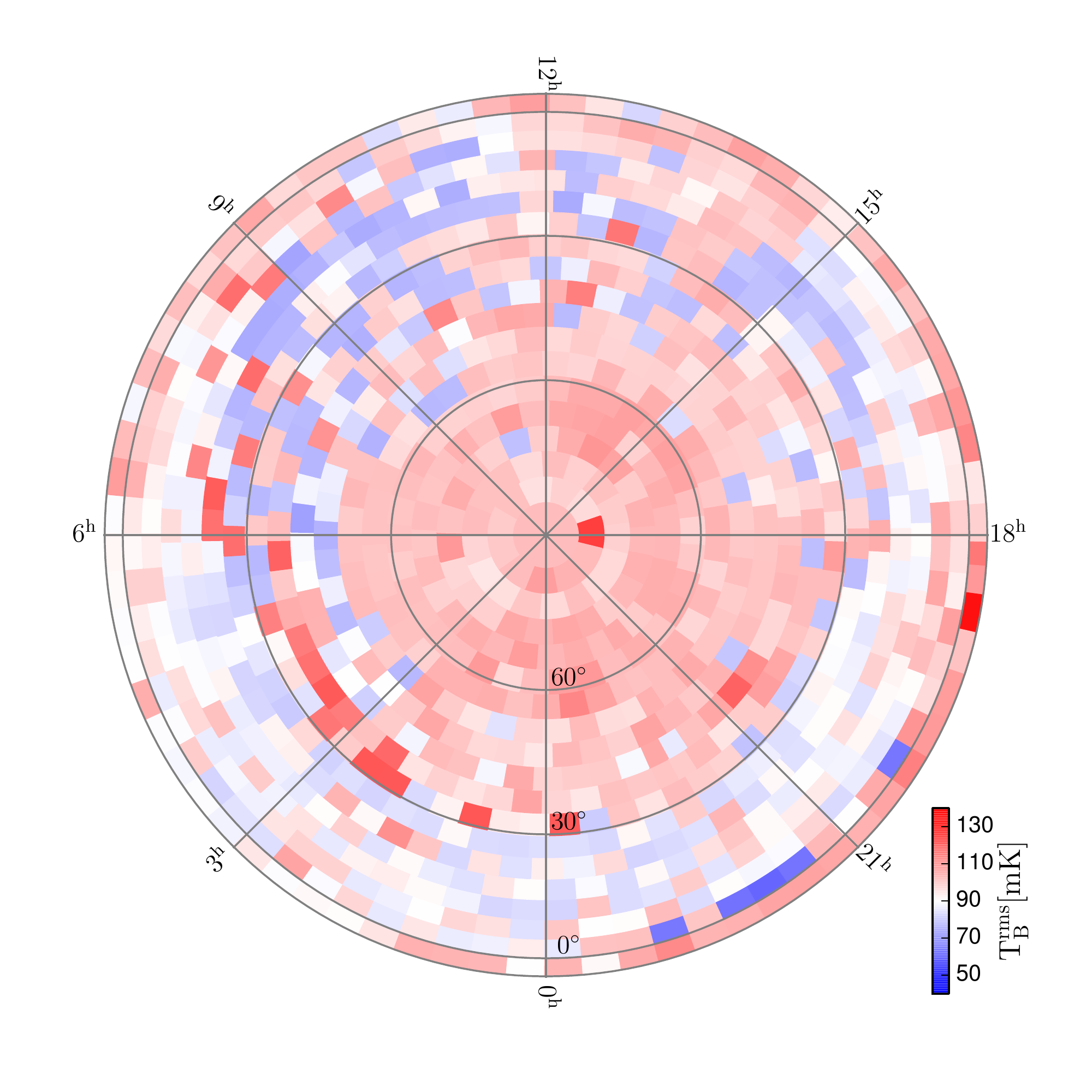}\\[0ex]
\caption{EBHIS noise, $\sigma_\mathrm{rms}$, at full resolution per observed field.}%
\label{fig:allsky_rms_ZEA_G10_patches}%
\end{figure}

EBHIS data are taken at all seasons, thus at different ambient temperatures, elevation angles, and system temperatures. This means that final RMS noise levels in the spectral data can differ substantially. Even though the MW covers a substantial fraction of the Galactic EBHIS data, each individual spectrum contains enough emission line-free spectral channels to robustly compute the RMS from the data itself. We applied an iterative procedure of repeatedly computing the standard deviation $\sigma_\mathrm{rms}$ in a spectrum and subsequently clipping outliers in excess of 3$\sigma_\mathrm{rms}$. Usually, this scheme converges after three or four iterations. In Fig.~\ref{fig:allsky_rms_ZEA_G10_patches} we display the average RMS, measured this way, per observed field. The overall median RMS value is $\sigma_\mathrm{rms}=90~\mathrm{mK}$.

The RMS in spectral channels containing \ion{H}{i} line emission will be higher, because line emission contributes to the system temperature and as such increases noise. Likewise, the baseline system temperature is not constant such that the RMS is slightly frequency-dependent (typically $\lesssim$1\% over the MW velocity interval).

\subsection{Uncertainties from EBHIS--LAB comparison}\label{subsec:ebhis_lab_uncertainties}

\begin{figure}[!t]
\centering
\includegraphics[width=0.48\textwidth,viewport=22 15 553 570,clip=]{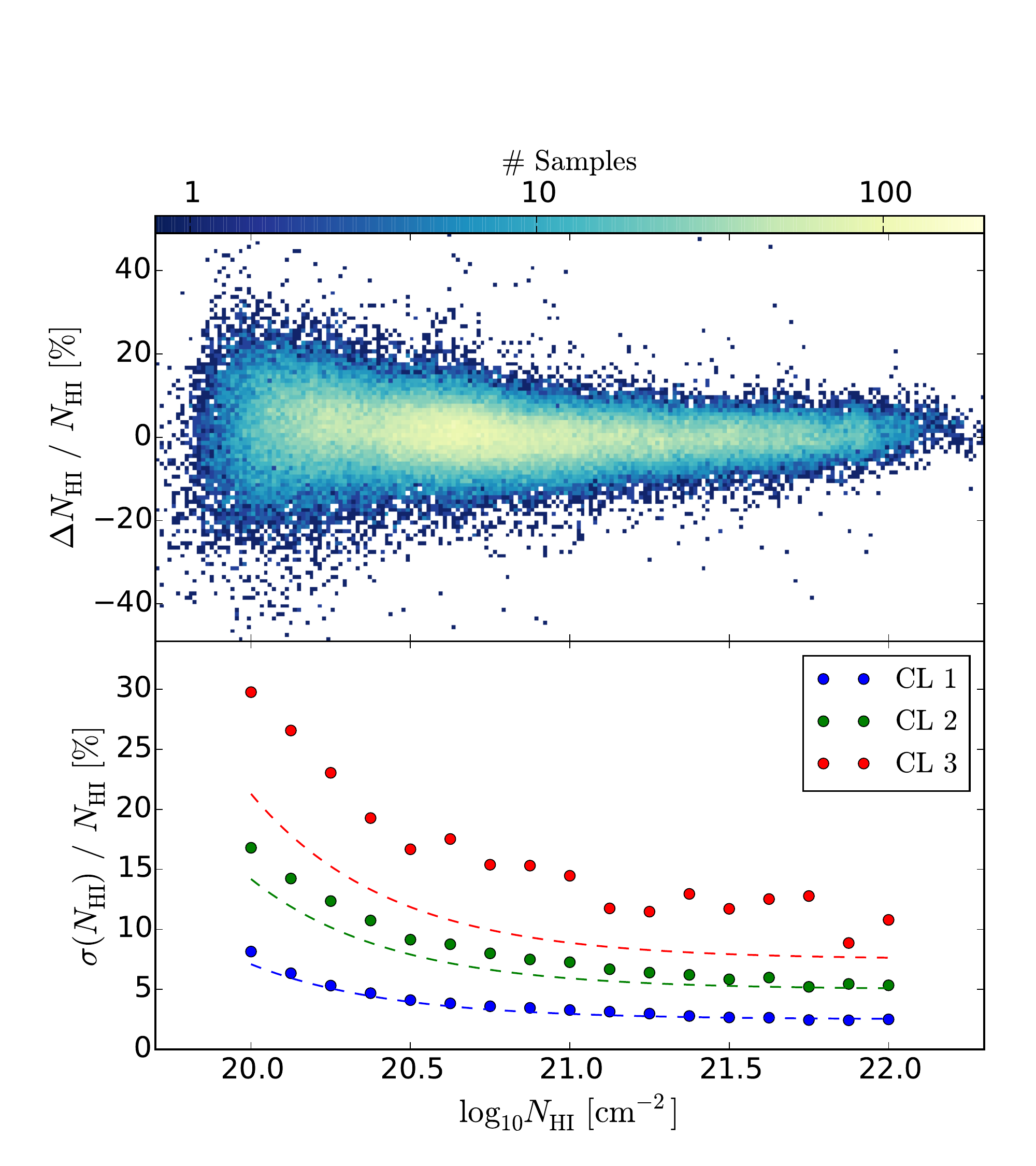}\\[0ex]
\caption{Uncertainty of the EBHIS--LAB column density distribution. \textit{Top panel:} relative column density difference of EBHIS with respect to LAB. \textit{Bottom panel:} the 68.3\% (CL-1), 95.4\% (Cl-2), and 99.7\% (CL-3) confidence intervals from the distribution in the top panel. Because the difference between two distributions has a wider spread than the two constituents, we rescale the difference $\Delta N_\ion{H}{i}$ to the equivalent standard deviation $\sigma(N_\ion{H}{i}) = \Delta N_\ion{H}{i} / \sqrt{2}$. The dashed lines represent the theoretical expectation, for the case where the error is caused by pure thermal noise, as measured in emission-line free regions, plus an additional contribution of 2.5\% that represents the typical intensity calibration uncertainties.}
\label{fig:nhi_uncertainties_ebhis_lab}
\end{figure}

In the previous sections, we have already discussed how difficult it is to disentangle the various uncertainties in the final data products (\ion{H}{i} column densities or brightness temperature spectra). In particular, the stray-radiation contribution can hardly be separated from baseline-fit errors. Therefore, we attempt to quantify the uncertainty of the column density distribution as a whole using empirical methods.

One method to do this was already utilized in Section~\ref{subsec:calibrationaccuracy}. By comparing EBHIS column densities to the LAB survey we not only can check on the correct flux-density scale, but the relative ensemble error can also be inferred from the scatter about the best-fit linear
relation in the $N_\ion{H}{i}^\mathrm{EBHIS}$--$N_\ion{H}{i}^\mathrm{LAB}$
or $T_\mathrm{b}^\mathrm{EBHIS}$--$T_\mathrm{b}^\mathrm{LAB}$ plots (Fig.~\ref{fig:ebhis_vs_lab}), for example, as a function of column density or brightness temperature. As an example, in Fig.~\ref{fig:nhi_uncertainties_ebhis_lab} (top panel), a histogram of the percentaged column density differences, $\Delta N_\ion{H}{i}\equiv N_\ion{H}{i}^\mathrm{EBHIS}-N_\ion{H}{i}^\mathrm{LAB}$, is shown as a function of $\log N_\ion{H}{i}$.

Because we work with the difference of two noisy distributions, the uncertainties of the two contributing data sets have to be smaller. If both distributions had the same standard deviation, scaling the measured width of the resulting distribution by $\sqrt{2}$ would provide a good estimator of the original width. The LAB and EBHIS surveys both have different noise levels. Taking the ratio of RMS values into account, one could easily correct for this. However, systematic uncertainties need not necessarily be distributed in the same way, for instance, their ratio could be different. It is not even clear whether the uncertainty distributions of both surveys have a similar shape. Therefore, we assume that both data sets share the same magnitude of systematic effects. This is justified further by the RMS values of LAB and EBHIS, which are comparable, and in our experience the quality of the data reduction scales with the noise level.

\subsubsection{Column density uncertainties}\label{subsubsec:coldens_uncertainties}

Applying the aforementioned scaling by $\sqrt{2}$, we can estimate the standard deviation of the EBHIS column density distribution, again as a function of $\log N_\ion{H}{i}$. This is plotted in the lower panel of Fig.~\ref{fig:nhi_uncertainties_ebhis_lab} (filled circles), as calculated in bins of width 0.125~dex. We show three confidence intervals, 68.3\% (CL-1), 95.4\% (Cl-2), and 99.7\% (CL-3), as calculated from the distribution in the top panel. The dashed lines represent the theoretical expectation, in the case where the uncertainties are caused by thermal noise in addition to a 2.5\% increase. This additional contribution reflects the typical S7-based intensity calibration uncertainties \citep{winkel10}.

The CL-1 theoretical curve matches the measured distribution width very well. Only the lowest two data points ($\log N_\ion{H}{i} \lesssim20.25$) show a slight excess, most likely caused by the imperfect SR- and baseline correction. For the two other confidence levels, the estimated distributions are wider than predicted, meaning that the systematic errors are not perfectly normal-distributed.

\subsubsection{Brightness temperature uncertainties}\label{subsubsec:tb_uncertainties}

\begin{figure}[!t]
\centering
\includegraphics[width=0.48\textwidth,viewport=22 15 553 570,clip=]{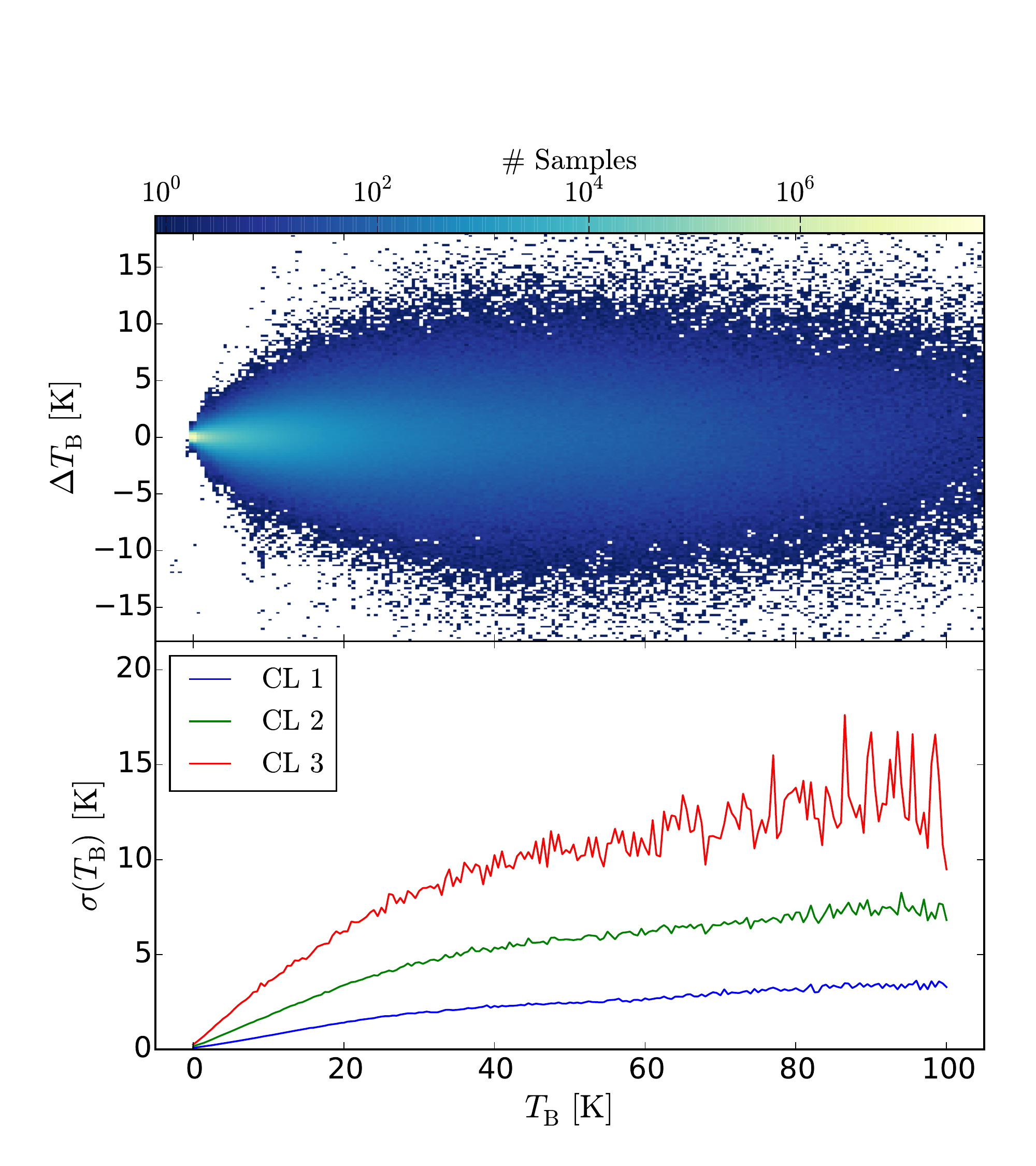}\\[0ex]
\caption{Uncertainty of the EBHIS--LAB brightness temperature distribution. \textit{Top panel:} brightness temperature difference of EBHIS with respect to LAB. \textit{Bottom panel:} the 68.3\% (CL-1), 95.4\% (Cl-2), and 99.7\% (CL-3) confidence intervals from the distribution in the top panel. Similar to Fig.~\ref{fig:nhi_uncertainties_ebhis_lab}, we use $\sigma(T_\mathrm{B}) = \Delta T_\mathrm{B} / \sqrt{2}$.}
\label{fig:tb_uncertainties_ebhis_lab}
\end{figure}

Following the same approach, we evaluated the brightness temperature uncertainties (see Fig.\,\ref{fig:tb_uncertainties_ebhis_lab}). For very low values of $T_\mathrm{B}$, the errors are consistent with the RMS noise level in EBHIS. At the highest $T_\mathrm{B}$ values, the uncertainty approaches about 2.5\% of the signal brightness temperature, which is compatible with the column density case.

\subsection{Uncertainties from EBHIS--EBHIS comparison}\label{subsec:ebhis_uncertainties}

\begin{figure}[!t]
\centering
\includegraphics[width=0.48\textwidth,viewport=22 15 553 570,clip=]{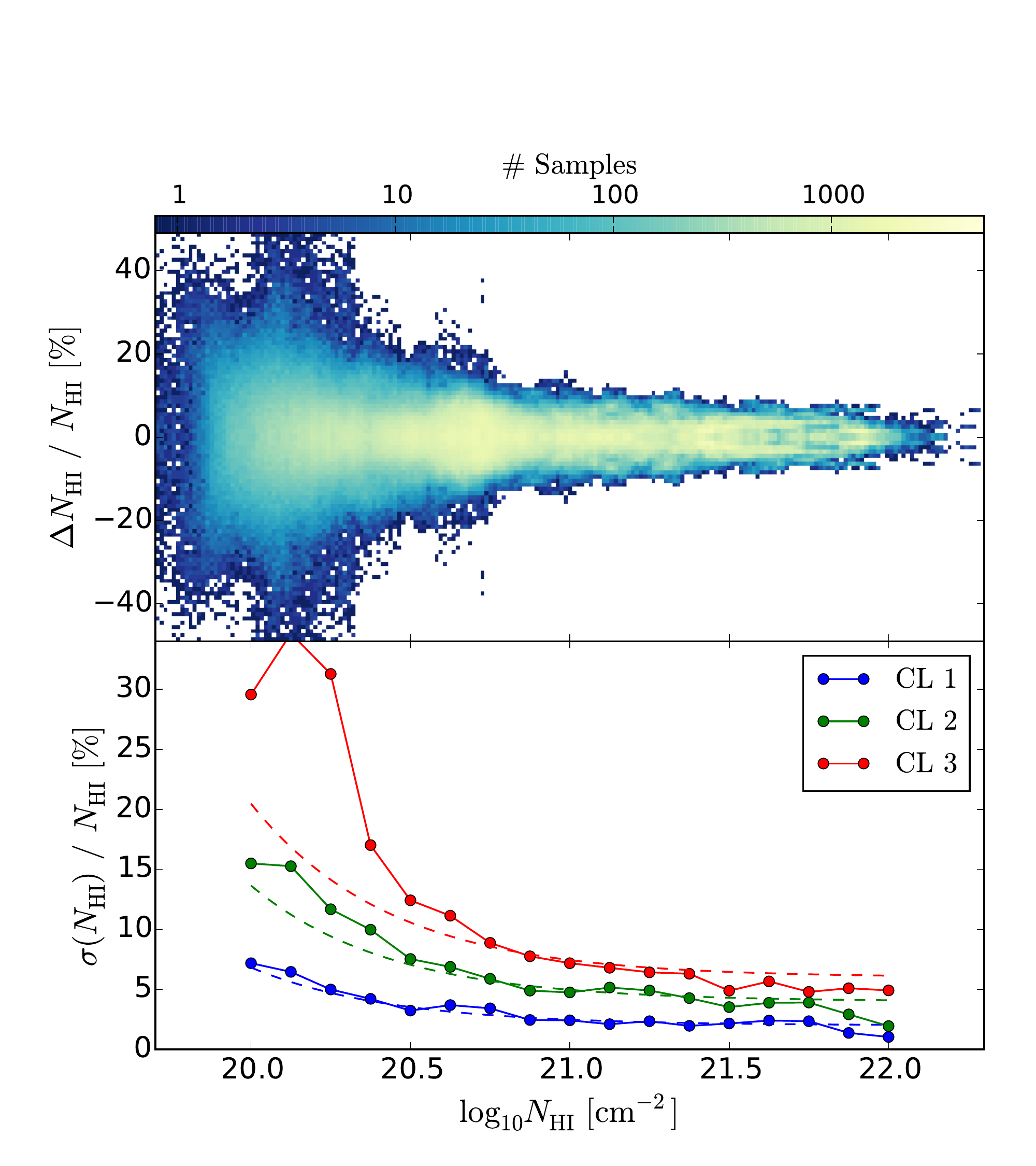}\\[0ex]
\caption{As in Fig.~\ref{fig:nhi_uncertainties_ebhis_lab} but computed from column density differences of EBHIS in the overlap regions of the observed fields.}
\label{fig:nhi_uncertainties_ebhis_ebhis_300}
\end{figure}

\begin{figure}[!t]
\centering
\includegraphics[width=0.48\textwidth,viewport=22 15 553 570,clip=]{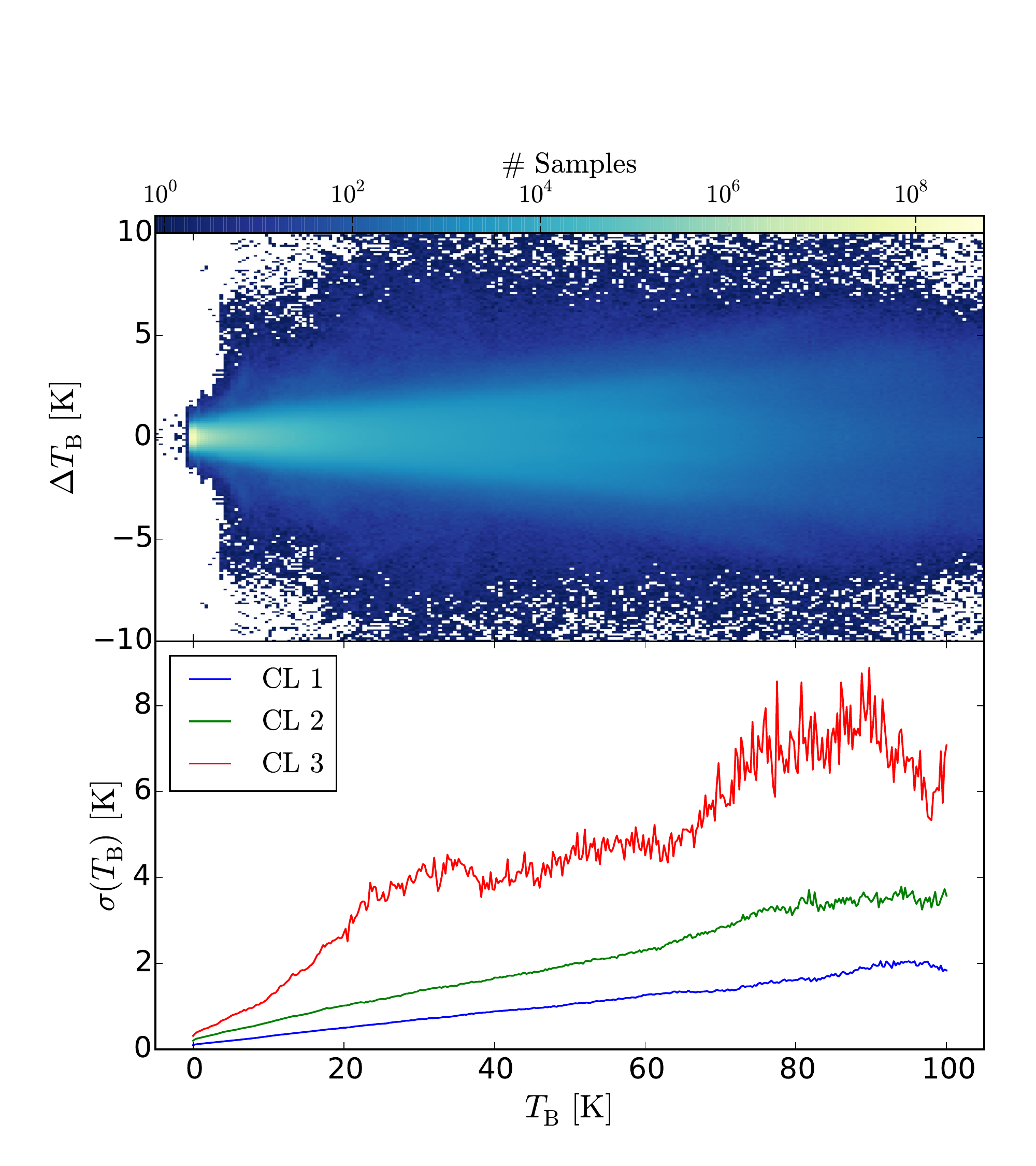}\\[0ex]
\caption{As in Fig.~\ref{fig:tb_uncertainties_ebhis_lab} but computed from brightness temperature differences of EBHIS in the overlap regions of the observed fields.}
\label{fig:tb_uncertainties_ebhis_ebhis_300}
\end{figure}

A particular drawback of the EBHIS--LAB comparison is the impossibility of getting distinct systematic errors for each of the two surveys. It remains unclear, whether EBHIS or LAB or both surveys together determine the scatter in the uncertainty distributions. To circumvent this problem, we tried to use EBHIS data alone for a similar study. Utilizing the overlaps of the 915 observed fields, a scatter histogram can again be inferred. One particular difficulty is that the beam weights in the overlap regime are lower than in the inside of the fields, and the RMS is likewise higher\footnote{If two neighboring observations are combined, both of them contribute to the overlap so that the RMS in the overlap is even slightly smaller than usual.}. Furthermore, by using overlap areas alone, the sampling of data points on the sky becomes highly irregular.

This effect is visible in Figs.~\ref{fig:nhi_uncertainties_ebhis_ebhis_300} and~\ref{fig:tb_uncertainties_ebhis_ebhis_300} where the distribution of $N_\ion{H}{i}$ and $T_\mathrm{B}$ deviations is much less Gaussian-like than in Figs.~\ref{fig:nhi_uncertainties_ebhis_lab} and~\ref{fig:tb_uncertainties_ebhis_lab}. Nevertheless, the confidence intervals (lower panels) are even slightly narrower than inferred from the EBHIS--LAB comparison. In particular, the $\sigma(T_\mathrm{B})$ inferred from EBHIS alone is lower and discloses a roughly linear behavior, $\sigma(T_\mathrm{B})\approx 0.02\cdot T_\mathrm{B}$ for $T_\mathrm{B}\gg\sigma_\mathrm{rms}$.

\subsection{Discussion of derived uncertainty curves}
Among the published data products, the Milky Way column density map is probably of highest scientific interest. In this section, we presented the first-ever attempt to quantify the systematic and statistical uncertainties of the $N_\ion{H}{i}$ distribution.

Although the measured $N_\ion{H}{i}$ distributions are not perfectly normal distributed, it is probably fair to say that uncertainties are mostly determined by the 2.5\% intensity calibration error already reported in \citet{winkel10}. This is a typical value for single-dish \ion{H}{i} observations. Only the lowest column densities, $N_\ion{H}{i}\lesssim2\cdot10^{20}~\mathrm{cm}^{-2}$, show a slight excess, which we attribute to baseline and SR-correction problems. However, compared to the pure thermal noise and calibration uncertainties, their impact is minor.

Also the $T_\mathrm{B}$ error distributions reveal interesting results. While $\sigma(T_\mathrm{B})$ from EBHIS alone is approximately 2\% over the full range, it shows a completely different behavior in the EBHIS--LAB plot, where it is higher at medium values of $T_\mathrm{B}$. We attribute this to the LAB survey, but the reason for this remains unclear.

It is important to note that for small $T_\mathrm{B}$, the predictions from the uncertainty distributions should not be used. Here, one should work with the local brightness temperature RMS that can easily be measured for each sight line and with high accuracy.

The results indicate that EBHIS performs somewhat better than LAB, but to be conservative, it can be safely assumed that EBHIS is at least on a par with LAB in terms of systematic uncertainties. \textit{With the completion of the second sky coverage of EBHIS data, it will be possible to assess the ensemble uncertainties to much better precision from EBHIS alone.}

\subsection{Polynomial parametrization of uncertainties}\label{subsec:uncertainty_parametrization}
\begin{table}
\caption{Coefficients for a piece-wise polynomial parametrization of $N_\ion{H}{i}$ uncertainty curves in Figs.~\ref{fig:nhi_uncertainties_ebhis_lab} and ~\ref{fig:nhi_uncertainties_ebhis_ebhis_300}.}
\label{tab:nhi_polyfit}
\centering
\begin{tabular}{l r r r r r}
\hline\hline
\rule{0ex}{3ex}CL & $\log N^\mathrm{l}$ & $\log N^\mathrm{u}$ & $a_0$ & $a_1$ & $a_2$ \\
\hline
\multicolumn{6}{l}{\rule{0ex}{3ex}LAB--EBHIS (Fig.~\ref{fig:nhi_uncertainties_ebhis_lab})} \\
\hline
\rule{0ex}{3ex}1 & 19.0 & 20.5 & 7851.92 &  -768.187  &   18.800\\
1 & 20.5 & 23.0 & 279.89  &  -24.953 &     0.561\\
2 & 19.0 & 20.5  & 6490.52 &  -625.672 &    15.099\\
2 & 20.5 & 23.0  & 723.14 &   -64.890  &    1.466\\
3 & 19.0 & 20.5  & -3167.04 &   344.222 &    -9.219\\
3 & 20.5 & 23.0  & 1090.50 &   -96.962  &    2.176\\
\hline
\multicolumn{6}{l}{\rule{0ex}{3ex}EBHIS--EBHIS (Fig.~\ref{fig:nhi_uncertainties_ebhis_ebhis_300})} \\
\hline
\rule{0ex}{3ex}1 & 19.0 & 20.5 & -181.71  &   26.872 &    -0.871\\
1 & 20.5 & 23.0 & 109.16 &    -8.740  &    0.175\\
2 & 19.0 & 20.5  & -9280.02 &   936.857  &  -23.603\\
2 & 20.5 & 23.0  & 324.47 &   -27.141 &     0.569\\
3 & 19.0 & 20.5  & -125472.84 & 12466.934 &  -309.591\\
3 & 20.5 & 23.0  & 2112.36 &  -193.710  &    4.451\\
\hline
\end{tabular}
\end{table}

\begin{table}
\caption{Coefficients for a piece-wise polynomial parametrization of $T_\mathrm{B}$ uncertainty curves in Figs.~\ref{fig:tb_uncertainties_ebhis_lab} and ~\ref{fig:tb_uncertainties_ebhis_ebhis_300}.}
\label{tab:tb_polyfit}
\centering
\begin{tabular}{l r r r r r}
\hline\hline
\rule{0ex}{3ex}CL & $T_\mathrm{B}^\mathrm{l}$ & $T_\mathrm{B}^\mathrm{u}$ & $a_0$ & $a_1$ & $a_2$ \\
\hline
\multicolumn{6}{l}{\rule{0ex}{3ex}LAB--EBHIS (Fig.~\ref{fig:tb_uncertainties_ebhis_lab})} \\
\hline
\rule{0ex}{3ex}1 & 0 & 3 & 0.082 &     4.217 &    51.244\\
1 & 3 & 10 & 0.086 &     5.255 &    12.418\\
1 & 10 & 40 & -0.123 &     9.470 &    -8.800\\
1 & 40 & 150 & 1.391 &     2.165 & --\\
2 & 0 & 3  & 0.168 &    10.207 &   157.108\\
2 & 3 & 10  & 0.116 &    16.675 &    -0.965\\
2 & 10 & 40  & -0.203 &    21.741 &   -19.675\\
2 & 40 & 150  & 3.765 &     3.951 & --\\
3 & 0 & 3  & 0.257  &   28.598  &  176.797\\
3 & 3 & 10  &  0.218 &    35.736  &  -17.435\\
3 & 10 & 40  &  0.256  &   36.996  &  -33.836\\
3 & 40 & 150  &  6.743  &    7.246 & --\\
\hline
\multicolumn{6}{l}{\rule{0ex}{3ex}EBHIS--EBHIS (Fig.~\ref{fig:tb_uncertainties_ebhis_ebhis_300})} \\
\hline
\rule{0ex}{3ex}1 & 0 & 3 & 0.104 &     0.031 &    -0.003\\
1 & 3 & 150 & 0.108 &     0.019 &  --\\
2 & 0 & 3  & 0.211 &     0.070  &   -0.009\\
2 & 3 & 150  & 0.235 &     0.036 & --\\
3 & 0 & 3  & 0.331 &     0.119  &   -0.013\\
3 & 3 & 150  &  0.337 &     0.079 & --\\
\hline
\end{tabular}
\end{table}

For convenience, we calculated parametrizations of the uncertainty values displayed in Figs.~\ref{fig:nhi_uncertainties_ebhis_lab}~to~\ref{fig:tb_uncertainties_ebhis_ebhis_300} based on piece-wise polynomials. For the $N_\ion{H}{i}$ errors, Table~\ref{tab:nhi_polyfit} contains column density intervals with respective polynomial coefficients. It is
\begin{equation}
\frac{\sigma(N_\ion{H}{i})}{N_\ion{H}{i}} \left[\%\right] = \sum_{k=0}^2 a_k \left(\log N_\ion{H}{i}\right)^k
\end{equation}
for each bin $\log N^\mathrm{l}\leq \log N_\ion{H}{i} < \log N^\mathrm{u}$.

Likewise, Table~\ref{tab:tb_polyfit} provides coefficients for the brightness temperature uncertainties, with
\begin{equation}
\sigma(T_\mathrm{B}) = \sum_{k=0}^2 a_k \left(\frac{T_\mathrm{B}}{100~\mathrm{K}}\right)^k
\end{equation}
and bin intervals $T_\mathrm{B}^\mathrm{l}\leq T_\mathrm{B} < T_\mathrm{B}^\mathrm{u}$.

\section{Data products}\label{sec:dataproducts}
We make EBHIS freely available to the astronomical community via CDS. Several different data products are provided for convenience.

\subsection{Column density maps}
For many purposes one only desires the total \ion{H}{i} column density for a (set of) given position(s), e.g., to correct X-ray flux densities for foreground absorption. Full-(northern-)sky column density maps are published in FITS format \citep[][and references therein]{hanisch01} in two sky projections, zenith-equal-area \citep[ZEA; see][compare Fig.\,\ref{fig:allsky_nhi}]{greisen02,calabretta02}, and HEALPix \citep{gorski05}, which is widely used for all-sky datasets. An inofficial version also exists of the LAB column density map in the HEALPix representation\footnote{See \url{http://lambda.gsfc.nasa.gov/product/foreground/}.}. It was reprojected from the original plate-carr\'{e}e projection to HEALPix using nearest-neighbor resampling \citep{land07}. For this, we find significant deviations compared to our own calculations, which were derived from the original LAB observations rather than from an intermediary data product. We recommend not using the unofficial LAB HEALPix map any longer.

\subsection{Spectral data}
The full spectral data will be provided for download in FITS binary table format on a HEALPix grid (nside-1024, Galactic coordinates are used for indexing). Because the EBHIS Milky Way spectral data set has a size of about 50~GByte, we split it into smaller chunks, namely into the 192 HEALPix pixels of the nside-4 tessellation. Because EBHIS is only available for declinations above $-5\degr$, not all of the nside-4 pixels contain data, and these are therefore omitted. Additionally, smaller spectral data cubes of size $20\degr\times20\degr$ are available for download in FITS image format.

\section{Summary}

We presented the first data release of the Effelsberg--Bonn \ion{H}{i} Survey (EBHIS), elaborated on the recent changes to the data processing software and techniques compared to \citet{winkel10}, and studied the quality of the resulting data products in great detail. EBHIS is a 21-cm \ion{H}{i} line survey of the sky north of $\delta\geq-5\degr$, comprising full spatial sampling and high sensitivity. At full spectral resolution of $\delta v=1.44~\mathrm{km\,s},^{-1}$ EBHIS has an average noise level of $\sim90~\mathrm{mK}$. Its angular resolution is $10\farcm8$. About 1000~h of observing time with the Effelsberg 100-m telescope were spent until the completion of the first coverage that forms the basis for the first data release. Owing to the large bandwidth of 100 MHz, not only is the Milky Way emission covered in EBHIS, but also \ion{H}{i} in the Local Volume out to a redshift of $z\sim0.07$. The presented data products comprise the Milky Way emission. The analysis of the extragalactic part of the survey will be published elsewhere.

EBHIS was designed to replace the northern-hemisphere portion of the important Leiden/Argentine/Bonn Survey \citep[LAB;][]{kalberla05}, which has about a factor of 4 lower angular resolution and severely suffers from the insufficient spatial sampling \citep[beam-by-beam only; compare][e.g., their Fig. 7]{kerp11}.

Great effort went into developing new data reduction techniques, including but not limited to software radio-frequency-interference mitigation, a frequency-dependent flux-density calibration scheme, and robust baseline estimation in the 2-D time--frequency domain. Furthermore, the very successful stray-radiation prediction algorithms, which were also employed for the LAB and Galactic All-Sky Survey \citep[GASS;][]{mcclure09, kalberla10, kalberla15}, have been adapted to the Effelsberg telescope.

We carefully assessed the data quality of the EBHIS spectra and inferred column densities. Our investigation revealed a mismatch in intensity calibration of GASS with respect to EBHIS and LAB, which led to a revised calibration scale in the third data release of GASS \citep{kalberla15}. EBHIS and LAB show a near-perfect one-to-one relation in column densities and brightness temperatures. From a comparison of EBHIS and LAB, and from the overlaps of the EBHIS fields themselves, we estimated the uncertainties of the EBHIS column density distribution and brightness temperatures. These uncertainties do not necessarily represent those of individual lines of sight but rather the ensemble properties. Still, the $N_\ion{H}{i}$ distribution uncertainties -- calculated for the first time for a large-area single-dish survey of the neutral hydrogen -- increase the value of the published $N_\ion{H}{i}$ map, which will probably be the primarily used data product.

Column density maps and full spectral data have been made freely available to the public.

\section{Outlook}

At the moment, we are undertaking observations to map the sky north of $\delta\geq30\degr$ for a second time, and we will also ask to extend this campaign to cover the full northern hemisphere again. This second coverage will boost the EBHIS sensitivity by 30\%, will allow even more sophisticated data reduction methods like basket-weaving to be applied, and will help to suppress residual RFI and SR.

In \citet{winkel12b} we presented a fast linear least-squares approach to apply to the basket-weaving observing technique. The idea is to map a portion of the sky twice, with two orthogonal scanning directions. Since the underlying astronomical sky is basically constant, one can disentangle scanning effects from the desired original brightness distribution. With respect to \ion{H}{i}, apparent scanning effects are mostly caused by inaccurate baseline solutions, e.g., if (weak) \ion{H}{i} sources are not correctly flagged or windowed. This is a common problem in single-dish astronomy, as shown by \citet{calabretta14}, among others.

However, for the continuum flux densities inherently present in the EBHIS raw data, scanning effects are mostly given by changes in the underlying system temperature, especially since the different feeds have different RX temperatures. With basket-weaving it will be possible to remove these scanning effects very efficiently and produce a 1.4-GHz Stokes-I map as byproduct of our survey. The feasibility of this goal has been presented in \citet{winkel12b}. However, we remind the interested reader that our SR solution only applies for the \ion{H}{i} sky and not the 1.4~GHz continuum sky. This is because the \ion{H}{i} SR removal does not need to account for ground radiation, while this contribution is significant for continuum. Therefore, like the HIPASS continuum data set \citep{calabretta14}, the EBHIS continuum data will suffer substantially from SR effects.

\begin{acknowledgements}

EBHIS is based on observations with the 100-m telescope of the MPIfR (Max-Planck-Institut für Radioastronomie) at Effelsberg. We would like to thank the observatory staff for the substantial and continuous assistance carrying out the measurements. We are grateful to Axel Jessner for carefully proofreading the manuscript and for his valuable comments.

The authors thank the Deutsche Forschungsgemeinschaft (DFG) for support under grant numbers KE757/7-1, KE757/7-2, KE757/7-3, and KE757/9-1. B.W. was partially funded by the International Max Planck Research School for Astronomy and Astrophysics at the Universities of Bonn and Cologne (IMPRS Bonn/Cologne). L.F. is also a member of IMPRS Bonn/Cologne. D.L. is a member of the Bonn--Cologne Graduate School of Physics and Astronomy (BCGS).

This research has made use of NASA's Astrophysics Data System and the HyperLeda database \citep{makarov14}.

We would like to express our gratitude to the developers of the many C/C++ and Python libraries, made available as open-source software, which we have used: most importantly, NumPy \citep{NumPy} and SciPy \citep{SciPy}, Cython \citep{Cython}, Astropy \citep{Astropy}, and the GNU Scientific library \citep{GSL}. Figures have been prepared using matplotlib \citep{Matplotlib} and in part using the Kapteyn Package \citep{KapteynPackage}.

\end{acknowledgements}


\bibliographystyle{aa}
\bibliography{references}

\begin{thebibliography}{56}
\expandafter\ifx\csname natexlab\endcsname\relax\def\natexlab#1{#1}\fi

\bibitem[{{Arnal} {et~al.}(2000){Arnal}, {Bajaja}, {Larrarte}, {Morras}, \&
  {P{\"o}ppel}}]{arnal00}
{Arnal}, E.~M., {Bajaja}, E., {Larrarte}, J.~J., {Morras}, R., \& {P{\"o}ppel},
  W.~G.~L. 2000, \aaps, 142, 35

\bibitem[{{Astropy Collaboration} {et~al.}(2013){Astropy Collaboration},
  {Robitaille}, {Tollerud}, {Greenfield}, {Droettboom}, {Bray}, {Aldcroft},
  {Davis}, {Ginsburg}, {Price-Whelan}, {Kerzendorf}, {Conley}, {Crighton},
  {Barbary}, {Muna}, {Ferguson}, {Grollier}, {Parikh}, {Nair}, {Unther},
  {Deil}, {Woillez}, {Conseil}, {Kramer}, {Turner}, {Singer}, {Fox}, {Weaver},
  {Zabalza}, {Edwards}, {Azalee Bostroem}, {Burke}, {Casey}, {Crawford},
  {Dencheva}, {Ely}, {Jenness}, {Labrie}, {Lim}, {Pierfederici}, {Pontzen},
  {Ptak}, {Refsdal}, {Servillat}, \& {Streicher}}]{Astropy}
{Astropy Collaboration}, {Robitaille}, T.~P., {Tollerud}, E.~J., {et~al.} 2013,
  \aap, 558, A33

\bibitem[{{Baars}(2007)}]{baars07}
{Baars}, J.~W.~M., ed. 2007, Astrophysics and Space Science Library, Vol. 348,
  {The Paraboloidal Reflector Antenna in Radio Astronomy and Communication}

\bibitem[{{Bajaja} {et~al.}(2005){Bajaja}, {Arnal}, {Larrarte}, {Morras},
  {P{\"o}ppel}, \& {Kalberla}}]{bajaja05}
{Bajaja}, E., {Arnal}, E.~M., {Larrarte}, J.~J., {et~al.} 2005, \aap, 440, 767

\bibitem[{{Bajaja} {et~al.}(1985){Bajaja}, {Cappa de Nicolau}, {Cersosimo},
  {Martin}, {Loiseau}, {Morras}, {Olano}, \& {Poeppel}}]{bajaja85}
{Bajaja}, E., {Cappa de Nicolau}, C.~E., {Cersosimo}, J.~C., {et~al.} 1985,
  \apjs, 58, 143

\bibitem[{{Barnes} {et~al.}(2005){Barnes}, {Briggs}, \&
  {Calabretta}}]{barnes05}
{Barnes}, D.~G., {Briggs}, F.~H., \& {Calabretta}, M.~R. 2005, Radio Science,
  40, 5

\bibitem[{{Barnes} {et~al.}(2001){Barnes}, {Staveley-Smith}, {de Blok},
  {Oosterloo}, {Stewart}, {Wright}, {Banks}, {Bhathal}, {Boyce}, {Calabretta},
  {Disney}, {Drinkwater}, {Ekers}, {Freeman}, {Gibson}, {Green}, {Haynes}, {te
  Lintel Hekkert}, {Henning}, {Jerjen}, {Juraszek}, {Kesteven}, {Kilborn},
  {Knezek}, {Koribalski}, {Kraan-Korteweg}, {Malin}, {Marquarding}, {Minchin},
  {Mould}, {Price}, {Putman}, {Ryder}, {Sadler}, {Schr{\"o}der}, {Stootman},
  {Webster}, {Wilson}, \& {Ye}}]{barnes01}
{Barnes}, D.~G., {Staveley-Smith}, L., {de Blok}, W.~J.~G., {et~al.} 2001,
  \mnras, 322, 486

\bibitem[{{Barr} {et~al.}(2013){Barr}, {Champion}, {Kramer}, {Eatough},
  {Freire}, {Karuppusamy}, {Lee}, {Verbiest}, {Bassa}, {Lyne}, {Stappers},
  {Lorimer}, \& {Klein}}]{barr13}
{Barr}, E.~D., {Champion}, D.~J., {Kramer}, M., {et~al.} 2013, \mnras, 435,
  2234

\bibitem[{Behnel {et~al.}(2011)Behnel, Bradshaw, Citro, Dalcin, Seljebotn, \&
  Smith}]{Cython}
Behnel, S., Bradshaw, R., Citro, C., {et~al.} 2011, Computing in Science
  Engineering, 13, 31

\bibitem[{{Boggs} \& {Rogers}(1990)}]{boggs90}
{Boggs}, P.~T. \& {Rogers}, J.~E. 1990, in Contemporary Mathematics, Vol. 112,
  Statistical analysis of measurement error models and applications, 186

\bibitem[{{Boothroyd} {et~al.}(2011){Boothroyd}, {Blagrave}, {Lockman},
  {Martin}, {Pinheiro Gon{\c c}alves}, \& {Srikanth}}]{boothroyd11}
{Boothroyd}, A.~I., {Blagrave}, K., {Lockman}, F.~J., {et~al.} 2011, \aap, 536,
  A81

\bibitem[{{Bracewell} \& {Roberts}(1954)}]{bracewell54}
{Bracewell}, R.~N. \& {Roberts}, J.~A. 1954, Australian Journal of Physics, 7,
  615

\bibitem[{{Calabretta} \& {Greisen}(2002)}]{calabretta02}
{Calabretta}, M.~R. \& {Greisen}, E.~W. 2002, \aap, 395, 1077

\bibitem[{{Calabretta} {et~al.}(2014){Calabretta}, {Staveley-Smith}, \&
  {Barnes}}]{calabretta14}
{Calabretta}, M.~R., {Staveley-Smith}, L., \& {Barnes}, D.~G. 2014, \pasa, 31,
  7

\bibitem[{{Condon} {et~al.}(1998){Condon}, {Cotton}, {Greisen}, {Yin},
  {Perley}, {Taylor}, \& {Broderick}}]{condon98}
{Condon}, J.~J., {Cotton}, W.~D., {Greisen}, E.~W., {et~al.} 1998, \aj, 115,
  1693

\bibitem[{{Fl{\"o}er} {et~al.}(2010){Fl{\"o}er}, {Winkel}, \&
  {Kerp}}]{floeer10}
{Fl{\"o}er}, L., {Winkel}, B., \& {Kerp}, J. 2010, in RFI Mitigation Workshop,
  42

\bibitem[{{Giovanelli} {et~al.}(2005){Giovanelli}, {Haynes}, {Kent},
  {Perillat}, {Saintonge}, {Brosch}, {Catinella}, {Hoffman}, {Stierwalt},
  {Spekkens}, {Lerner}, {Masters}, {Momjian}, {Rosenberg}, {Springob},
  {Boselli}, {Charmandaris}, {Darling}, {Davies}, {Garcia Lambas}, {Gavazzi},
  {Giovanardi}, {Hardy}, {Hunt}, {Iovino}, {Karachentsev}, {Karachentseva},
  {Koopmann}, {Marinoni}, {Minchin}, {Muller}, {Putman}, {Pantoja}, {Salzer},
  {Scodeggio}, {Skillman}, {Solanes}, {Valotto}, {van Driel}, \& {van
  Zee}}]{giovanelli05}
{Giovanelli}, R., {Haynes}, M.~P., {Kent}, B.~R., {et~al.} 2005, \aj, 130, 2598

\bibitem[{{G{\'o}rski} {et~al.}(2005){G{\'o}rski}, {Hivon}, {Banday},
  {Wandelt}, {Hansen}, {Reinecke}, \& {Bartelmann}}]{gorski05}
{G{\'o}rski}, K.~M., {Hivon}, E., {Banday}, A.~J., {et~al.} 2005, \apj, 622,
  759

\bibitem[{Gough(2009)}]{GSL}
Gough, B. 2009, GNU Scientific Library Reference Manual - Third Edition, 3rd
  edn. (Network Theory Ltd.)

\bibitem[{{Greisen} \& {Calabretta}(2002)}]{greisen02}
{Greisen}, E.~W. \& {Calabretta}, M.~R. 2002, \aap, 395, 1061

\bibitem[{{Hanisch} {et~al.}(2001){Hanisch}, {Farris}, {Greisen}, {Pence},
  {Schlesinger}, {Teuben}, {Thompson}, \& {Warnock}}]{hanisch01}
{Hanisch}, R.~J., {Farris}, A., {Greisen}, E.~W., {et~al.} 2001, \aap, 376, 359

\bibitem[{{Hartmann} \& {Burton}(1997)}]{hartmann97}
{Hartmann}, D. \& {Burton}, W.~B. 1997, {Atlas of Galactic Neutral Hydrogen}

\bibitem[{{Hartmann} {et~al.}(1996){Hartmann}, {Kalberla}, {Burton}, \&
  {Mebold}}]{hartmann96}
{Hartmann}, D., {Kalberla}, P.~M.~W., {Burton}, W.~B., \& {Mebold}, U. 1996,
  \aaps, 119, 115

\bibitem[{{Hartsuijker} {et~al.}(1972){Hartsuijker}, {Baars}, {Drenth}, \&
  {Gelato-Volders}}]{hartsuijker72}
{Hartsuijker}, A.~P., {Baars}, J.~W.~M., {Drenth}, S., \& {Gelato-Volders}, L.
  1972, IEEE Transactions on Antennas and Propagation, 20, 166

\bibitem[{{Haynes} {et~al.}(2011){Haynes}, {Giovanelli}, {Martin}, {Hess},
  {Saintonge}, {Adams}, {Hallenbeck}, {Hoffman}, {Huang}, {Kent}, {Koopmann},
  {Papastergis}, {Stierwalt}, {Balonek}, {Craig}, {Higdon}, {Kornreich},
  {Miller}, {O'Donoghue}, {Olowin}, {Rosenberg}, {Spekkens}, {Troischt}, \&
  {Wilcots}}]{haynes11}
{Haynes}, M.~P., {Giovanelli}, R., {Martin}, A.~M., {et~al.} 2011, \aj, 142,
  170

\bibitem[{{Higgs}(1967)}]{higgs67}
{Higgs}, L.~A. 1967, Bulletin of the Astronomical Institutes of the Netherlands
  Supplement Series, 2, 59

\bibitem[{{Higgs} \& {Tapping}(2000)}]{higgs00}
{Higgs}, L.~A. \& {Tapping}, K.~F. 2000, \aj, 120, 2471

\bibitem[{Hunter(2007)}]{Matplotlib}
Hunter, J. 2007, Computing in Science Engineering, 9, 90

\bibitem[{Jones {et~al.}(2001)Jones, Oliphant, Peterson, {et~al.}}]{SciPy}
Jones, E., Oliphant, T., Peterson, P., {et~al.} 2001, {SciPy}: Open source
  scientific tools for {Python}, [Online; accessed 2015-06-27]

\bibitem[{{Kalberla} {et~al.}(2005){Kalberla}, {Burton}, {Hartmann}, {Arnal},
  {Bajaja}, {Morras}, \& {P{\"o}ppel}}]{kalberla05}
{Kalberla}, P.~M.~W., {Burton}, W.~B., {Hartmann}, D., {et~al.} 2005, \aap,
  440, 775

\bibitem[{{Kalberla} \& {Haud}(2015)}]{kalberla15}
{Kalberla}, P.~M.~W. \& {Haud}, U. 2015, \aap, 578, A78

\bibitem[{{Kalberla} {et~al.}(2010){Kalberla}, {McClure-Griffiths}, {Pisano},
  {Calabretta}, {Ford}, {Lockman}, {Staveley-Smith}, {Kerp}, {Winkel},
  {Murphy}, \& {Newton-McGee}}]{kalberla10}
{Kalberla}, P.~M.~W., {McClure-Griffiths}, N.~M., {Pisano}, D.~J., {et~al.}
  2010, \aap, 521, A17+

\bibitem[{{Kalberla} {et~al.}(1980{\natexlab{a}}){Kalberla}, {Mebold}, \&
  {Reich}}]{kalberla80a}
{Kalberla}, P.~M.~W., {Mebold}, U., \& {Reich}, W. 1980{\natexlab{a}}, \aap,
  82, 275

\bibitem[{{Kalberla} {et~al.}(1980{\natexlab{b}}){Kalberla}, {Mebold}, \&
  {Velden}}]{kalberla80b}
{Kalberla}, P.~M.~W., {Mebold}, U., \& {Velden}, L. 1980{\natexlab{b}}, \aaps,
  39, 337

\bibitem[{Keller {et~al.}(2013)Keller, Dohlus, \& Hauth}]{keller13}
Keller, R., Dohlus, M., \& Hauth, W. 2013, in Electromagnetics in Advanced
  Applications (ICEAA), 2013 International Conference on, 1323--1326

\bibitem[{{Keller} {et~al.}(2006){Keller}, {Nalbach}, {M\"{u}ller}, {Teuber},
  {Sch\"{a}fer}, {Klein}, {Kr\"{a}mer}, {Bell}, \& {Meyer}}]{keller06}
{Keller}, R., {Nalbach}, M., {M\"{u}ller}, K., {et~al.} 2006, Multi-Beam
  Receiver for Beam-Park Experiments and Data Collection Unit for Beam Park
  Experiments with Multi-Beam Receivers, Tech. rep., Max-Planck-Institut
  f\"{u}r Radioastronomie

\bibitem[{{Kerp} {et~al.}(2011){Kerp}, {Winkel}, {Ben Bekhti}, {Fl{\"o}er}, \&
  {Kalberla}}]{kerp11}
{Kerp}, J., {Winkel}, B., {Ben Bekhti}, N., {Fl{\"o}er}, L., \& {Kalberla},
  P.~M.~W. 2011, Astronomische Nachrichten, 332, 637

\bibitem[{{Klein} {et~al.}(2012){Klein}, {Hochg{\"u}rtel}, {Kr{\"a}mer},
  {Bell}, {Meyer}, \& {G{\"u}sten}}]{klein12}
{Klein}, B., {Hochg{\"u}rtel}, S., {Kr{\"a}mer}, I., {et~al.} 2012, \aap, 542,
  L3

\bibitem[{Kramer \& Champion(2013)}]{kramer13}
Kramer, M. \& Champion, D.~J. 2013, Classical and Quantum Gravity, 30, 224009

\bibitem[{{Land} \& {Slosar}(2007)}]{land07}
{Land}, K. \& {Slosar}, A. 2007, \prd, 76, 087301

\bibitem[{{Lockman} {et~al.}(1986){Lockman}, {Jahoda}, \&
  {McCammon}}]{lockman86}
{Lockman}, F.~J., {Jahoda}, K., \& {McCammon}, D. 1986, \apj, 302, 432

\bibitem[{{Makarov} {et~al.}(2014){Makarov}, {Prugniel}, {Terekhova},
  {Courtois}, \& {Vauglin}}]{makarov14}
{Makarov}, D., {Prugniel}, P., {Terekhova}, N., {Courtois}, H., \& {Vauglin},
  I. 2014, \aap, 570, A13

\bibitem[{{Martin} {et~al.}(2015){Martin}, {Blagrave}, {Pinheiro Goncalves},
  {Lockman}, {Boothroyd}, {Miville-Deschenes}, {Joncas}, \&
  {Stephan}}]{martin15}
{Martin}, P.~G., {Blagrave}, K.~P.~M., {Pinheiro Goncalves}, D., {et~al.} 2015,
  ArXiv e-prints

\bibitem[{{McClure-Griffiths} {et~al.}(2009){McClure-Griffiths}, {Pisano},
  {Calabretta}, {Ford}, {Lockman}, {Staveley-Smith}, {Kalberla}, {Bailin},
  {Dedes}, {Janowiecki}, {Gibson}, {Murphy}, {Nakanishi}, \&
  {Newton-McGee}}]{mcclure09}
{McClure-Griffiths}, N.~M., {Pisano}, D.~J., {Calabretta}, M.~R., {et~al.}
  2009, \apjs, 181, 398

\bibitem[{{Minchin} {et~al.}(2007){Minchin}, {Auld}, {Davies}, {Catinella},
  {Cortese}, {Linder}, {Momjian}, {Muller}, {O'Neil}, {Rosenberg}, {Sabatini},
  {Schneider}, {Stage}, {van Driel}, \& {AGES Team}}]{minchin07}
{Minchin}, R.~F., {Auld}, R., {Davies}, J.~I., {et~al.} 2007, in IAU Symposium,
  Vol. 235, IAU Symposium, ed. F.~{Combes} \& J.~{Palou{\v s}}, 227--229

\bibitem[{{Peek} {et~al.}(2011){Peek}, {Heiles}, {Douglas}, {Lee}, {Grcevich},
  {Stanimirovi{\'c}}, {Putman}, {Korpela}, {Gibson}, {Begum}, {Saul},
  {Robishaw}, \& {Kr{\v c}o}}]{peek11}
{Peek}, J.~E.~G., {Heiles}, C., {Douglas}, K.~A., {et~al.} 2011, \apjs, 194, 20

\bibitem[{{Prestage} {et~al.}(2009){Prestage}, {Constantikes}, {Hunter},
  {King}, {Lacasse}, {Lockman}, \& {Norrod}}]{prestage09}
{Prestage}, R.~M., {Constantikes}, K.~T., {Hunter}, T.~R., {et~al.} 2009, IEEE
  Proceedings, 97, 1382

\bibitem[{Shannon(1949)}]{shannon49}
Shannon, C.~E. 1949, Proc. Institute of Radio Engineers, 37, 10

\bibitem[{{Stanko} {et~al.}(2005){Stanko}, {Klein}, \& {Kerp}}]{stanko05}
{Stanko}, S., {Klein}, B., \& {Kerp}, J. 2005, \aap, 436, 391

\bibitem[{{Stark} {et~al.}(1992){Stark}, {Gammie}, {Wilson}, {Bally}, {Linke},
  {Heiles}, \& {Hurwitz}}]{stark92}
{Stark}, A.~A., {Gammie}, C.~F., {Wilson}, R.~W., {et~al.} 1992, \apjs, 79, 77

\bibitem[{{Terlouw} \& {Vogelaar}(2015)}]{KapteynPackage}
{Terlouw}, J.~P. \& {Vogelaar}, M.~G.~R. 2015, {Kapteyn Package, version 2.3},
  {Kapteyn Astronomical Institute}, Groningen, available from
  \url{http://www.astro.rug.nl/software/kapteyn/}

\bibitem[{van~der Walt {et~al.}(2011)van~der Walt, Colbert, \&
  Varoquaux}]{NumPy}
van~der Walt, S., Colbert, S., \& Varoquaux, G. 2011, Computing in Science
  Engineering, 13, 22

\bibitem[{{van Woerden}(1962)}]{vanwoerden62}
{van Woerden}, H. 1962, {De neutrale waterstof in Orion}

\bibitem[{{Winkel} {et~al.}(2012{\natexlab{a}}){Winkel}, {Fl{\"o}er}, \&
  {Kraus}}]{winkel12b}
{Winkel}, B., {Fl{\"o}er}, L., \& {Kraus}, A. 2012{\natexlab{a}}, \aap, 547,
  A119

\bibitem[{{Winkel} {et~al.}(2010){Winkel}, {Kalberla}, {Kerp}, \&
  {Fl{\"o}er}}]{winkel10}
{Winkel}, B., {Kalberla}, P.~M.~W., {Kerp}, J., \& {Fl{\"o}er}, L. 2010, \apjs,
  188, 488

\bibitem[{{Winkel} {et~al.}(2012{\natexlab{b}}){Winkel}, {Kraus}, \&
  {Bach}}]{winkel12}
{Winkel}, B., {Kraus}, A., \& {Bach}, U. 2012{\natexlab{b}}, \aap, 540, A140

\end{thebibliography}

\appendix

\section{Supplementary figures}\label{appsec:supp_figures}

\subsection{Additional sky maps}\label{appsubsec:fullskymaps}

\begin{figure*}[!t]
\centering%
\includegraphics[width=0.95\textwidth,viewport=10 5 1060 1000,clip=]{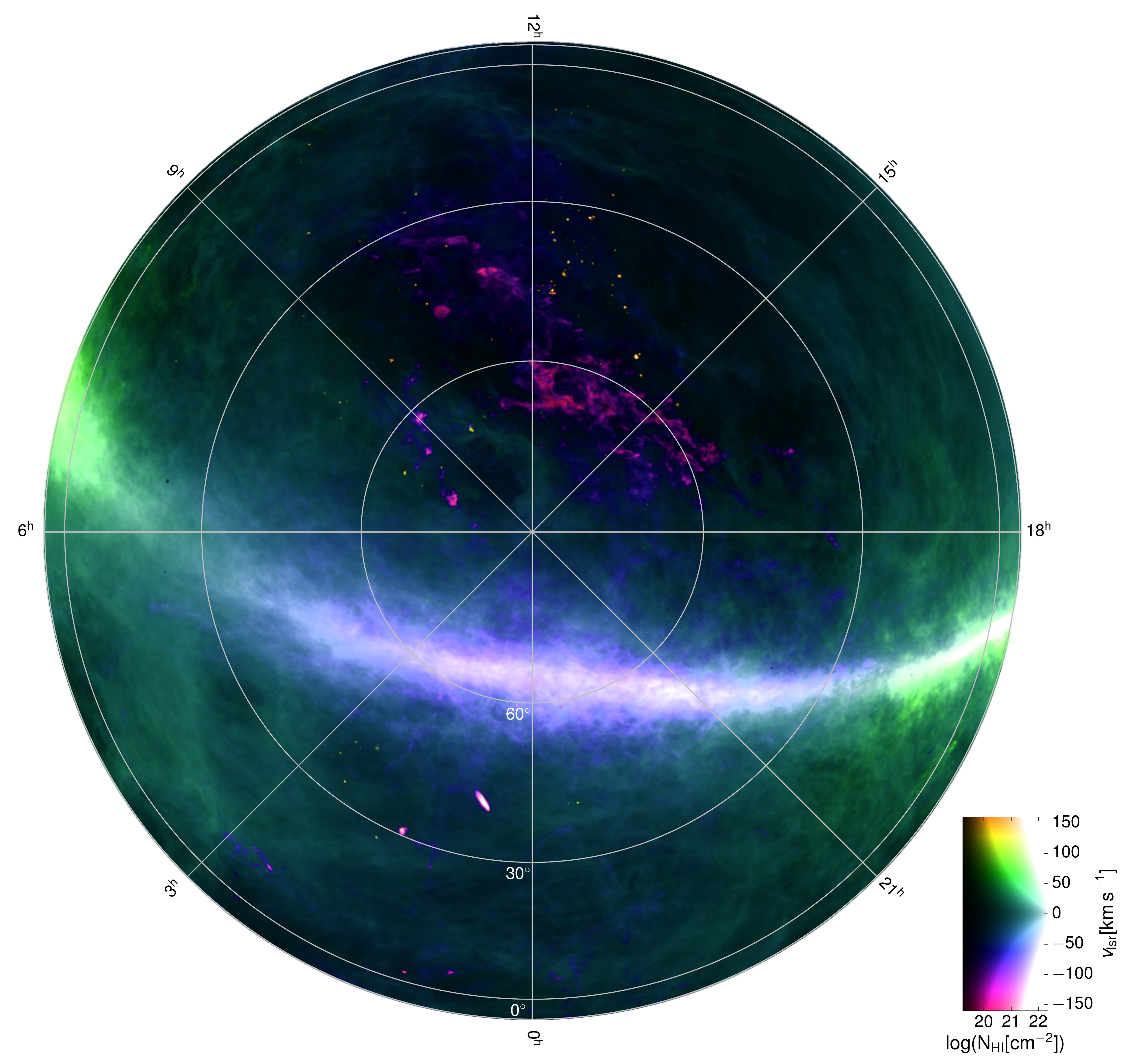}\\[0ex]
\caption{Composite image displaying $N_\ion{H}{i}$ column densities (encoded as brightness) and intensity-weighted velocities (Moment-1, encoded as different colors) for EBHIS. To make the weaker intermediate- and high-velocity gas more visible -- compared to the bright MW disk emission -- we modified the brightness scale as a function of velocity, as indicated in the color wedge.}%
\label{fig:allsky_composite}%
\end{figure*}

Figure~\ref{fig:allsky_composite} contains a composite image combining $N_\ion{H}{i}$ column densities and intensity-weighted velocities (Moment-1). For the column density part, the same FITS map as in Fig.~\ref{fig:allsky_nhi} was used. For the Moment-1 image, first a mask was produced by Gaussian-filtering the original data cube and applying an intensity threshold. Otherwise, low-column density spectra would lead to numerical artifacts in the Moment-1 calculation. Intensity-weighted velocities are encoded as colors in Figure~\ref{fig:allsky_composite}. The resulting image is converted to the HSV color space and the saturation- and value subchannels (the ``brightness'') are modified according to the column density in each pixel. To account for the high density contrast (two orders of magnitude) in the $N_\ion{H}{i}$ map, the brightness is modified differently as a function of velocity. Otherwise, the relatively fainter intermediate- and high-velocity gas would hardly be visible. The two-dimensional color wedge depicts the HSV transfer function.

\begin{figure*}[!t]
\centering%
\includegraphics[width=0.45\textwidth,viewport=40 50 1040 1050,clip=]{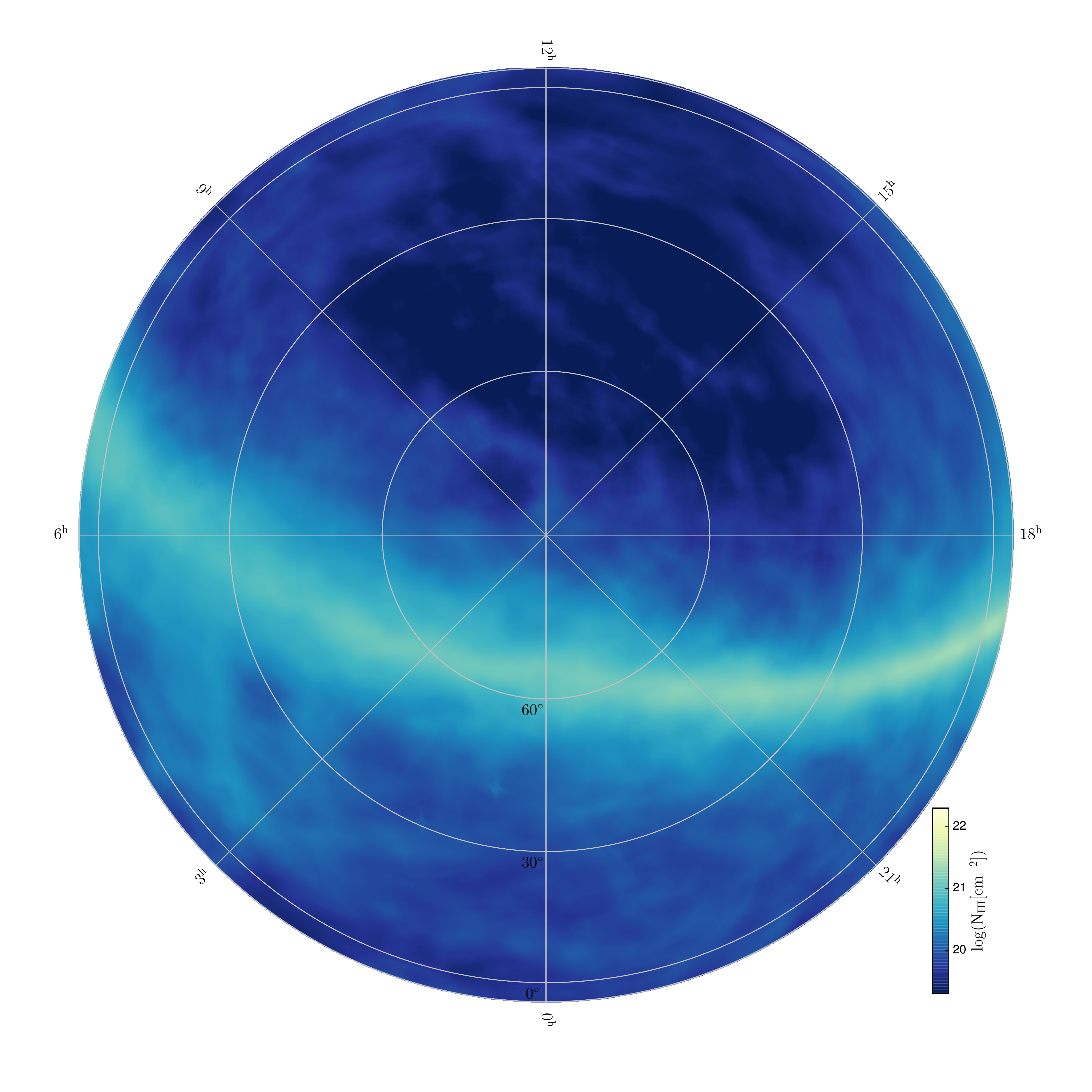}\hfill
\includegraphics[width=0.45\textwidth,viewport=40 50 1040 1050,clip=]{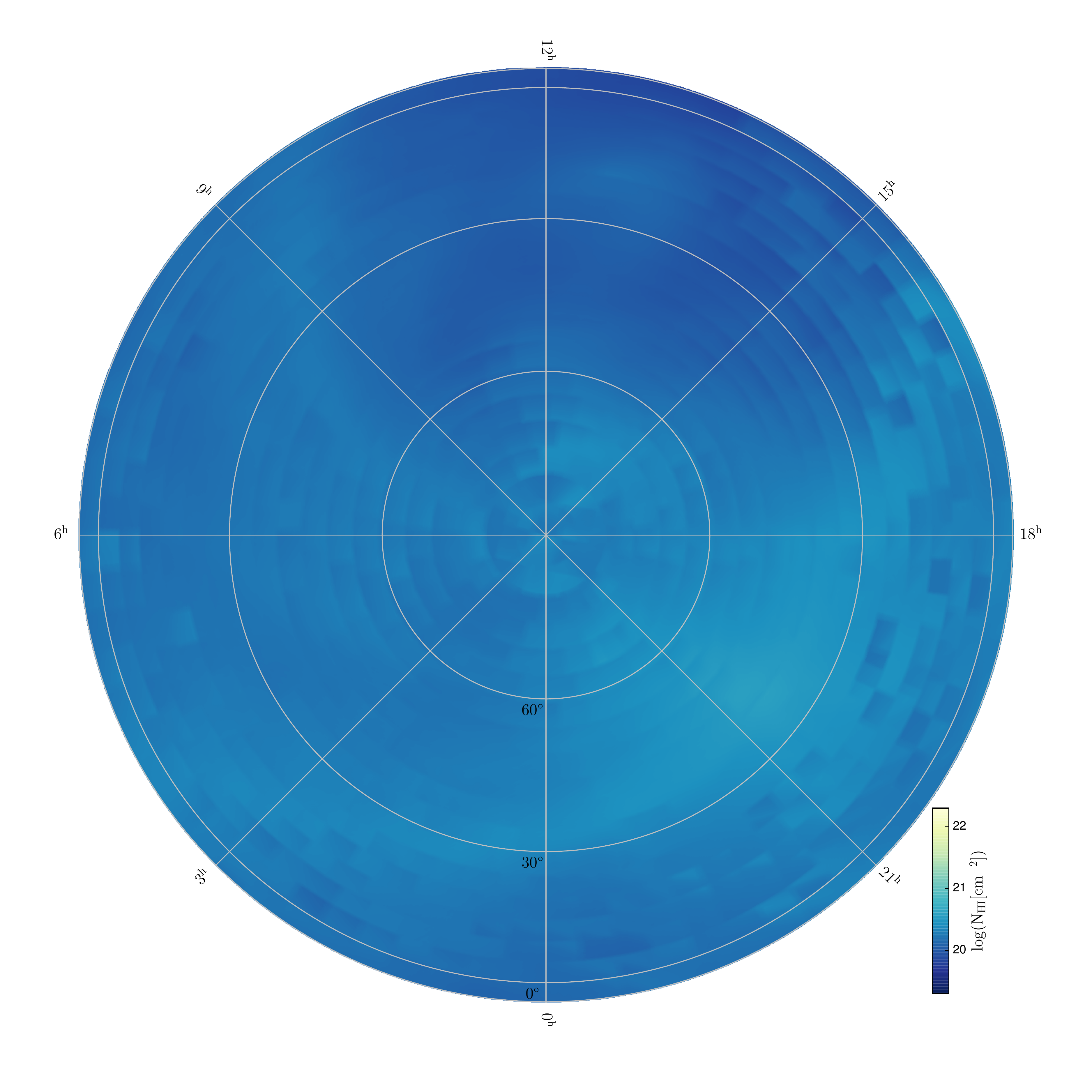}\\[-2ex]
\includegraphics[width=0.85\textwidth,viewport=40 50 1040 1050,clip=]{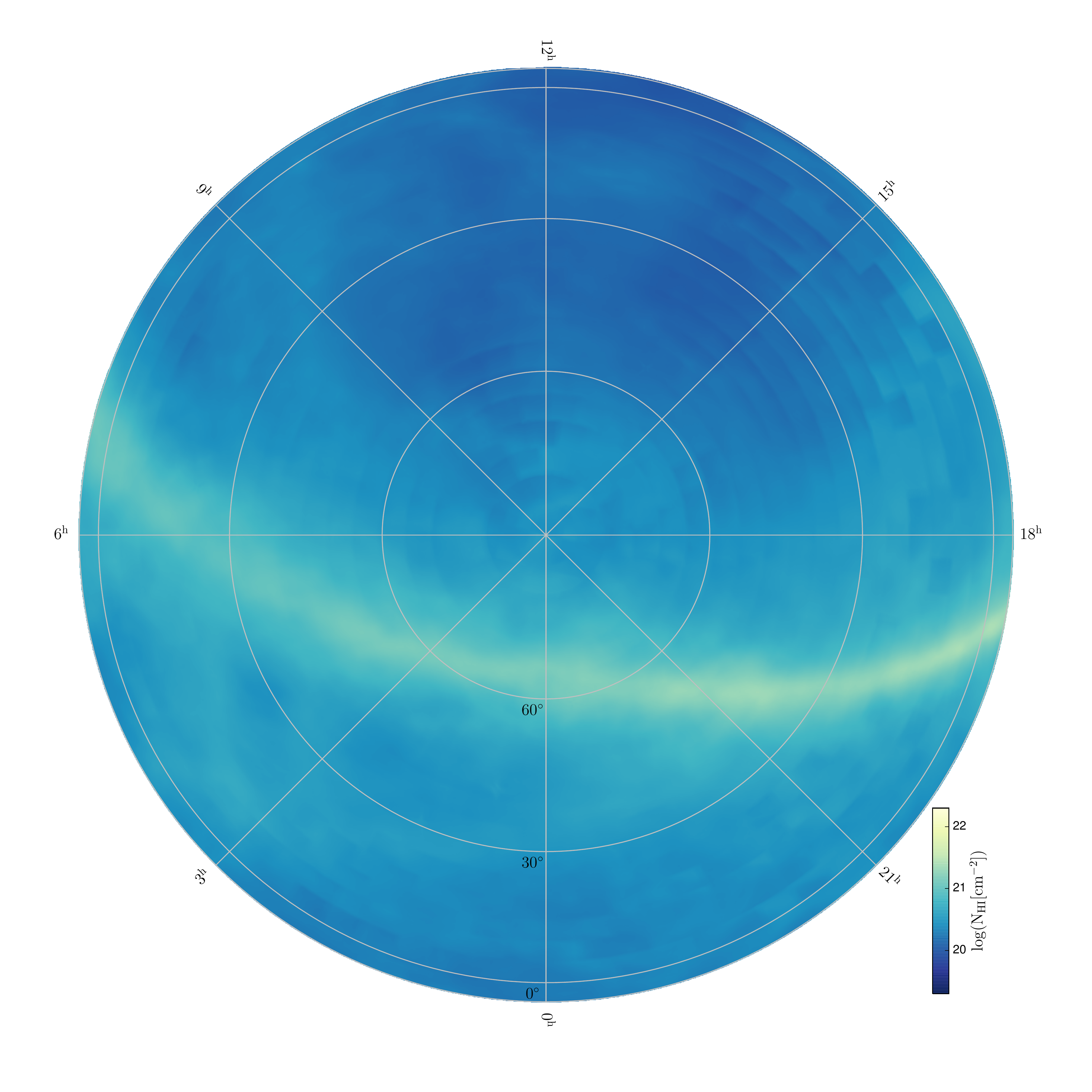}\\[0ex]
\caption{Stray-radiation correction maps showing the near- and far-side-lobe contribution (top panels), as well as the total amount of SR (bottom panel). The color scale matches the one used in Fig.~\ref{fig:allsky_nhi} to allow direct comparison.}%
\label{fig:allsky_SRtot}%
\end{figure*}

In Fig.~\ref{fig:allsky_SRtot} the SR correction, marginalized to column densities, is visualized. The bottom panel is the total SR contribution as predicted by our SR model. The upper panels contain near- (left) and far-side-lobe (right) contributions. The maps are on the same grid as the EBHIS $N_\ion{H}{i}$ map in Fig.~\ref{fig:allsky_nhi} and share the same intensity scale to allow comparison. Toward higher Galactic latitudes, i.e., lower column densities, the SR contribution of the far side lobes gets increasingly important and can even be larger than the true column density in extreme cases.

\subsection{Flux-density calibration and velocity consistency}\label{appsubsec:flux_comparison}

\begin{figure*}[!t]
\centering%
\includegraphics[width=0.32\textwidth,viewport=35 54 555 605,clip=]{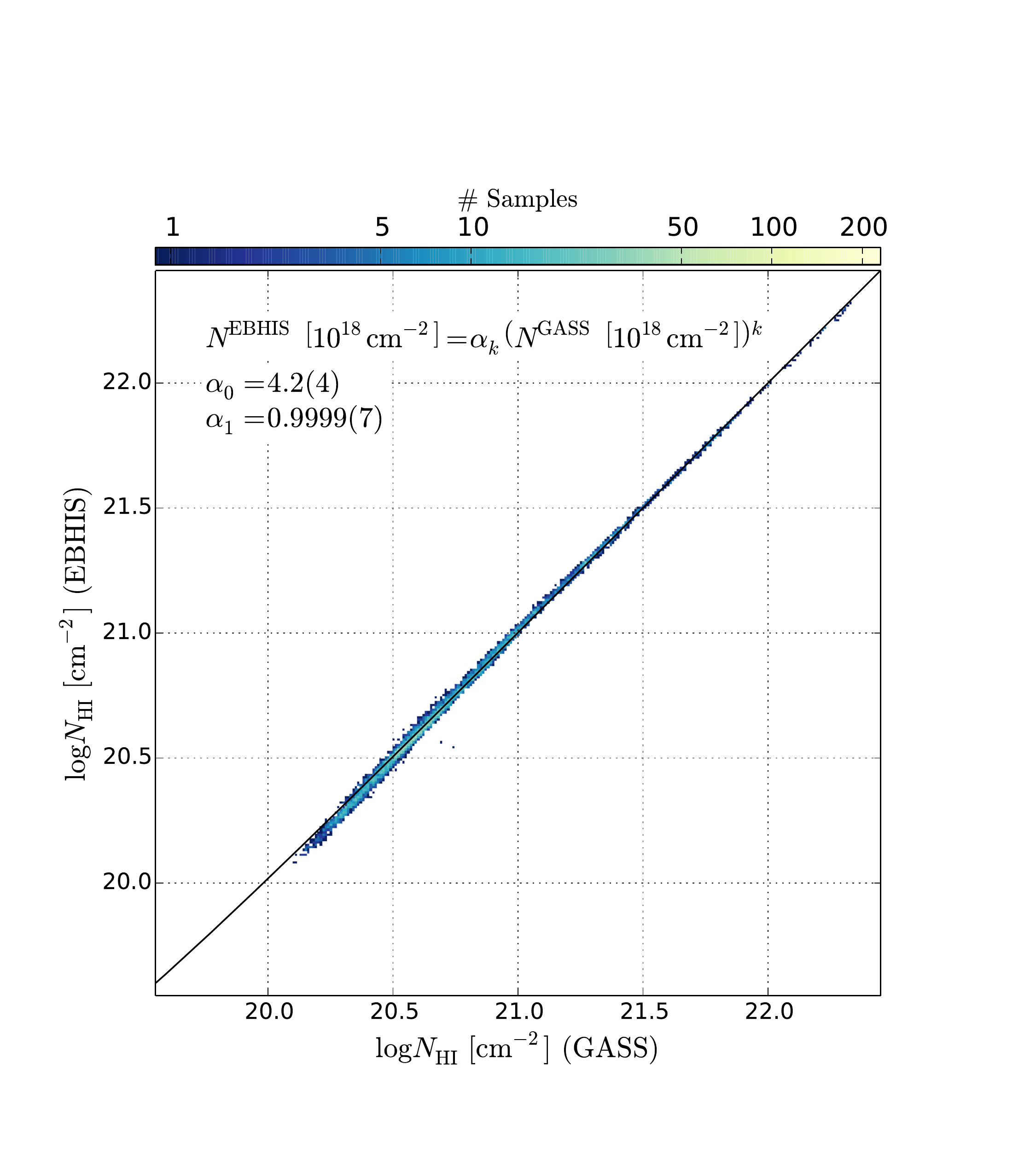}~
\includegraphics[width=0.32\textwidth,viewport=35 54 555 605,clip=]{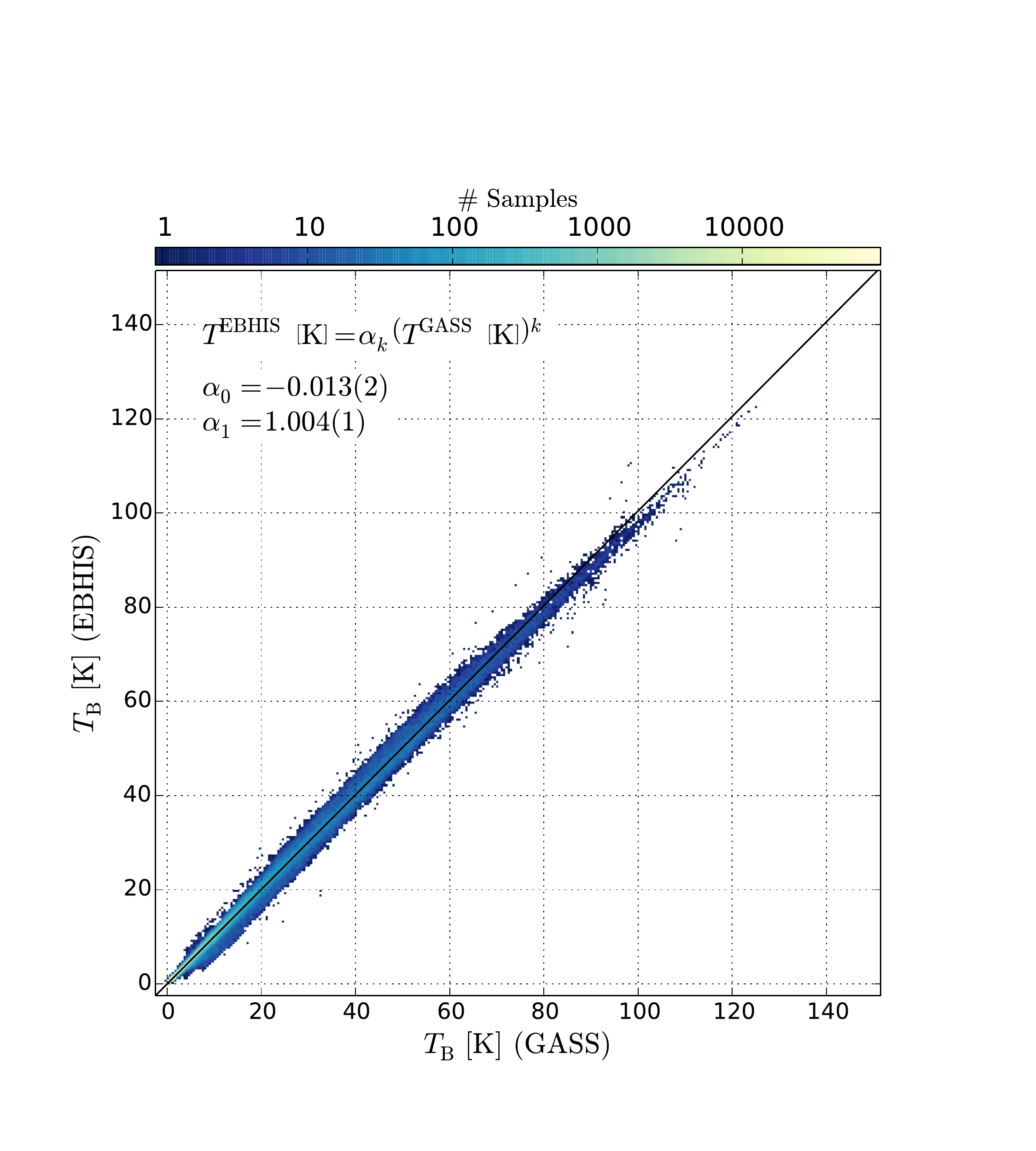}~
\includegraphics[width=0.32\textwidth,viewport=35 54 555 605,clip=]{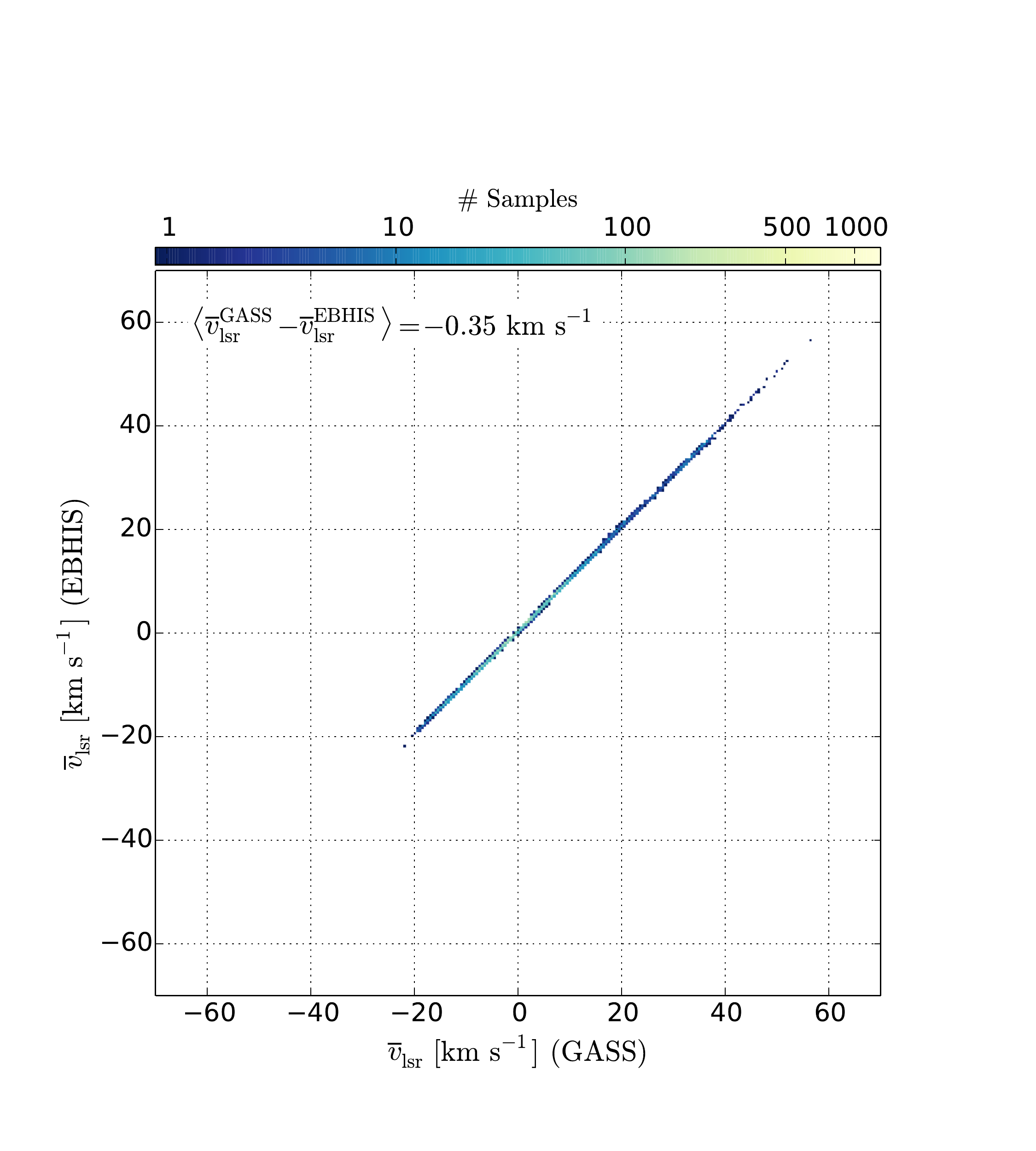}\\[0ex]
\includegraphics[width=0.32\textwidth,viewport=35 54 555 605,clip=]{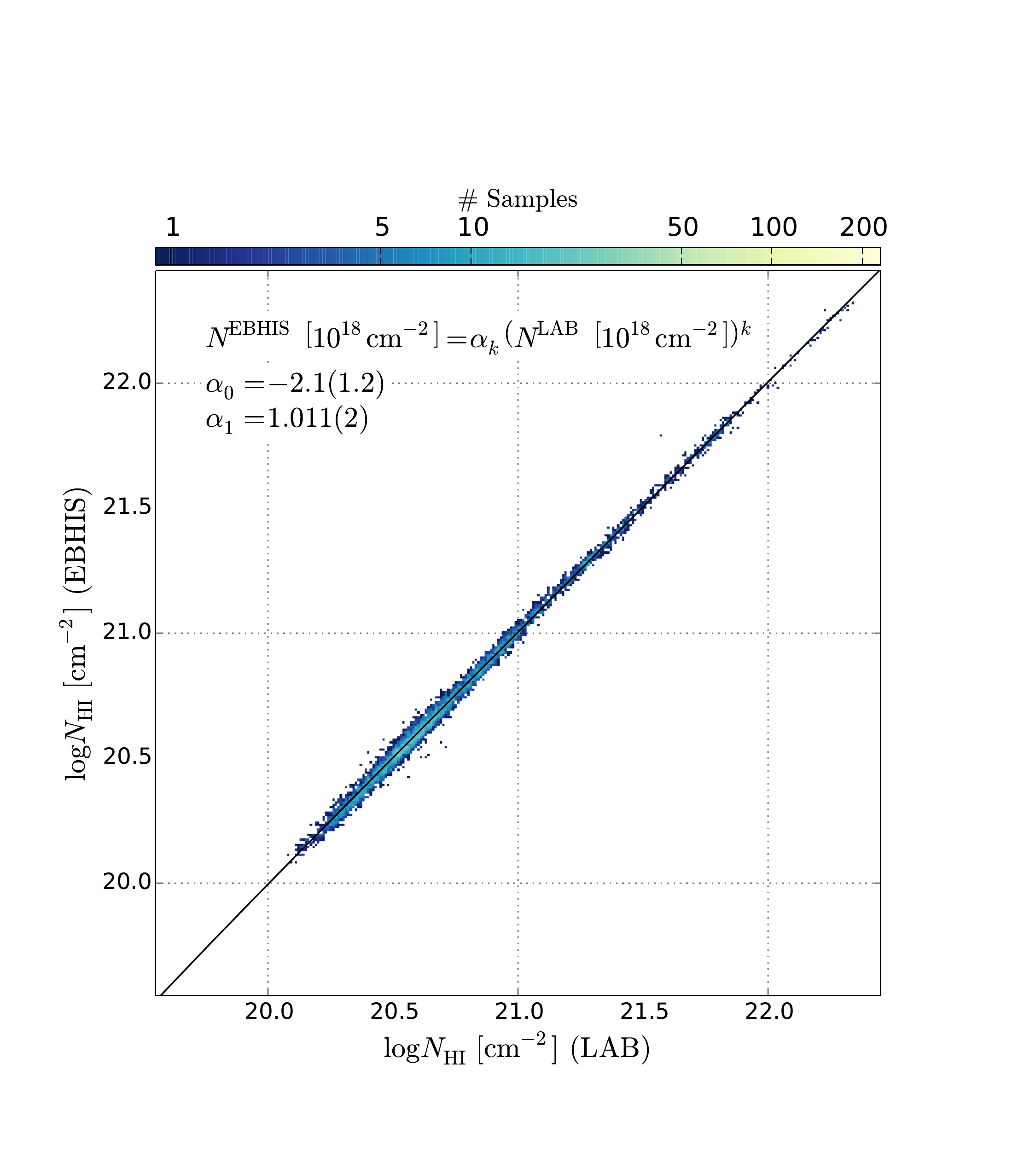}~
\includegraphics[width=0.32\textwidth,viewport=35 54 555 605,clip=]{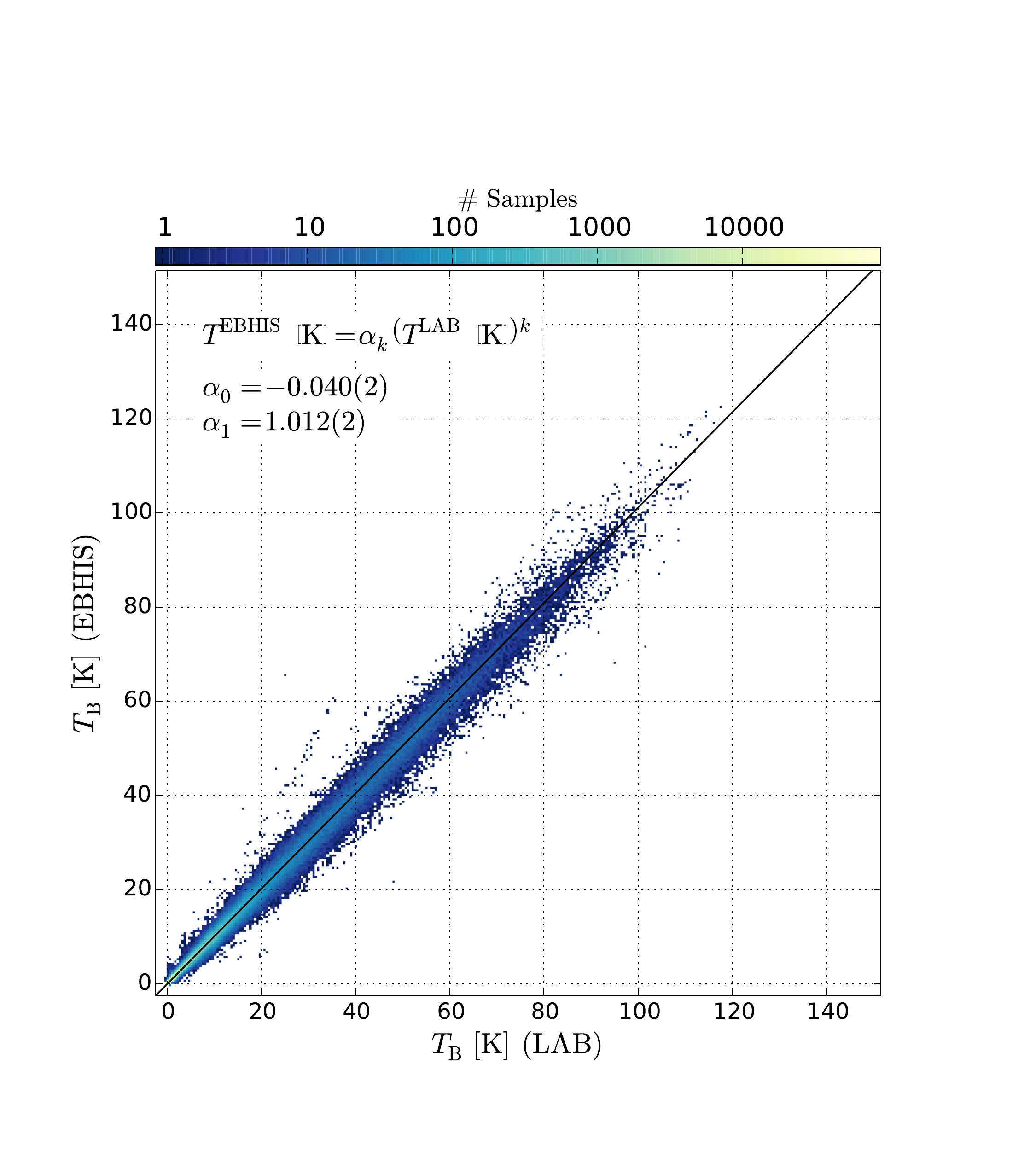}~
\includegraphics[width=0.32\textwidth,viewport=35 54 555 605,clip=]{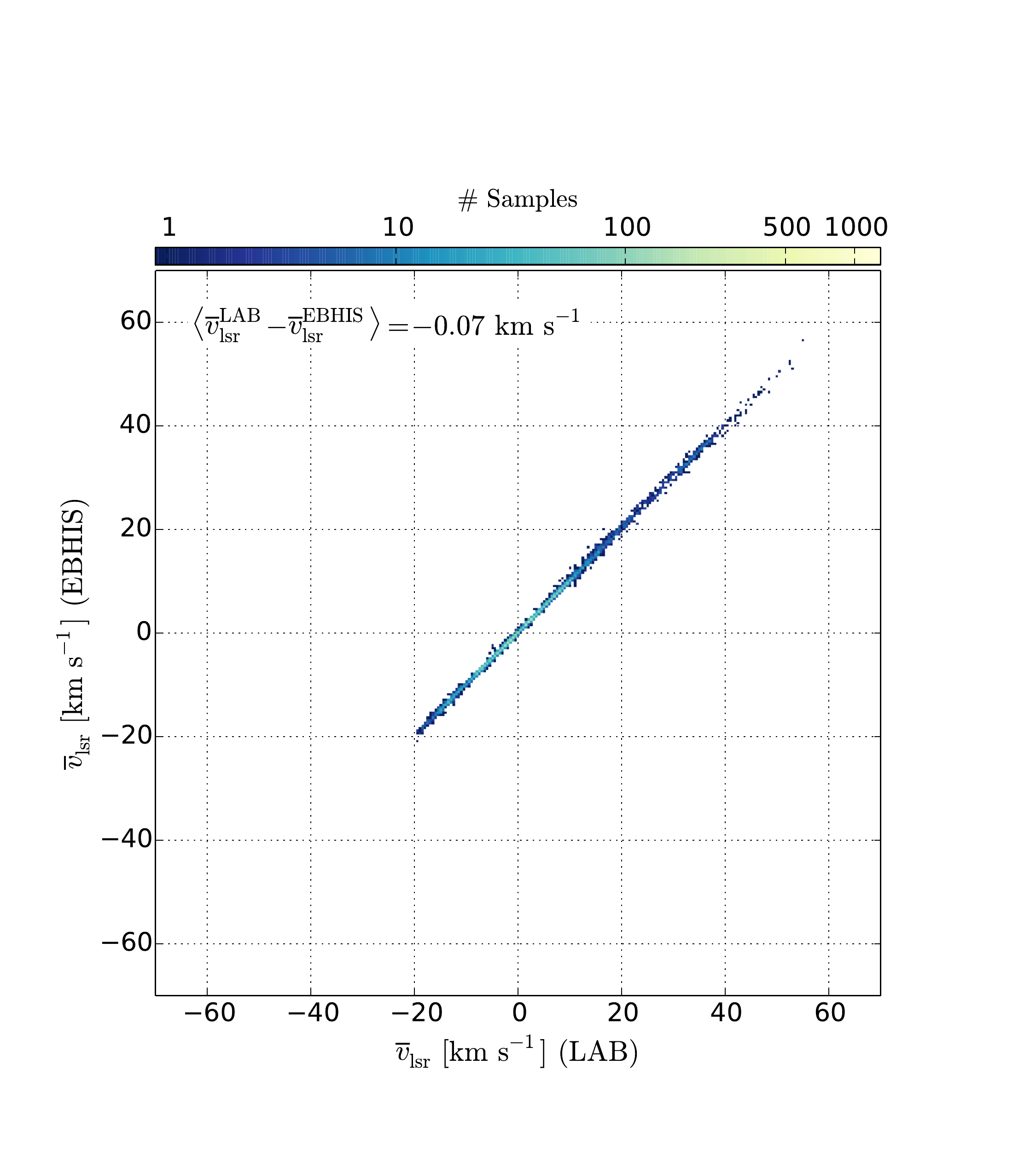}\\[0ex]
\includegraphics[width=0.32\textwidth,viewport=35 54 555 605,clip=]{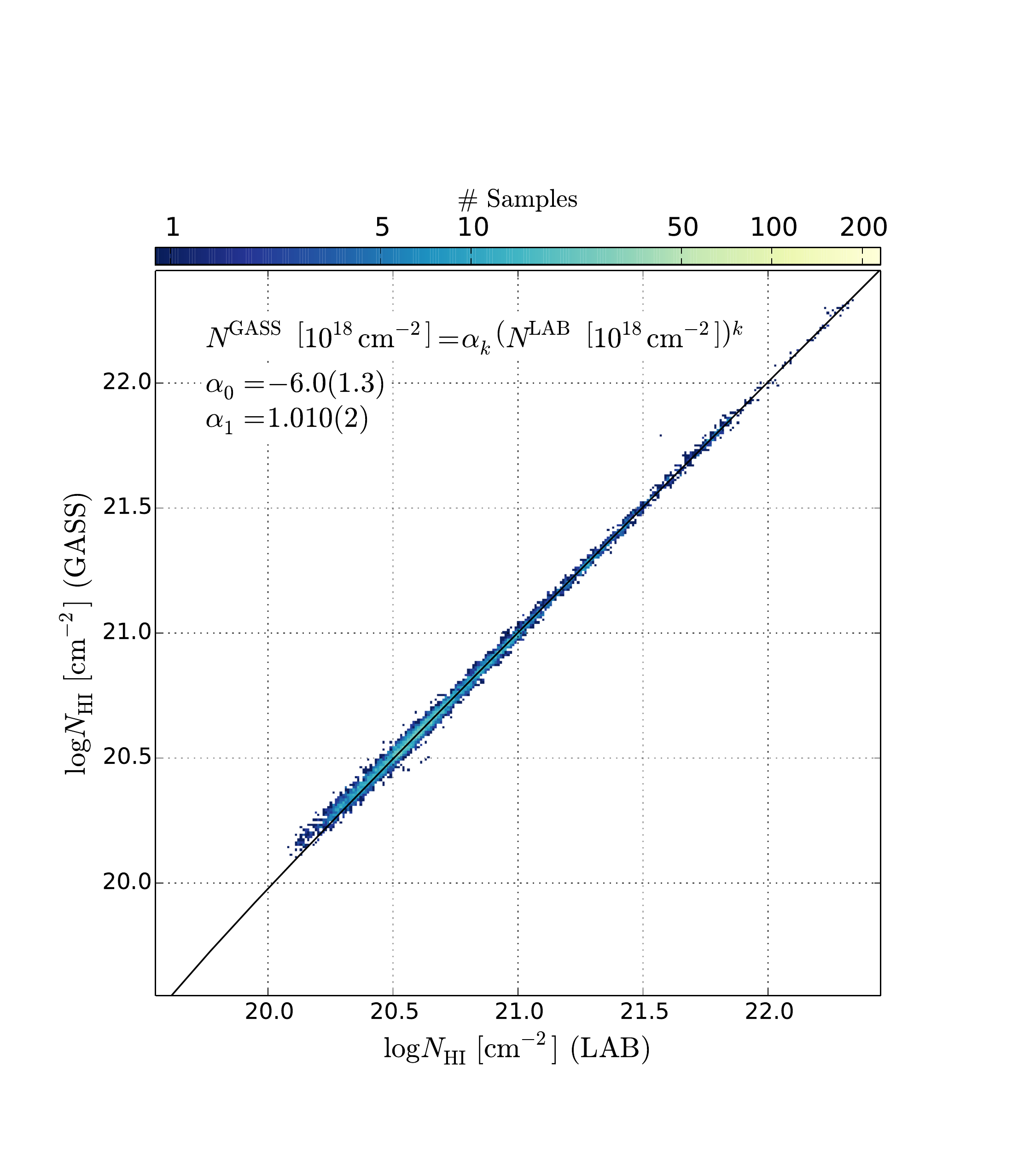}~
\includegraphics[width=0.32\textwidth,viewport=35 54 555 605,clip=]{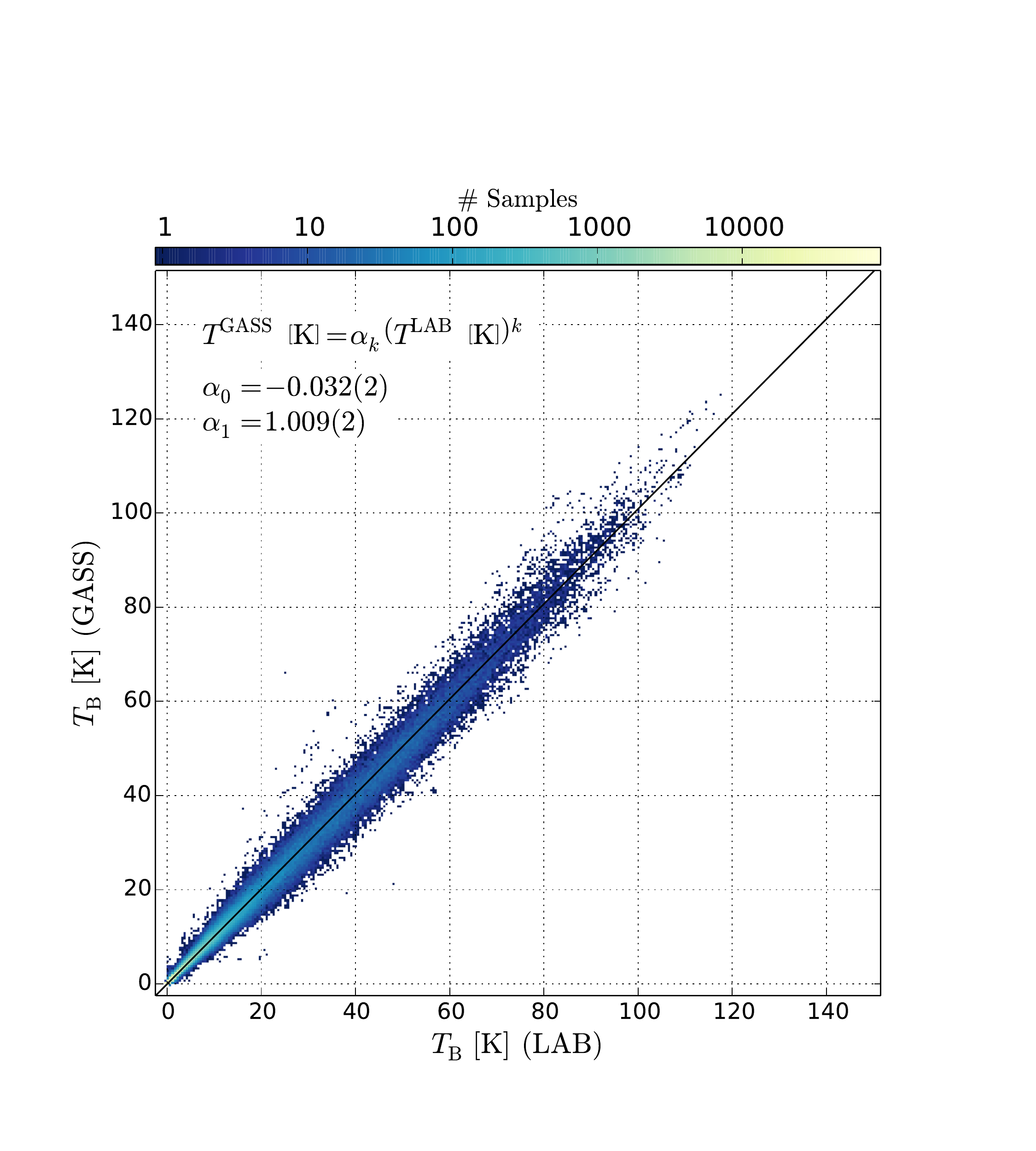}~
\includegraphics[width=0.32\textwidth,viewport=35 54 555 605,clip=]{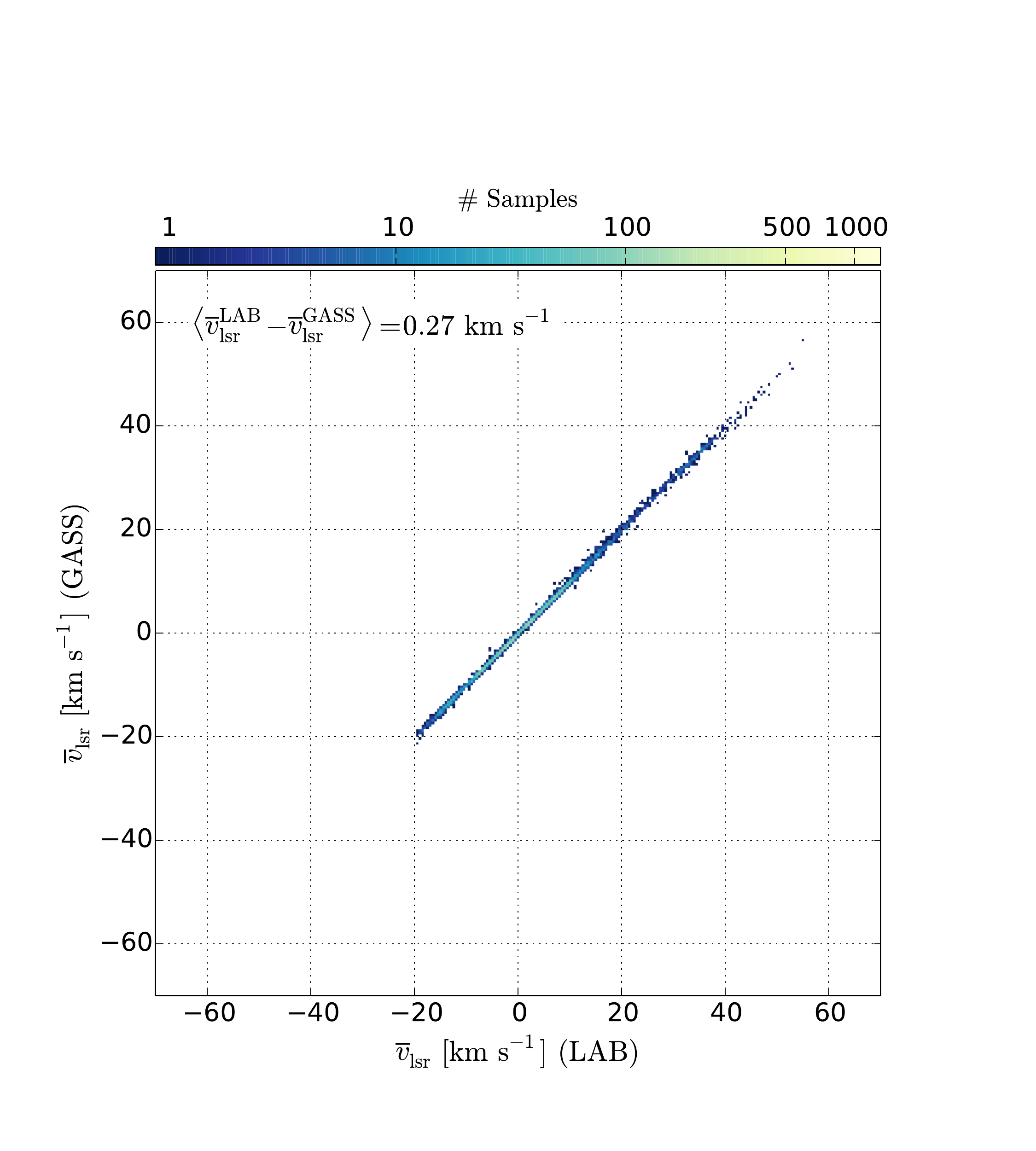}\\[0ex]
\caption{Comparison between EBHIS, GASS, and LAB in the overlap region $-4.5\degr\leq\delta\leq-0.2\degr$. Left panels: column densities; center panels: brightness temperatures; and right panels: intensity-weighted velocities (Moment-1). }%
\label{fig:ebhis_vs_gass_vs_lab_16.2}%
\end{figure*}

\begin{figure*}[!t]
\centering%
\includegraphics[width=0.32\textwidth,viewport=35 54 555 605,clip=]{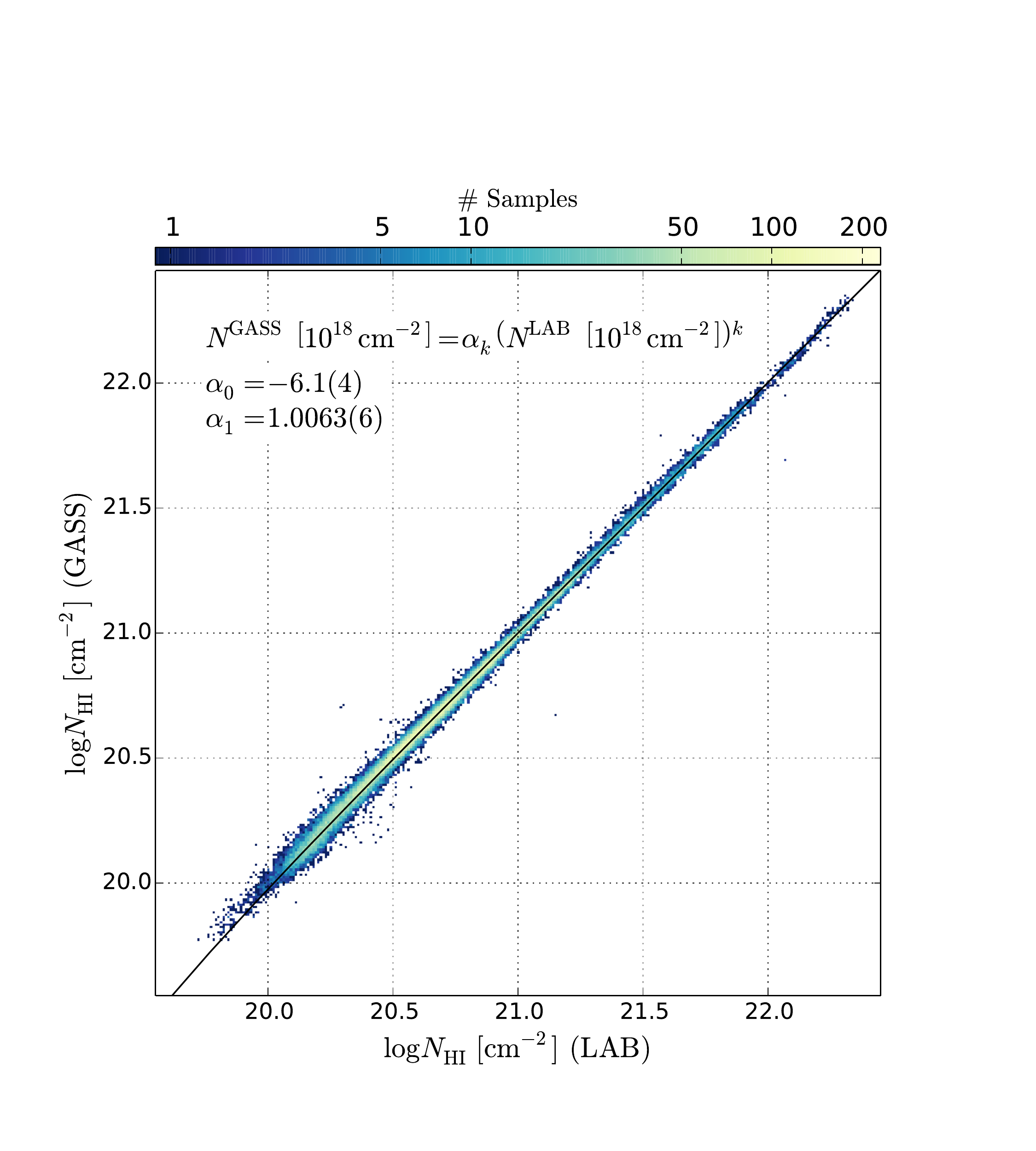}~
\includegraphics[width=0.32\textwidth,viewport=35 54 555 605,clip=]{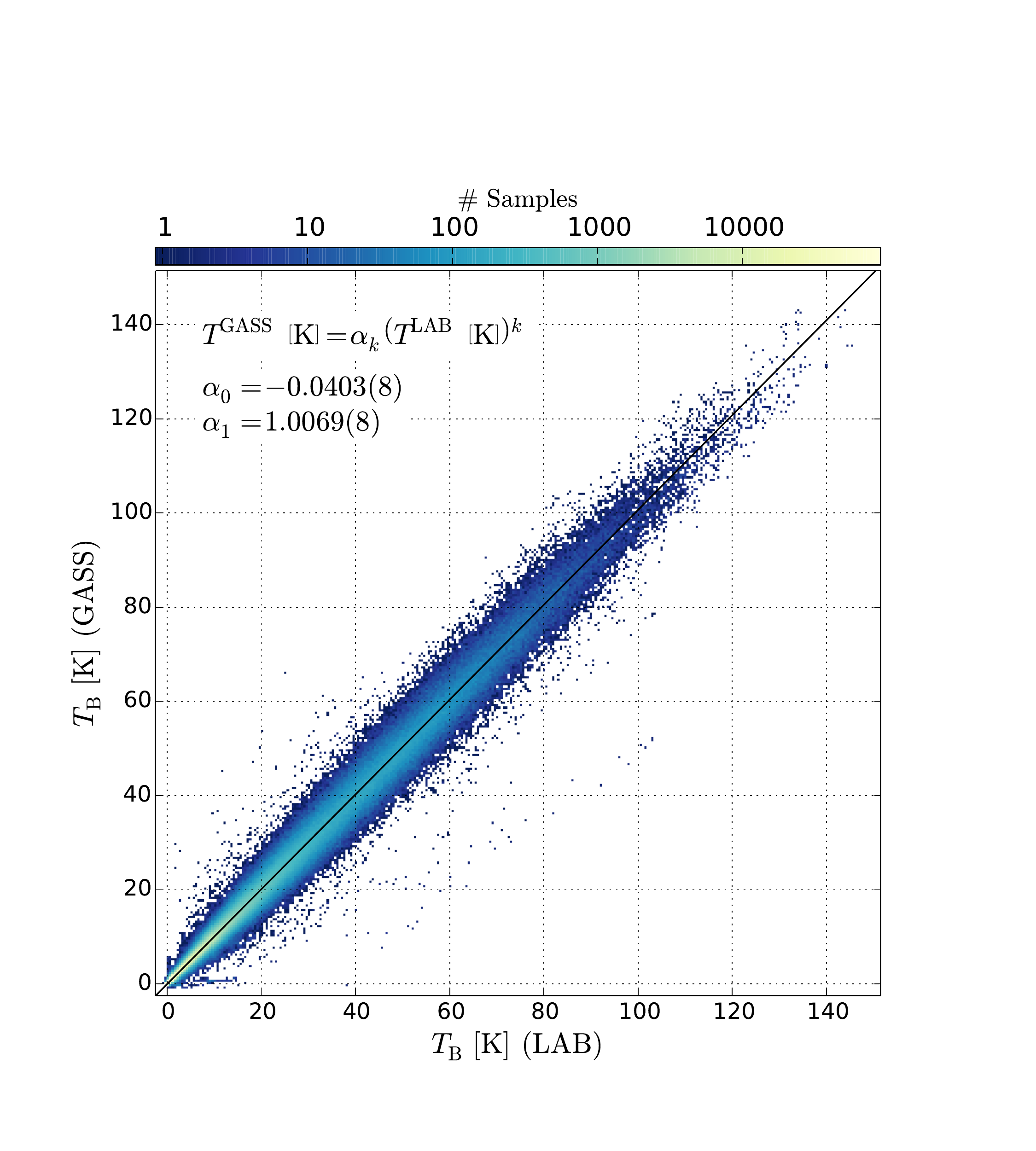}~
\includegraphics[width=0.32\textwidth,viewport=35 54 555 605,clip=]{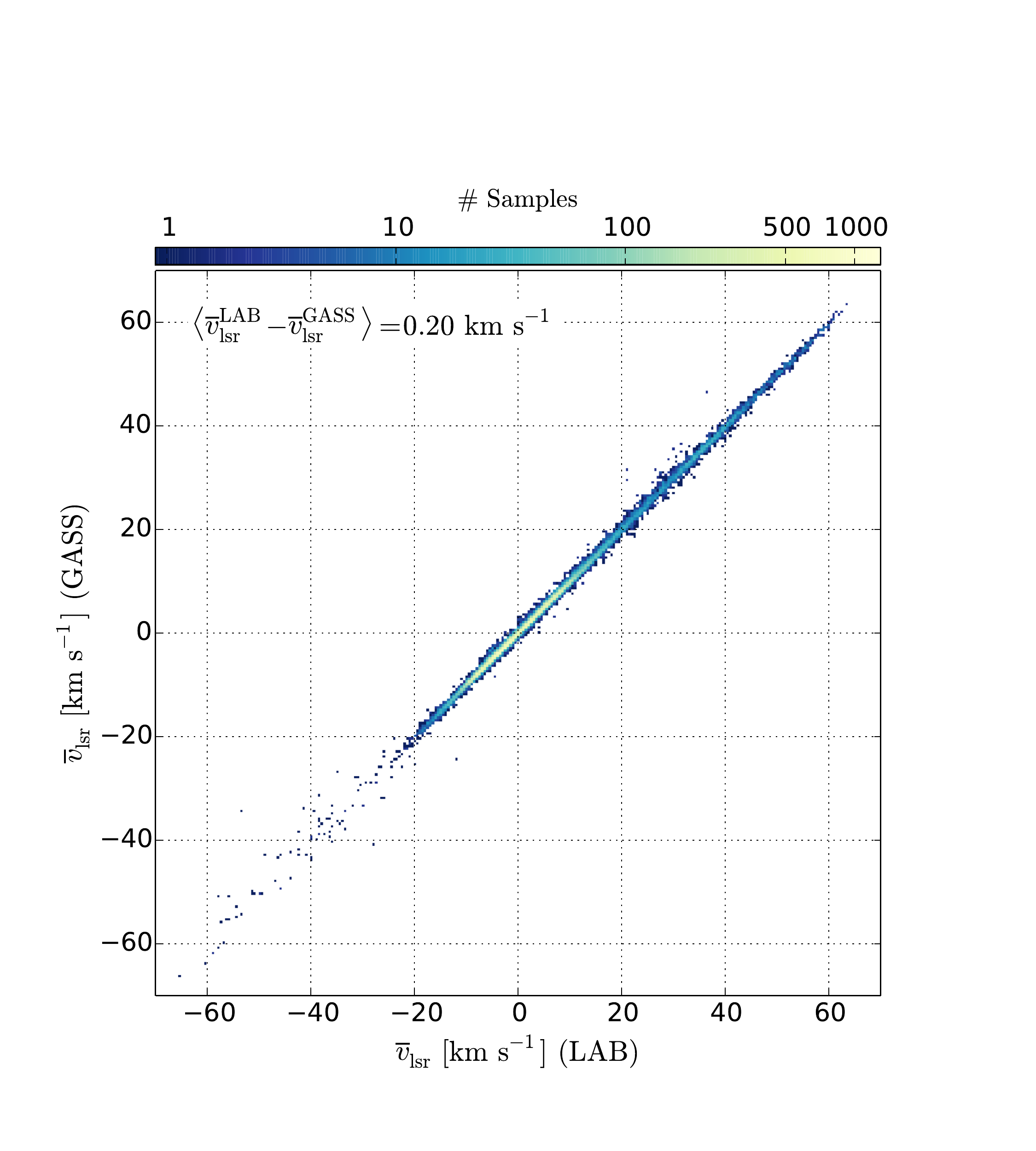}\\[0ex]
\caption{Comparison between GASS and LDS (LAB with $\delta>-30\degr$). \textit{Left panel:} column densities; \textit{center panel:} brightness temperatures; and \textit{right panel:} intensity-weighted velocities (Moment-1).}%
\label{fig:gass_vs_lds_16.2}%
\end{figure*}

\begin{figure*}[!t]
\centering%
\includegraphics[width=0.32\textwidth,viewport=35 54 555 605,clip=]{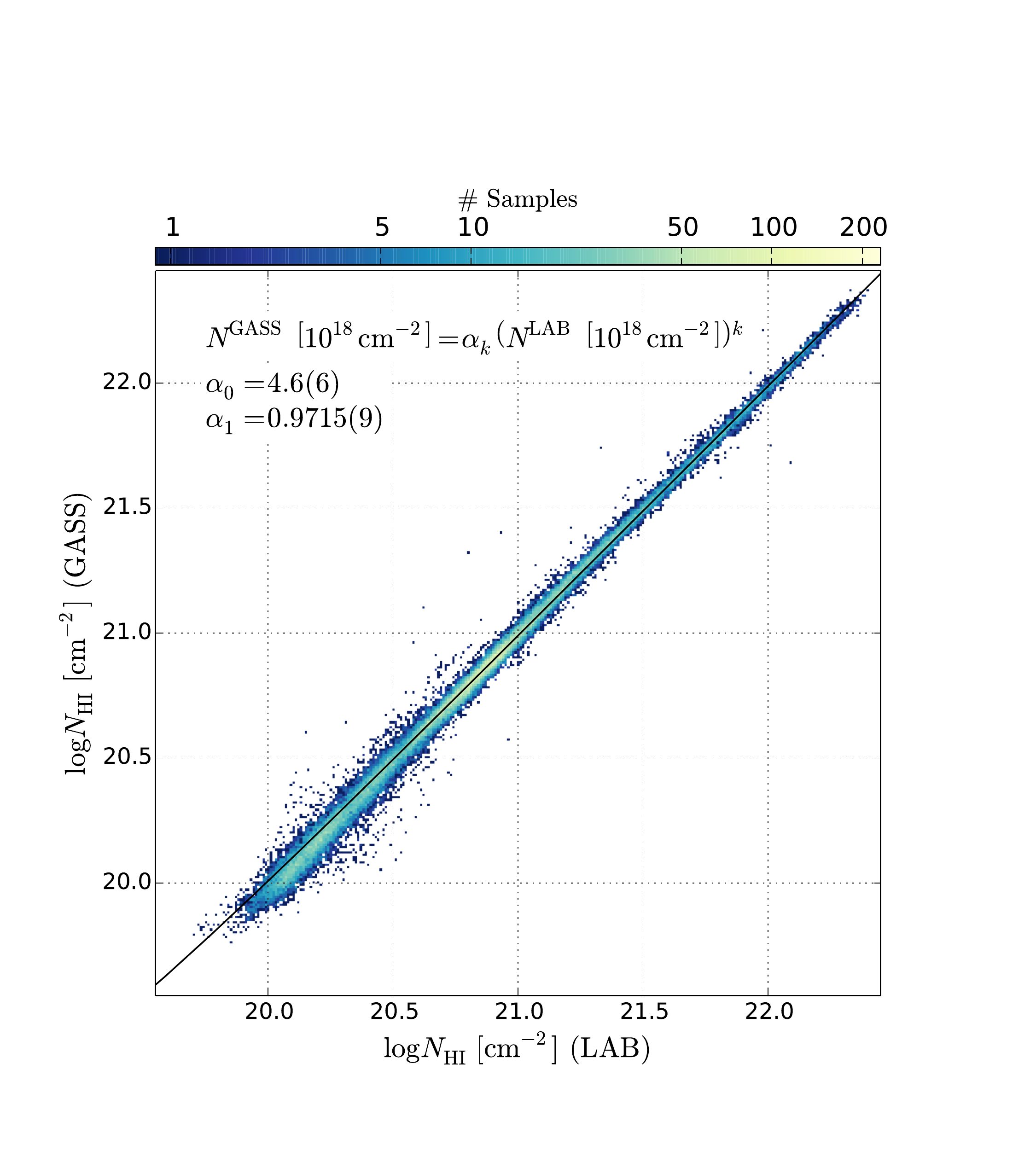}~
\includegraphics[width=0.32\textwidth,viewport=35 54 555 605,clip=]{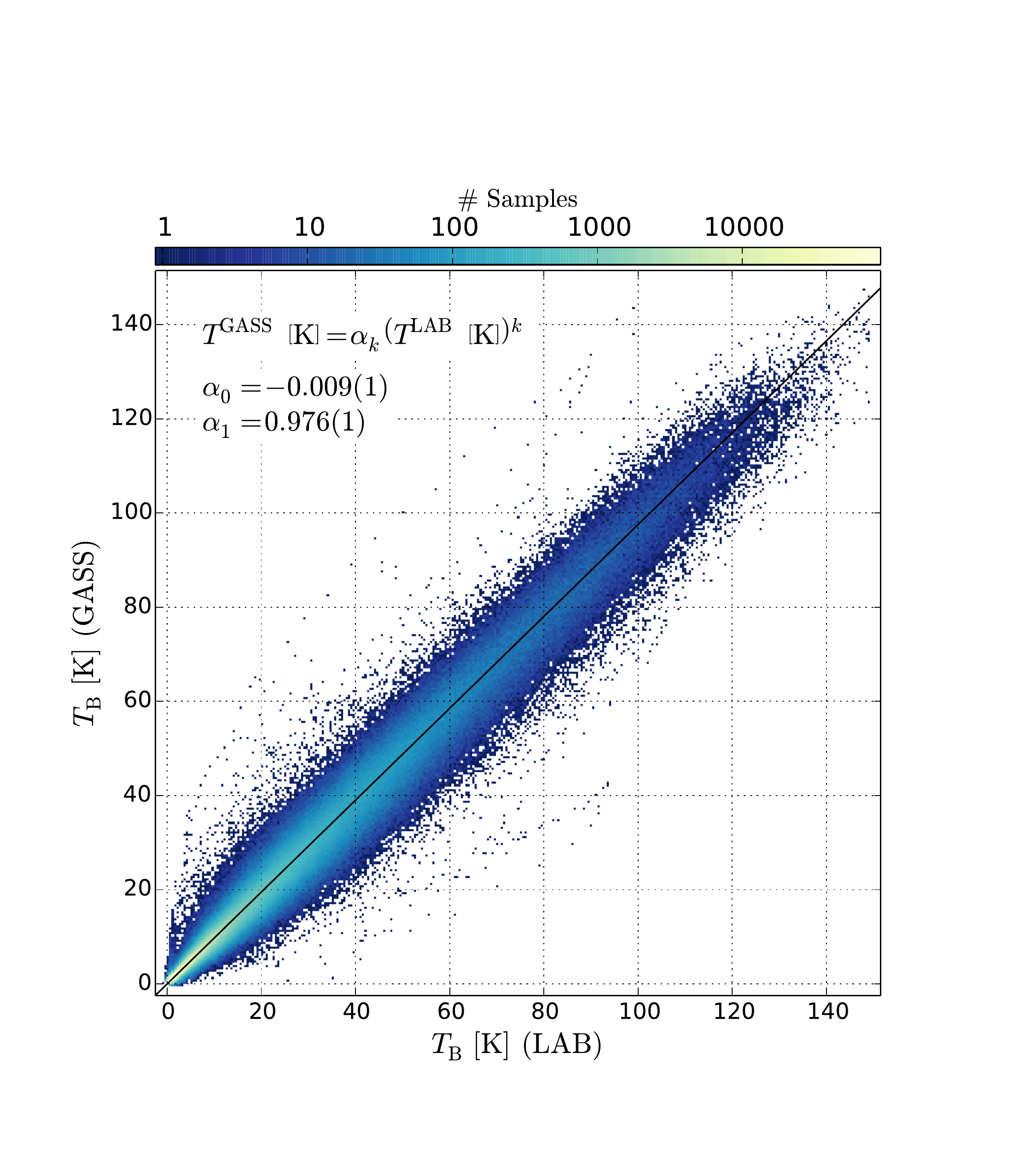}~
\includegraphics[width=0.32\textwidth,viewport=35 54 555 605,clip=]{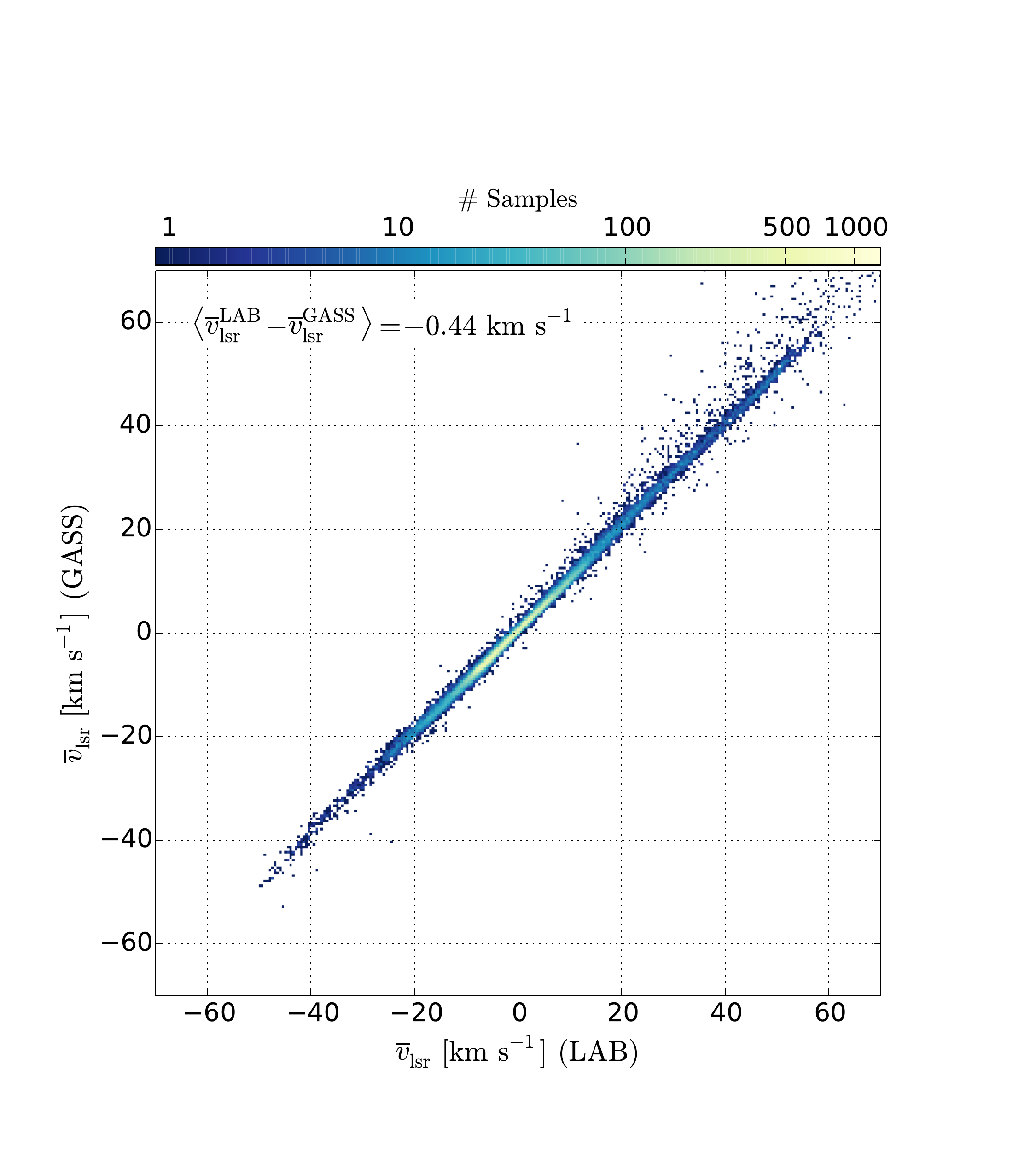}\\[0ex]
\caption{Comparison between GASS and IAR survey (LAB with $\delta<-30\degr$). \textit{Left panel:} column densities; \textit{center panel:} brightness temperatures; and \textit{right panel:} intensity-weighted velocities (Moment-1).}%
\label{fig:gass_vs_iar_16.2}%
\end{figure*}

In Figs.~\ref{fig:ebhis_vs_gass_vs_lab_16.2} to \ref{fig:gass_vs_iar_16.2} we present linear correlation diagrams for consistency checks of the intensity calibration (for the spectral integral, i.e., $N_\ion{H}{i}$ value, and brightness temperatures per voxel) and velocity axis accuracies. To avoid the side effects caused by the insufficient spatial sampling in LAB data, we took the original pointing positions of LAB and regridded EBHIS and GASS data to match. There is a common slice close to zero declination, $-4.5\degr\leq\delta\leq-0.2\degr$, where all three surveys provide data. For the comparison of LAB and GASS we split LAB into its constituents LDS (northern part, $\delta>-27\fdg5$) and IAR survey ($\delta<-27\fdg5$). Both subsamples show different behavior in terms of intensity calibration and velocity zero point.

\end{document}